\documentclass{aa}
\usepackage{graphicx}
\usepackage{latexsym}
\usepackage{color}
\usepackage{ulem}

\usepackage[utf8]{inputenc}
\usepackage[english]{babel}

\usepackage[dvipsnames]{xcolor}
\usepackage{amssymb}
\usepackage{longtable}    
\usepackage{lscape}
\usepackage{url}
\usepackage{natbib}
\bibpunct{(}{)}{;}{a}{}{,} % to follow the A&A style
\newcommand{\lamost}{{\sf LAMOST}}

\newcommand{\kepler}{{\it Kepler}}
\newcommand{\ktwo}{{\it K2}}
\newcommand{\degree}{$^{\circ}$}
\newcommand{\teff}{$T_{\rm eff}$}

\newcommand{\logg}{$\log g$}
\newcommand{\vsini}{$v\sin i$}

\newcommand{\feh}{[Fe/H]}

\newcommand{\rotfit}{{\sf ROTFIT}}

\newcommand{\lasp}{{\sf LASP}}
\newcommand{\kms}{km\,s$^{-1}$}
\newcommand{\halpha}{H$\alpha$}
\newcommand{\Whalpha}{$W^{\rm res}_{\rm H\alpha}$}
\newcommand{\WLi}{$W_{\rm Li}$}
\newcommand{\gaia}{{\it Gaia}}
\newcommand{\tess}{{\it TESS}}
\newcommand{\prot}{$P_{\rm rot}$}
\newcommand{\rha}{$R^{'}_{\rm H\alpha}$}
\newcommand{\fha}{$F_{\rm H\alpha}$}
\newcommand{\erg}{erg\,cm$^{-2}$s$^{-1}$}

\newcommand{\eagles}{{\sf EAGLES}}

\begin{document}

\title{LAMOST medium-resolution observations of the Pleiades}

\author{A. Frasca\inst{1}\and 
        J.~Y. Zhang\inst{2,3,1}\and
        J. Alonso-Santiago\inst{1}\and
        J.~N. Fu\inst{2,3}\and
        J. Molenda-\.Zakowicz\inst{4}\and         
        P. De Cat\inst{5}\and
        G. Catanzaro\inst{1}
 }

\offprints{A. Frasca\\ \email{antonio.frasca@inaf.it}}

\institute{INAF - Osservatorio Astrofisico di Catania, via S. Sofia, 78, 95123 Catania, Italy
\and
Institute for Frontiers in Astronomy and Astrophysics, Beijing Normal University, Beijing~102206, P.~R.~China
\and
School of Physics and Astronomy, Beijing Normal University, Beijing~100875, P.~R.~China
\and
University of Wroc{\l}aw, Faculty of Physics and Astronomy, Astronomical Institute, ul.~Kopernika 11, 51-622 Wroc{\l}aw, Poland
\and
Royal observatory of Belgium, Ringlaan 3, B-1180 Brussel, Belgium
}

\date{Received 20/01/2025 / Accepted 07/04/2025}

% \abstract{}{}{}{}{} 
% 5 {} token are mandatory
 
\abstract 
{} %leave it empty if necessary  
{In this work we present the results of our analysis of medium-resolution LAMOST spectra of late-type candidate members of the Pleiades with the aim of determining the stellar parameters, activity level, and lithium abundance.} 
{We have used the code \rotfit\ to determine the atmospheric parameters (\teff, \logg, and \feh), radial velocity (RV), and projected rotation velocity (\vsini). Moreover, for late-type stars (\teff$\le 6500$\,K), we also calculated the \halpha\ and \ion{Li}{i}$\lambda$6708 net equivalent width by means of the subtraction of inactive photospheric templates. We have also used rotation periods available in the literature and we have purposely determined them for 89 stars by analyzing the available \tess\ photometry.}
{We have derived the RV, \vsini, and atmospheric parameters for 1581 spectra of 283 stars. Literature data were used 
to assess the accuracy of the derived parameters. 
The RV distribution of the cluster members peaks at 5.0\,\kms\ with a dispersion of 1.4\,\kms, while the average metallicity is [Fe/H]=$-$0.03$\pm$0.06, in line with previous determinations.
Fitting empirical isochrones of Li depletion to EW measures of stars with \teff$\le 6500$\,K,
we obtain a reliable age for the Pleiades of 118$\pm$6\,Myr, in agreement with the recent literature. 
The activity indicators H$\alpha$ line flux (\fha) and luminosity ratio (\rha) show the hottest stars to be less active, on average, than the coldest ones, as expected for a 100-Myr old cluster. When plotted against the Rossby number $R_{\rm O}$, our \rha\ values display the typical activity-rotation trend with a steep decay for $R_{\rm O}\geq$\,0.2 and a nearly flat (saturated) activity level for smaller values.  
However, we still see a slight dependence on $R_{\rm O}$ in the saturated regime which is well fitted by a power law with a slope of $-0.18\pm0.02$, in agreement with some previous work. 
For three sources with multi-epoch data we have found \lamost\ spectra acquired during flares, which are characterized by strong and broad H$\alpha$ profiles and the presence of the \ion{He}{i}\,$\lambda$6678\AA\ emission line.   
Among our targets we identify 39 possible SB1 and ten SB2 systems.
We have also shown the potential of the \lamost-MRS spectra, which allowed us to refine the orbital solution of some binary and to discover a new double-lined binary.
}
{}

% MAXIMUM 6 KEYWORDS !! 
   \titlerunning{The Pleiades cluster as seen by LAMOST}
      \authorrunning{A. Frasca et al.}

\maketitle

\section{Introduction}
\label{Sec:Intro}

Open clusters (OCs) are the result of the gravitational collapse of molecular clouds, which leave, after the gas dispersal occurring during the first $\approx$\,100 Myr, stellar populations with homogeneous ages and chemical compositions \citep[see][for a recent review]{Krumholz2019}.
In the course of their evolution, stars belonging to open clusters are gradually dispersed within the Galaxy by tidal disruption and N-body evaporation, becoming field stars. Therefore, open clusters can be considered as the building blocks of our and other galaxies.
The members of OCs, which are bound by mutual gravitation, 
provide ideal samples for studying stellar properties like rotation and  magnetic activity as a function of the stellar mass, unaffected by variations in age or metallicity \citep[e.g.,][]{Barnes2003,Fang2018,Frasca2015}. 
Young OCs, particularly those of $\approx$100\,Myr, have already removed most of the gas and recently stabilized in nuclear fusion processes, making them excellent candidates for these studies.

Among the OCs of the Milky Way, the Pleiades hold a unique position, serving as an ideal laboratory in which to investigate stellar evolution. Its proximity to Earth ($\approx$\,130\,pc) and the estimated age of 125 Myr, based on the lithium depletion boundary \citep{Stauffer1998}, make it the closest young OC and a prototype for studying young solar-type stars in cluster environments. In this regard, the works of \citet{Soderblom1993c, Soderblom1993d}, based on the analysis of high-resolution spectra of a relevant number of members, are of fundamental importance for the study of the rotation, activity and abundance of lithium. 

Research on the Pleiades cluster over the past few decades has advanced our understanding of the relationships between stellar rotation, magnetic activity, and radius inflation. Early studies identified correlations between radius inflation and magnetic activity \citep[e.g.,][]{Torres2006, Morales2007,Morales2008,Stassun2012}, which was further investigated using theoretical models \citep[e.g.,][]{Feiden2013,Feiden2014,Jackson2014a,Jackson2014b,Somers2014,Somers2015b,Somers2015a}. \cite{Barrado2016} found a strong connection between lithium depletion, rotation, and activity in Pleiades F, G, and K stars, particularly for stars with luminosities in the range $0.5$--$0.2 \,L_\mathrm{\odot}$. \cite{Cao2022} measured starspot filling fractions for 240 stars in the Pleiades, showing that active stars reach a saturation level of $0.248 \pm 0.005$, with slower rotators showing a decline. These theoretical insights set the foundation for future observational studies. For example, \cite{Somers2017} employed a fitting of the spectral energy distribution to confirm the magnetic origin of the radius inflation in the Pleiades by analyzing 83 stars, demonstrating a clear connection between rapid rotation, magnetic activity, and enhanced lithium abundance. Building on this, \cite{Jackson2018} used maximum likelihood modeling to investigate the inflated radii of low-mass stars in the Pleiades, highlighting the roles of magnetic inhibition of convection and stellar spots in driving radius inflation, in comparison with earlier findings by \cite{Rebull2016}.

More recent research has expanded on these foundational studies by incorporating new data and methodologies. \cite{Wanderley2024} analyzed APOGEE spectra to measure the magnetic fields of 62 M-type dwarfs in the Pleiades, revealing a strong correlation between magnetic field strength, rotation, and radius inflation. Other investigations have broadened the focus to stellar properties and dynamics. For instance, \cite{Heyl2022} utilized $Gaia$ EDR3 data to identify 289 stars that have escaped the cluster, including three white dwarfs, providing insights into the cluster's dynamical evolution. Additionally, \cite{Alfonso2023} identified 958 cluster members using \gaia\ DR3 data \citep{GaiaDR3}, confirming the consistency of the cluster's distance, age, and metallicity with previous studies \citep{gaia2018, Bossini2019}. These efforts, along with studies on lithium-rotation correlations \citep{Bouvier2018}, spot activity \citep{Fang2016}, and the influence of cool starspots on stellar evolution \citep{Guo2018}, continue to deepen our understanding of stellar evolution in young OCs. 

The Large Sky Area Multi-Object Fiber Spectroscopic Telescope \citep[\lamost;][]{Cui12} is a National Major Scientific Project undertaken 
by the Chinese Academy of Science. It is a unique instrument, located at the Xinglong station and situated south of the main peak of the 
Yanshan mountains in Hebei province (China).\ \lamost\  combines a large aperture (4-meter telescope) with a wide field of view 
(circular region with a diameter of 5 degrees on the sky) that is covered by 4000 optical fibers. These fibers are connected to 16 
multi-object optical spectrometers with 250 fibers each \citep{wang1996,xing1998}, making this instrument the ideal tool for obtaining 
spectroscopic observations for a large number of targets in an efficient way. The data acquired with the \lamost\ instrument allow 
multi-fold analyses of the observed objects to be conducted, including a homogeneous determination of the atmospheric parameters (APs): 
the effective temperature \teff, surface gravity \logg, metallicity \feh, as well as the radial velocity RV and projected rotational velocity \vsini.
Leveraging unique LAMOST medium-resolution spectroscopic data, this study aims to address critical questions about stellar activity, rotation, and lithium abundance in young stars.

This paper is organized as follows. Section \ref{Sec:Data} presents the selection of the targets that make up our sample and the photometric and spectroscopic data used in this work. In Sect. \ref{Sec:Analysis} we describe the data analysis performed with the code \rotfit. 
We also discuss the data quality by comparing our results with values from the literature. Section \ref{Sec:Prot} concerns the rotation periods used in this work, part of which we derived by analyzing {\it TESS} light curves.
In Sect. \ref{Sec:chrom_lithium} we introduce the procedure for measuring activity indicators and lithium abundance. 
Section~\ref{Sec:Flares} shows the multi-epoch monitoring of some stars thanks to which 
remarkable flares events were detected for a few sources while the age of the cluster is discussed in detail in Sect.~\ref{Sec:age}.
The analysis of RV curves for some spectroscopic binaries present in our sample is carried out in Sect. \ref{Sec:SB}. Finally, we provide a summary of our findings in Sect. \ref{Sec:Concl}.

%====================================================================
\section{Observations and sample selection}
\label{Sec:Data}

\begin{figure*}[ht]
\begin{center}
\includegraphics[width=6.15cm,viewport= 30 0 380 330]{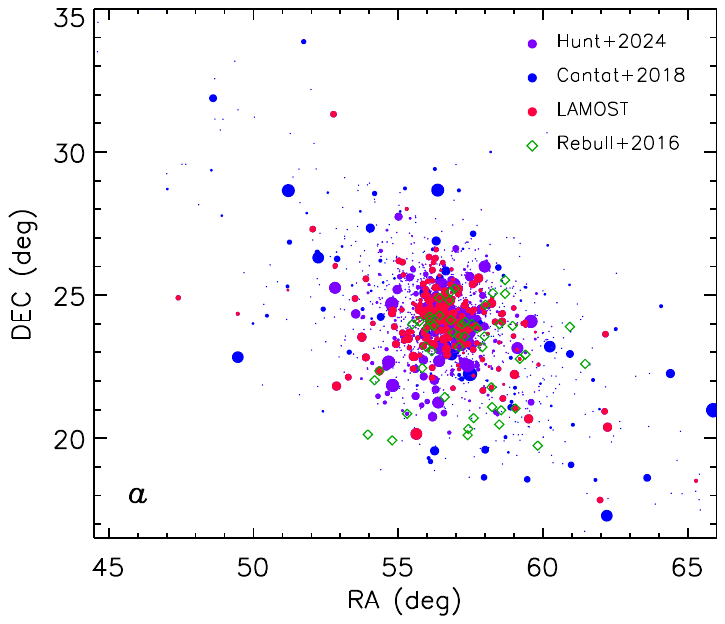}
\hspace{-.4cm}
\includegraphics[width=6.45cm,viewport= 10 0 380 330]{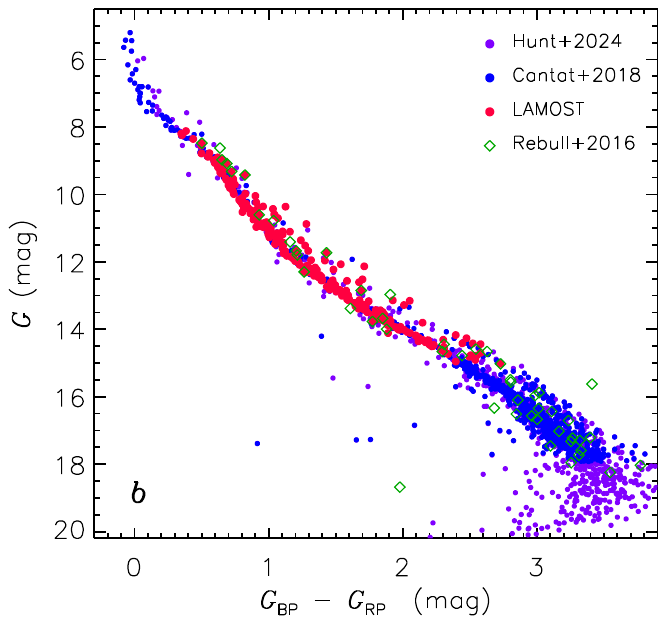}
\hspace{-.7cm}
\includegraphics[width=6.45cm,viewport= 10 0 380 330]{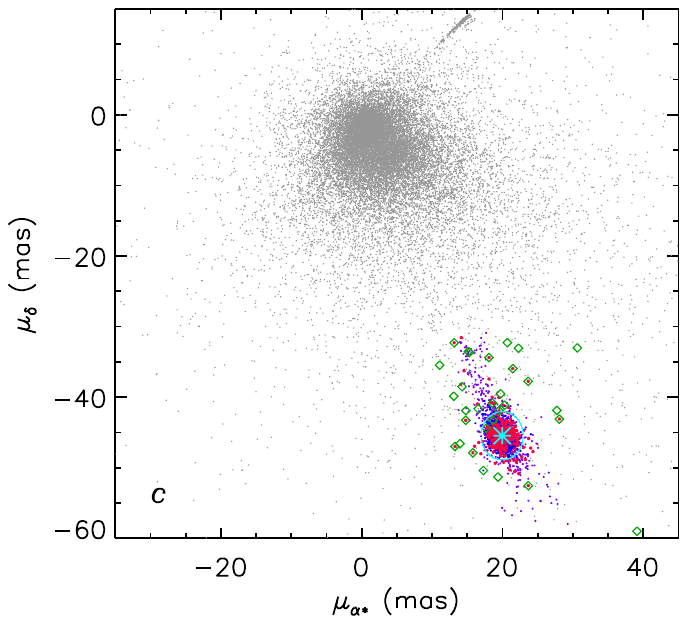}
\caption{Left panel: Spatial distribution of the Pleiades members according to \citet[][purple dots]{Hunt2024} and \citet[][blue dots]{Cantat2018}. The symbol size scales with the $G$ magnitude.
The green diamonds mark the \ktwo\ variable stars in  \citet{Rebull2016} that are not members according to either \citet{Hunt2024} or \citet{Cantat2018}. Red dots represent the stars observed with LAMOST MRS. The meaning of the symbols is also indicated in the legend.
Middle panel: Color--magnitude diagram of the same sources. Right panel: Proper motion diagram of all the \gaia\ DR3 sources with $G\leq 15$\,mag (small grey dots) in the field of the Pleiades (center coordinates RA(2000) = $03^{\rm h}46^{\rm m}$, DEC(2000) = $+24\degr06\arcmin$, radius 4$\degr$). The Pleiades members are highlighted with the same symbols as in the other panels.
The cyan asterisk denotes the average proper motion of the cluster according to \citet{Hunt2024} and the cyan ellipse is the 3$\sigma$ contour.
}
\label{Fig:Sky_CMD_PPM}
\end{center}
\end{figure*}

\subsection{Sample selection}
\label{Subsec:Sample}

The first step of our study is the selection of bona-fide members of the Pleiades. To this aim, we have used the list of 1061 members selected by  \citet{Cantat2018} on the basis of their astrometric and photometric properties. They applied an unsupervised membership assignment code (UPMASK) to the \gaia\ DR2 data down to $G=18$\,mag in the fields of a large number of OCs, including the Pleiades. 
Another important source of information for our study was the new catalog of members of galactic OCs compiled by \citet{Hunt2024}, who used a similar approach, but with a different algorithm applied to the \gaia\ DR3 data down to magnitude $G\approx 20$\,mag. 
They found 1721 Pleiades members, 1042 of which are in common
with \citet{Cantat2018}. The higher number of sources in \citet{Hunt2024}, apart from the different algorithm, is mostly the result of a larger sampled sky area and the deeper limiting magnitude. Their cluster membership lists include many new members of the already known clusters and often encompass tidal tails, as is likely the case of the Pleiades \citep[e.g.,][]{Dinnbier2020}.  
Another important source of information is the list of the 759 Pleiades members observed with \kepler-\ktwo\ space mission for which \citet{Rebull2016} reported periods and variation amplitudes. Many of these stars (548) are also members according to \citet{Cantat2018} and 709 are in the \citet{Hunt2024} catalog. Due to the different selection criteria adopted by \citet{Rebull2016}, further 50 sources are not included in any of the previous catalogs.
We compiled a final list of 1790 likely members selected from at least one of the aforementioned catalogs, which results from adding to the 1721 members of \citet{Hunt2024} the 19 \citet{Cantat2018} members not included in the previous catalog and the 50 \citet{Rebull2016} additional sources not contained in either of the two previous lists.
Figure~\ref{Fig:Sky_CMD_PPM} shows the spatial distribution, the position on the color-magnitude diagram (CMD), and the proper motion diagram. We have used different symbols to designate the stars classified in different papers as members of the Pleiades.
We remark that we have updated the photometric and kinematic data of the investigated sources with the \gaia\ DR3  in this figure and in the rest of the paper. 
We note a handful of candidate members that lie far below the cluster sequence in the CMD. However, as shown by the red dots, none of them have been observed with \lamost-MRS, while some members lie just above the main sequence (MS), but close to the binary sequence. Regarding their kinematic properties, almost all stars observed with \lamost-MRS fall within the 3$\sigma$ ellipse of the proper motion (PPM) distributions defined by \citet{Hunt2024} or just outside it. The largest scatter in the PPM diagram is shown by some members of \citet{Hunt2024} that outline a tail-like structure and by the few candidates of \citet{Rebull2016} that are not included in the other two lists. The latter sources can be considered low-probability members or probable contaminants. However, very few of them ($\approx$\,10) have been observed with \lamost-MRS and are included in our analysis.

\subsection{Spectroscopy}
\label{Subsec:Spectra}

\lamost\ started a five-year medium-resolution spectroscopic survey \citep[MRS, $R \sim 7500$, $4950\,\mathrm{\mathring{A}}<\lambda<5350 \, \mathrm{\mathring{A}}$ (blue arm) and $6300\,\mathrm{\mathring{A}}<\lambda <6800\,\mathrm{\mathring{A}}$ (red arm);][]{Liu2020} in 2017 September, after a five-year low-resolution spectroscopic survey (LRS; $R \sim 1800$, $3800 \, \mathrm{\mathring{A}} < \lambda < 9000 \, \mathrm{\mathring{A}}$). Each arm of each of the 16 spectrographs uses a 4K$\times$4K EEV CCD with 12 $\mu m$ square pixels as a detector. In the blue arm, the CCD pixel size corresponds to a sampling $\Delta\lambda\simeq 0.12\,\mathrm{\mathring{A}}$, while in the red arm, it corresponds to about $0.15\,\mathrm{\mathring{A}}$ \citep{Cui12,Wang2019}. 
\lamost\ MRS DR11, which we use in this paper, contains the data obtained from 2017 September to 2023 June (about 46 million spectra) and is currently available to the Chinese astronomical community only. The APs resulting from the automatic analysis of the raw spectra with the \lasp\ pipeline \citep{Luo2015,Wang2019}, implemented specifically for \lamost, are given in the \lamost\ MRS Parameter Catalog of DR11$\footnote{{https://www.lamost.org/dr11/v1.0/}}$.

The cross-match of the \citet{Cantat2018} sample with the \lamost\ DR11 catalog, adopting a radius of 3$\farcs$7 on the basis of the fiber pointing precision (0$\farcs$4) and the 3$\farcs$3 diameter of the fiber \citep[e.g.,][]{Zong2018}, produced 260 targets with MRS spectra, all in common with \citet{Hunt2024}, 233 of which have spectra with a sufficient signal-to-noise ratio per pixel in at least one arm (S/N$\geq$5) to be worthy of analysis.
None of the 19 Pleiades members according to \citet{Cantat2018} but not recovered by \citet{Hunt2024} was observed by \lamost\ MRS.
A further 37 \citet{Hunt2024} members have \lamost\ MRS data of sufficient quality. Among the additional 50 \citet{Rebull2016} members, 15 were observed with \lamost\ MRS and the spectra of 13 of them had a sufficient S/N for being analyzed with our tools.
We have also considered them in our analysis, which is therefore based on a total sample of 283 Pleiades members or candidates, which are highlighted with red dots in Fig.~\ref{Fig:Sky_CMD_PPM}. Due to the different selection criteria, we have also investigated in this work possible differences between these three subsamples that will be henceforth referred to as `Cantat', `Hunt' and `Rebull'.
In total we collected 1581 \lamost\ MRS co-added spectra that correspond to 283 different stars among our list of likely members of the Pleiades. We preferred to use co-added spectra, where each spectrum is the sum of all exposures obtained in a single observing night, to have the best possible S/N. Some objects were observed even in 14 different epochs.

\subsection{Photometry}
\label{Subsec:Obs_photo}

Space-born precise photometry for many targets was collected by the \kepler-\ktwo\ mission. For most of these sources the rotational periods, \prot, were derived by \citet{Rebull2016}. For the stars without \prot\ values in this catalog we have analyzed space-based photometry collected with the NASA's Transiting Exoplanet Survey Satellite ({\it TESS}; \citealt{Ricker2015}). Our targets were mostly observed by {\it TESS} in three consecutive sectors, namely 42 (from 2021-08-20 to 2021-09-16), 43 (from 2021-09-16 to 2021-10-12), and 44 (from 2021-10-12 to 2021-11-06).
The observations in these three consecutive sectors allowed us to obtain nearly uninterrupted sequences (with only a few gaps of $\approx$ 1 day) of high-precision 
photometry lasting about 77 days, with a cadence of two minutes.
For a sparse and nearby cluster like the Pleiades, there is no severe star crowding, also taken the large pixel size of {\it TESS} (21\arcsec) into account.
Therefore, for most sources, we do not expect a relevant flux contamination from nearby sources. In any case, we did not consider the data for members with a companion with a comparable (or lower) magnitude within 30\arcsec.
We downloaded the \tess\ light curves reduced by the MIT Quick Look Pipeline \citep[QLP,][]{Huang2020} from the MAST\footnote{\url{https://mast.stsci.edu/portal/Mashup/Clients/Mast/Portal.html}} archive and used the simple aperture photometry flux (SAP).

\section{Data analysis}
\label{Sec:Analysis}

We applied the code \rotfit\ \citep[e.g.,][]{Frasca2006,Frasca2015} to measure the RV, \vsini, and the APs (\teff, \logg, and \feh).
We adapted the code to fit with the \lamost\ MRS as we already did in a previous work \citep{Frasca2022}, in which thousands of MRS spectra in the \kepler\ field were analyzed.
The grid of templates consisted of high-resolution spectra of slowly rotating stars (\vsini\,$\leq3$\,\kms) with a low activity 
level that were retrieved from the ELODIE archive (R$\simeq$42,000; \citealt{Moultaka2004}). 
This is the same grid that is used in \citet{Frasca2022} and for the analysis of young stars within the \gaia-ESO survey by the OACT 
(Osservatorio Astrofisico di Catania) node \citep{Frasca2015}.
It contains spectra of 388 different stars, which sufficiently cover the space of the APs, especially at metallicity values near the solar 
ones, which are typical for the young OCs.
We prefer real-star spectra over synthetic ones because the former reproduce the photospheric features better than the latter, in which some lines may be missing or poorly reproduced due to uncertain values of strength, Land\'e factors, and broadening coefficients for the corresponding transitions. Moreover, the subtraction of the nonactive photospheric template from the target spectrum is of great help because it leaves as residuals the chromospheric core contribution and the \ion{Li}{i} line cleaned from blended neighbor lines. 

The analysis steps can be summarized as: {\rm i)} normalization of \lamost\ spectra (both the blue- and red-arm) by a fit of a low-order polynomial; {\rm ii)} 
measure of the RV by the cross-correlation with a few templates chosen from the ELODIE grid; {\rm iii)} determination of the APs and \vsini\
by $\chi^2$ minimization of the residuals of the differences observed--templates, with each template being rotationally broadened by the convolution with a linear-limb-darkened rotational profile of varying \vsini\footnote{We note that the template spectra of the ELODIE grid have been downgraded from their original resolution ($R_{\rm ELODIE}$=42,000) to that of \lamost-MRS ones ($R_{\rm MRS}=7500$) by convolution with  a Gaussian kernel of width $W=c\sqrt{1/R_{\rm MRS}^2 - 1/R_{\rm ELODIE}^2}$\,\kms.}; {\rm iv)} spectral type (SpT) classification by taking that 
of the template with the minimum $\chi^2$;  
{\rm v)} measure of the equivalent width of the emission in the H$\alpha$ core and residual absorption in the \ion{Li}{i}\,$\lambda$6708 line in the subtracted spectra.
For the details about the procedure, the reader is referred to \citet{Frasca2022}. 

For the stars with more than one \lamost-MRS spectrum we have computed, for each arm, the APs for the three best exposed spectra and averaged them using a variance-defined weight ($w_i=1/\sigma_i^2$), where $\sigma_i$ is the error of the given parameter in the $i$-th measure.  Therefore, we ended up with only two values of \teff, \logg, \feh, and \vsini\ per each target, one derived from the analysis of the blue-arm and the other of the red-arm spectra. The final parameters, which are reported in Table~\ref{Tab:APs}, are the weighted mean of the values derived from each arm, using again the variance-defined weights. Whenever the APs could only be measured in one arm (usually in the red arm) due to the low S/N or flaws in the spectrum of the other arm, we have taken these values as the final ones. 
The parameter errors calculated by \rotfit\ or the standard errors of the weighted mean, for the sources with multiple observations, have been also quoted in this table.

\setlength{\tabcolsep}{0.1cm}

\begin{table*}[ht]
  \caption{Stellar parameters of the investigated sources.}
\begin{center}
\begin{tabular}{ccclcrrccrrcr}
\hline\hline
\noalign{\smallskip}
RA   &  DEC  &  \gaia-DR3     & ~~~\teff  & \logg  &  \feh~~  &  \vsini~~   &  SpT        & Sub$^a$  &  N$^b$   & $<RV>$ & \prot$^c$ & Rem. \\ 
(J2000)  &  (J2000) &    & ~~~(K) &   &  &   (\kms) &  &  & & (\kms) & (d) &\\  
\noalign{\smallskip}
\hline
\noalign{\smallskip}
 58.615971 & 21.389748 &  \scriptsize{51742471745296768} &  5624$\pm$111 & 4.40$\pm$0.11  & -0.12$\pm$0.11  &   $\leq 8$ &     G5V  &  C  &      1  &    8.66$\pm$0.47 & 4.5480$^R$  &   ...	      \\
 59.845396 & 22.571372 & \scriptsize{53335045618385536} &   4360$\pm$ 42 & 4.66$\pm$0.04  &  0.01$\pm$0.05  &   $\leq 8$ &     K4V  &  C  &     12  &    5.35$\pm$3.59 & 1.8839$^R$  & RVvar	      \\
 53.274270 & 22.134250 & \scriptsize{61554650949438208} &  5196$\pm$ 58 & 4.54$\pm$0.08  &  0.00$\pm$0.09  &   $\leq 8$ &     K1V  &  C  &      1  &    6.87$\pm$0.62 &  5.7660$^R$  &  ...	      \\
 57.925378 & 21.668355 & \scriptsize{63730309584280960} &  5222$\pm$ 45 & 4.46$\pm$0.11  & -0.02$\pm$0.07  &   $\leq 8$ &    K0IV  &  C  &     10  &    7.04$\pm$1.33 & 5.4447$^R$  & RVvar	      \\
 58.644969 & 21.883909 & \scriptsize{63795455648004608} &  3918$\pm$ 44 & 4.65$\pm$0.05  & -0.15$\pm$0.06  &   9.0$\pm$1.7 &     K8V  &  C  &     13  &    3.56$\pm$2.16 & 2.7745$^R$  & RVvar	      \\
 57.849460 & 22.113383 & \scriptsize{63849572235829248} &   4640$\pm$ 38 & 4.62$\pm$0.05  & -0.03$\pm$0.06  &   $\leq 8$ &   K3.5V  &  C  &      2  &    6.83$\pm$0.49 &  6.4306$^R$  &  ...	      \\
 58.470090 & 22.418532 & \scriptsize{63876402894075904} &   \dots        & \dots          & \dots	      & \dots         &   \dots  &  C  &  \dots  &   \dots           &	6.3353$^R$  &   SB2 \\
 56.892334 & 21.746810 & \scriptsize{63916431989200256} &  4956$\pm$ 48 & 4.56$\pm$0.05  & -0.07$\pm$0.06  &  36.5$\pm$1.7 &   K2.5V  &  C  &      2  &    8.07$\pm$0.46 &  7.5554$^R$  &  ...	      \\
 57.399503 & 22.151421 & \scriptsize{63958801843006208} &  4052$\pm$ 74 & 4.68$\pm$0.11  & -0.03$\pm$0.10  &  66.3$\pm$2.9 &     K6V  &  C  &      1  &    3.38$\pm$1.84 & 0.2471$^R$  &   ...	      \\
\noalign{\smallskip}
\hline
\end{tabular}
\end{center}
{\bf Notes.} The full Table is available at the CDS.\\ $^{(a)}$ Subsample to which the target belongs: C\,=\,Cantat; H\,=\,Hunt; R\,=\,Rebull.$^{(b)}$ N is the number of observation epochs. $^{(c)}$ The rotation periods reported by \citet{Rebull2016} are flagged with $^R$; those measured in this work with \tess\ data are indicated with $^T$.
\label{Tab:APs}
\end{table*}

\subsection{Radial velocity}
\label{Sec:RV}

\begin{figure}[htb]
\includegraphics[width=8cm]{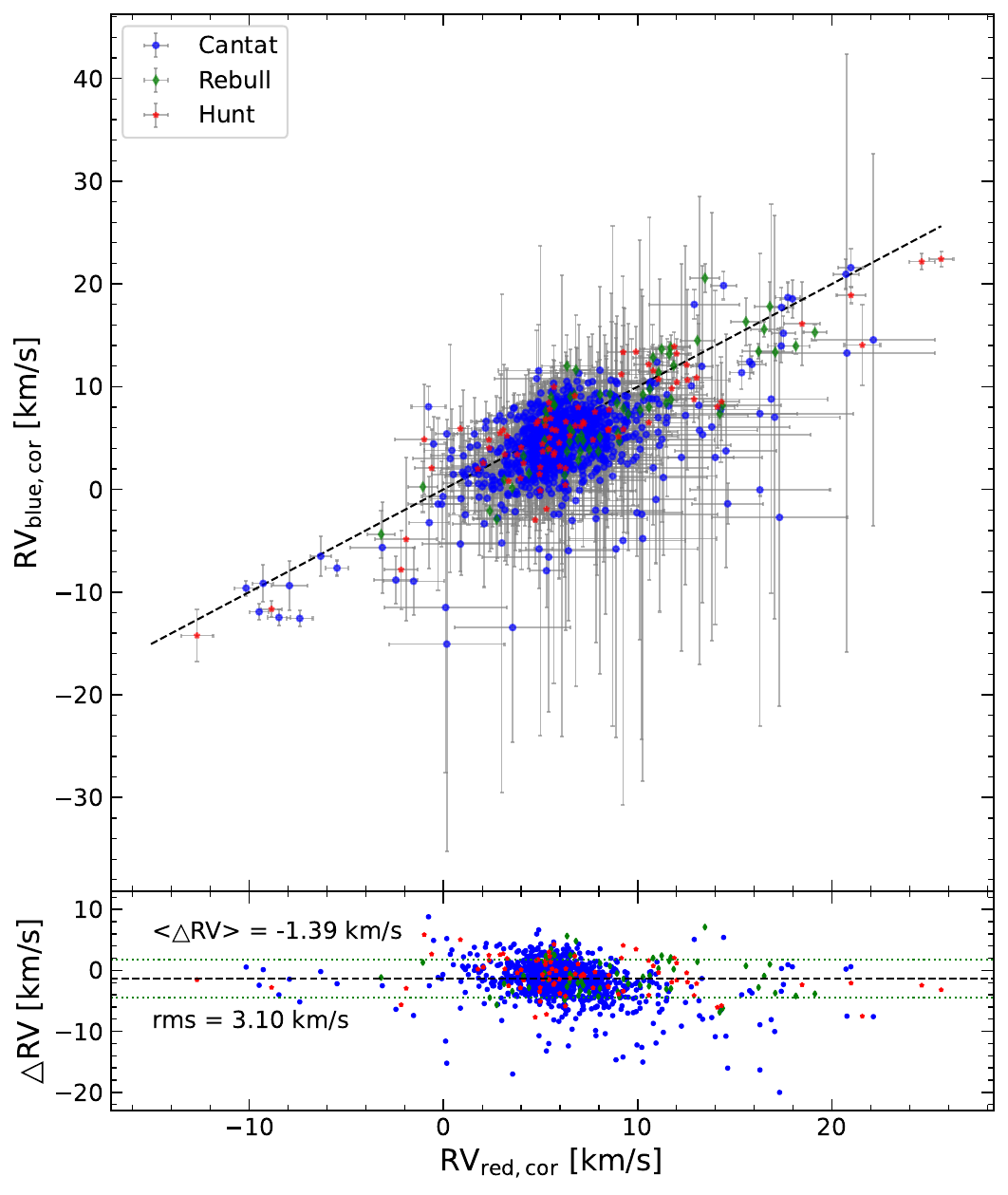}
\caption{Comparison between the RV measured in this work from the blue-arm and red-arm  \lamost\ MRS spectra (top panel). Different colors are used for the different subsamples of candidate members as indicated in the legend. The one-to-one relation is shown by the black dashed line. The RV differences (blue$-$red) displayed in the bottom panel show an average value of $-1.4$\,\kms\ (black dashed line) and a standard deviation of 3.1\,\kms\ (green dot-dashed lines).}
\label{Fig:RV_blue_red}
\end{figure}

As noted by \citet{Zong2020ApJS..251...15Z} and \citet{Frasca2022}, the RV measured on \lamost--MRS spectra can be affected by systematic offsets in different runs that are related to the wavelength calibration and can be different for the blue and red arm.  The largest offset of about 6.5 \kms\ is found for spectra acquired before May 2018 that were calibrated with Sc lamps, compared to the following ones for which Th-Ar lamps have been used. To account for these offsets and correct the RVs, we have used the spectra of RV standard stars from \citet{Soubiran2018} and \citet{Huang2018} falling in the same plates as those of our Pleiades candidates and taken at the same time (see Table~\ref{Tab:RV_Standards}).
For a few plates we found only one or no standard star. In these cases we have used the RVs of stars with $G\leq 12$\,mag in the \gaia-DR3 catalog with an RV error $\leq 0.5$\,\kms\  as reference, similar to what is done by \citet{Zong2020ApJS..251...15Z}. Although they are not RV standard stars in the strict sense, the use of a statistically significant number of them ensures a good determination of the RV offset. Indeed, we found a good agreement between the offset calculated with the RV standard stars listed in Table~\ref{Tab:RV_Standards} and those with RVs from \gaia.
We applied the proper corrections to the blue- and red-arm RVs.
The RV corrections for each plate and observing date are reported, for the blue- and red-arm spectra, in Table\,\ref{Tab:RV_corr}.

\setlength{\tabcolsep}{5pt}

\begin{table}[ht]
\caption{Radial velocity standard stars in the fields of the \lamost\ observations used in this paper.}
\begin{center}
\begin{tabular}{lrcrcc}
\hline
\hline
\noalign{\smallskip}
Star  & V~~~  &   SpT & RV~~  & $\sigma_{\rm RV}$ & Ref \\
      & (mag) &        & \multicolumn{2}{c}{(\kms)} &     \\ 
\hline
  \noalign{\smallskip}
\scriptsize{J03463726+2420366} &  12.16 &  K3V & 5.242 & 0.0111 & S18 \\
TYC1261-00573-1  & 11.48 & F8  & $-27.834$ & 0.0879 & S18 \\
TYC1800-01526-1  & 10.50 & F9  &   5.192 & 0.0220 & S18 \\
TYC1803-00234-1  &  8.85 & K2  &  15.640 & 0.0304 & S18 \\
TYC1803-00944-1  &  9.56 & G0V &   5.424 & 0.0336 & S18 \\
\scriptsize{J03435569+2425350} & 14.34 & M0e  & 5.099 & 0.0323 & H18 \\
\scriptsize{J03443541+2400048} & 14.45 & M1.0 & 6.280 & 0.0437 & H18 \\
\scriptsize{J03445017+2454400} & 13.44 & K5Ve & 5.740 & 0.0543 & H18 \\
\scriptsize{J03445895+2323202} & 13.77 & K7Ve & 5.846 & 0.0665 & H18 \\
\scriptsize{J03463532+2324424} & 14.77 & M1.7 & 6.248 & 0.0622 & H18 \\
\scriptsize{J03473800+2328051} & 13.51 & K6   & 5.630 & 0.0353 & H18 \\
\scriptsize{J03481729+2430160} & 14.33 & K7.5 & 5.249 & 0.0436 & H18  \\
\noalign{\smallskip}
\hline
\end{tabular}
\end{center}
{\bf Notes.} S18 = \citet{Soubiran2018}; H18 = \citet{Huang2018}.
\label{Tab:RV_Standards}
\end{table}

\begin{figure}[htb]
\includegraphics[width=9cm,viewport= 30 10 450 280]{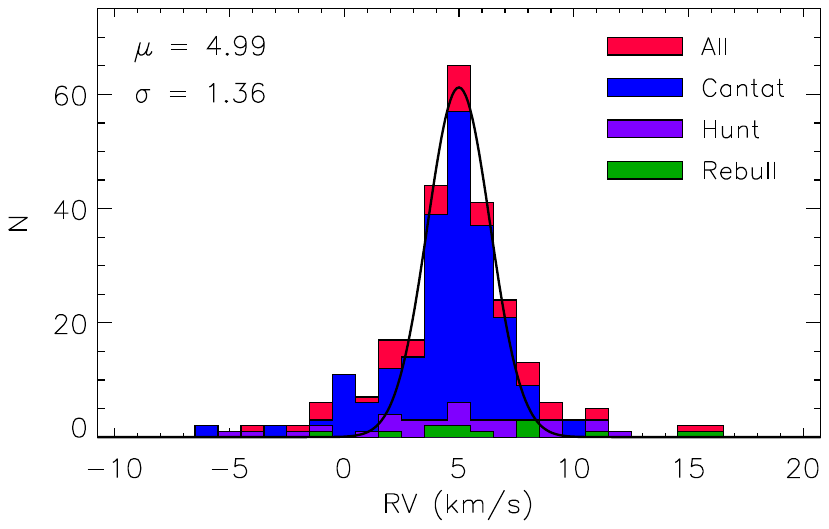}
\caption{RV distribution obtained with all the analyzed \lamost-MRS spectra of the cluster members (red histogram) and for the three subsamples, as indicated in the legend. The Gaussian fit is overplotted with a full black line; the center ($\mu$) and dispersion ($\sigma$) of the Gaussian are also marked.}
\label{Fig:RV_distr}
\end{figure}

An indicator of the precision of the RV measures that has been used in \citet{Frasca2022} is provided by the comparison of the blue- and red-arm results. In Fig.~\ref{Fig:RV_blue_red}, we compare the RV values 
obtained with the two arms, whose observations are done simultaneously, for all the spectra of the Pleiades members with an S/N$\geq10$. This suggests an independent error estimate of about 2.2\,\kms, which was derived by dividing the dispersion of the differences $\Delta RV=RV_{\rm blue}-RV_{\rm red}=3.1$\,\kms\ by $\sqrt2$. 
The rms dispersion is slightly larger than that found by \citet{Frasca2022} for a larger sample of MRS spectra in the \kepler\ field; this could be due to the quality of the spectra of the Pleiades members, which have a median S/N=74 and 32 in the red and blue arm, respectively. For comparison, the spectra of the \lamost-\kepler\ sample analyzed by \citet{Frasca2022} had a median S/N of 91 and 52 in the red and blue arm, respectively.  The errors we find in the present work are also in agreement with those of about 1\,\kms\ found by \citet{Liu2019RAA....19...75L} and \citet{Zhang2021} at S/N=20. 
In Table~\ref{Tab:RV_data} we report for all the analyzed \lamost-MRS spectra the individual RV values obtained for the blue and red arm before corrections along with their respective errors and the ``final'' RVs, which are obtained by a weighted average of the corrected blue- and red-arm values (not listed in Table~\ref{Tab:RV_data}), using $w=1/\sigma_{\rm RV}^2$ as the weight. In this table we report individual RV values, even for the stars with repeated observations, whenever S/N\,$\geq10$ in at least one arm and the peak centroid was measurable by the Gaussian fit. Whenever S/N$\,\geq10$ for one arm only (usually for the red arm where the cool stars have a higher flux) we took this value, corrected for the offset, as the final RV. 
We have chosen to report all the individual RV values in Table~\ref{Tab:RV_data} because there are objects with genuine RV variations caused by binarity or pulsations  among our sources. To spot them, we calculated the reduced $\chi^2$ and the probability $P(\chi^2)$ that the RV variations have a random occurrence \citep[e.g.,][]{Press1992}. Whenever $P(\chi^2)<0.05$ we considered the RV variation as significant and flagged the corresponding source with `RVvar' in Table\,\ref{Tab:APs}.
For the stars that are already known as single-lined spectroscopic binaries from the literature, we have added an `SB1' flag to the remarks.
With the cutoff S/N$>10$, we have RV values for 273 single-lined sources. For each of them we have calculated the weighted average of the individual values (with $w=1/\sigma_{\rm RV}^2$ as the weight), which are also reported in Table\,\ref{Tab:APs} along with the error, which is the largest between the standard error and the weighted standard deviation of individual values. With these average values, we built
the RV distribution, which is depicted as a red histogram in Fig.~\ref{Fig:RV_distr}. The center $\mu=4.99\pm0.03$\,\kms\ and the dispersion $\sigma=1.36\pm0.04$\,\kms\ of this distribution have been found by means of a Gaussian fitting. Some of the values in the wings that are in excess with respect to the Gaussian can be in part due to binaries or pulsating stars, but this could be the fingerprint of different kinematic groups. To investigate possible kinematic differences between the three subsamples, we have overplotted their RV distributions with different colors in Fig.\,\ref{Fig:RV_distr}. Despite the limited number of sources, the Hunt and Rebull subsamples show a flatter distribution compared to the Cantat one. The distribution of the  latter subsample (224 out of 273 sources) is very similar to that of the full sample.
If we merge the Hunt and Rebull subsamples (39 sources in total) and fit the corresponding RV distribution with a Gaussian we find $\mu=4.75\pm0.16$\,\kms\ and $\sigma=2.26\pm0.20$\,\kms, which indicates a similar mean RV but a  broader distribution. This is in line with the more scattered distribution displayed by these sources in the PPM diagram (Fig.\,\ref{Fig:Sky_CMD_PPM}).

For some spectra we noted two CCF peaks clearly above the noise (more than 5 times the CCF noise $\sigma_{\rm CCF}$) that we considered as significant. 
We classified the objects for which this occurred for at least one observation as double-lined spectroscopic binaries (SB2s) and flagged them accordingly in Table~\ref{Tab:APs}.
For these systems we do not provide the RVs and APs in that table.

\subsection{Atmospheric parameters and projected rotation velocity}
\label{Sec:APs}

\begin{figure}[htb]
\includegraphics[width=8cm]{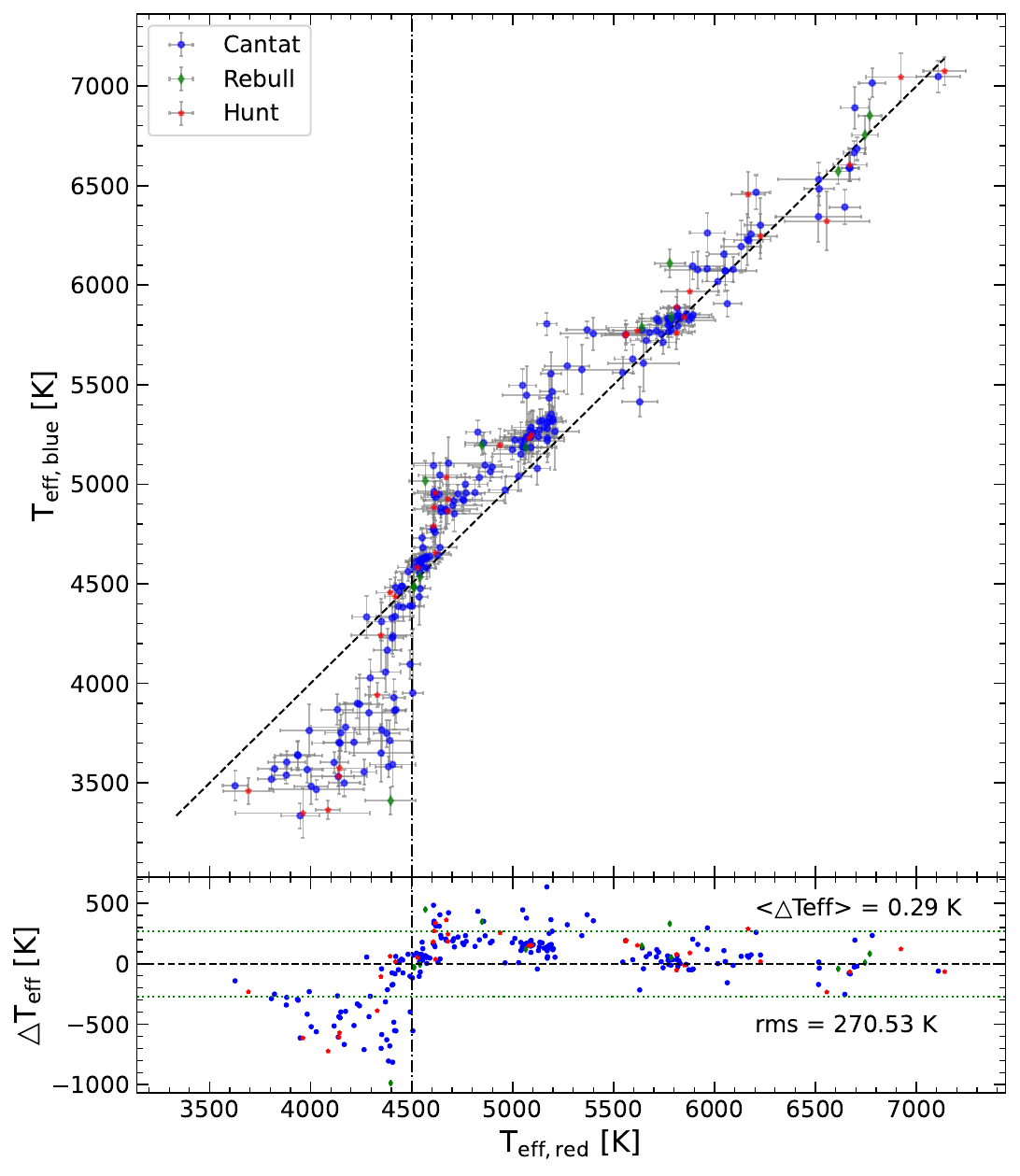}
\caption{Comparison of the \teff\ values derived from the blue- and red-arm LAMOST-MRS spectra with ROTFIT. The meaning of lines and symbols is the same as in Fig.~\ref{Fig:RV_blue_red}. We note the systematic discrepancy for \teff$_{\rm,red}\leq 4500$\,K. This boundary is given by the dash-dotted vertical line to guide the eye.}
\label{Fig:Teff_blue_red}
\end{figure}

The comparison of blue- and red-arm values for the other stellar parameters (\teff, \logg, \feh, and \vsini) allows us to get reliable estimates of their errors and to highlight specific issues with them.
As explained above, before obtaining the final parameters, we produced two sets of APs for each source from the analysis of the blue- and red-arm spectra, which can be compared as we did for the RVs.

For \teff, we observe a discrepancy between blue- and red-arm values for \teff$_{\rm,red}\leq 4500$\,K (Fig.~\ref{Fig:Teff_blue_red}). As the coldest targets in this cluster are also the faintest, we suspect this can be the result of the low S/N in their blue-arm spectra.
Another possibility is that the presence of spectral features strongly sensitive to the cold matter typical of starspots ($T\leq 4000$\,K) influences the determination of the temperature in the blue arm, leading to values lower than those of the pristine photosphere.
In fact, it has been found that effective temperatures measured in young stars from near-infrared spectra (7000--10\,000\,\AA) dominated by TiO bands, are systematically lower than those derived at optical wavelengths \citep [e.g.,][]{Cottaar2014,Flores2022}. As noted by \citet{Gangi2022}, this offset is particularly noticeable in the temperature range between about 4000 and 4500\,K, where it can even reach values of 500 K. This is the \teff\ range where the discrepancy is most evident in our data. A two-component model with synthetic spectra reproducing starspots and the photosphere is able to fit the spectral energy distribution of active pre-main sequence (PMS) stars \citep{Gangi2022} and explain this discrepancy. Such a detailed modeling for a large source number is beyond the scope of our work. 
However, when calculating the weighted average of \teff\ based on S/N, $\chi^2$, or variance, the blue-arm results usually weigh less for these cold objects, so the effect of starspots is mitigated. 
We note that neglecting stars colder than 4500\,K the dispersion decreases from 277\,K to 143\,K, which is a value in agreement with the results of \citet{Frasca2022}. These average \teff\ agree reasonably well with literature values (Sect.~\ref{Subsec:external}).

Regarding gravity, we find an overall good agreement between the blue-arm and red-arm results, with a handful of cases for which the blue-arm spectra gave rise to low gravity (\logg\,$<4.0$) when the red-arm \logg\ is
larger than 4.5 (see Fig.~\ref{Fig:logg_blue_red}). This corresponds again to cold stars with low S/N ratios in the blue arm that provided uncertain results, as also indicated by the large errors. The typical \logg\ errors derived by ROTFIT but also by this comparison (after dividing the rms scatter by $\sqrt2$) are approximately 0.15 dex.

\begin{figure}[htb]
\includegraphics[width=9cm,viewport= 30 10 450 280]{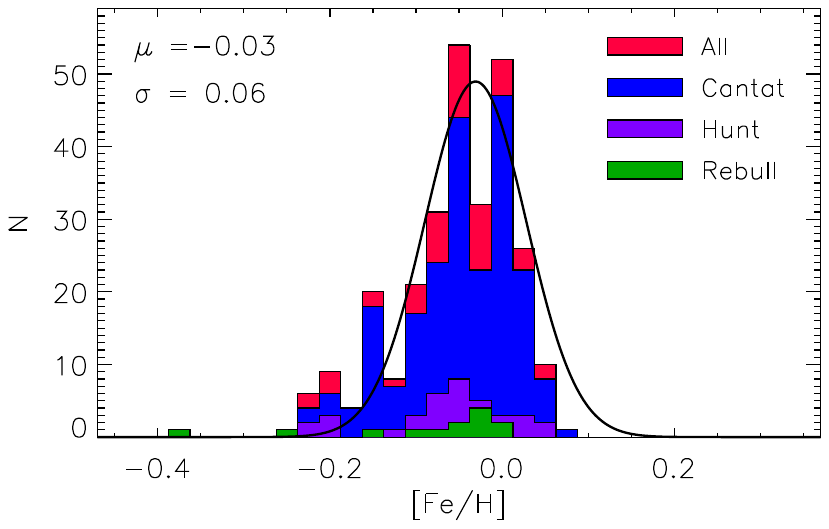}
\caption{[Fe/H] distribution for all the Pleiades members (red histogram) and for the three subsamples, as indicated in the legend. The Gaussian fit is overlaid with a  black line; the center ($\mu$) and dispersion ($\sigma$) of the Gaussian are also marked. }
\label{Fig:histo_feh_comb}
\end{figure}

As for the metallicity, the agreement of blue- and red-arm values is very good (Fig.~\ref{Fig:feh_blue_red}). As expected for the members of a young cluster, all the targets display a near-solar metallicity with very few ``extreme" values of [Fe/H]\,$\simeq-0.4$, mostly in the red arm. 
As for the other APs, the final metallicity reported in Table~\ref{Tab:APs} is the weighted average of the values measured in the blue and red arm. The [Fe/H] distribution is shown in Fig.~\ref{Fig:histo_feh_comb}, where we distinguished the different subsamples as we did for the RV. The distribution peaks at small negative values of [Fe/H] and displays a slightly asymmetric shape with a tail towards negative [Fe/H]. 
The average metallicity of the cluster, [Fe/H]\,=\,$-0.03\pm$0.06, is determined 
by fitting the [Fe/H] distribution with a Gaussian function, whose dispersion $\sigma$ has been adopted as the error of the cluster metallicity.
There is no clear distinction between the different subsamples, as also suggested by the values of $\mu$ and $\sigma$ found for the more kinematically scattered population of the Hunt and Rebull subsamples, which are the same (within the errors) as those of the Cantat subsample.  

\begin{figure}[htb]
\includegraphics[width=8cm]{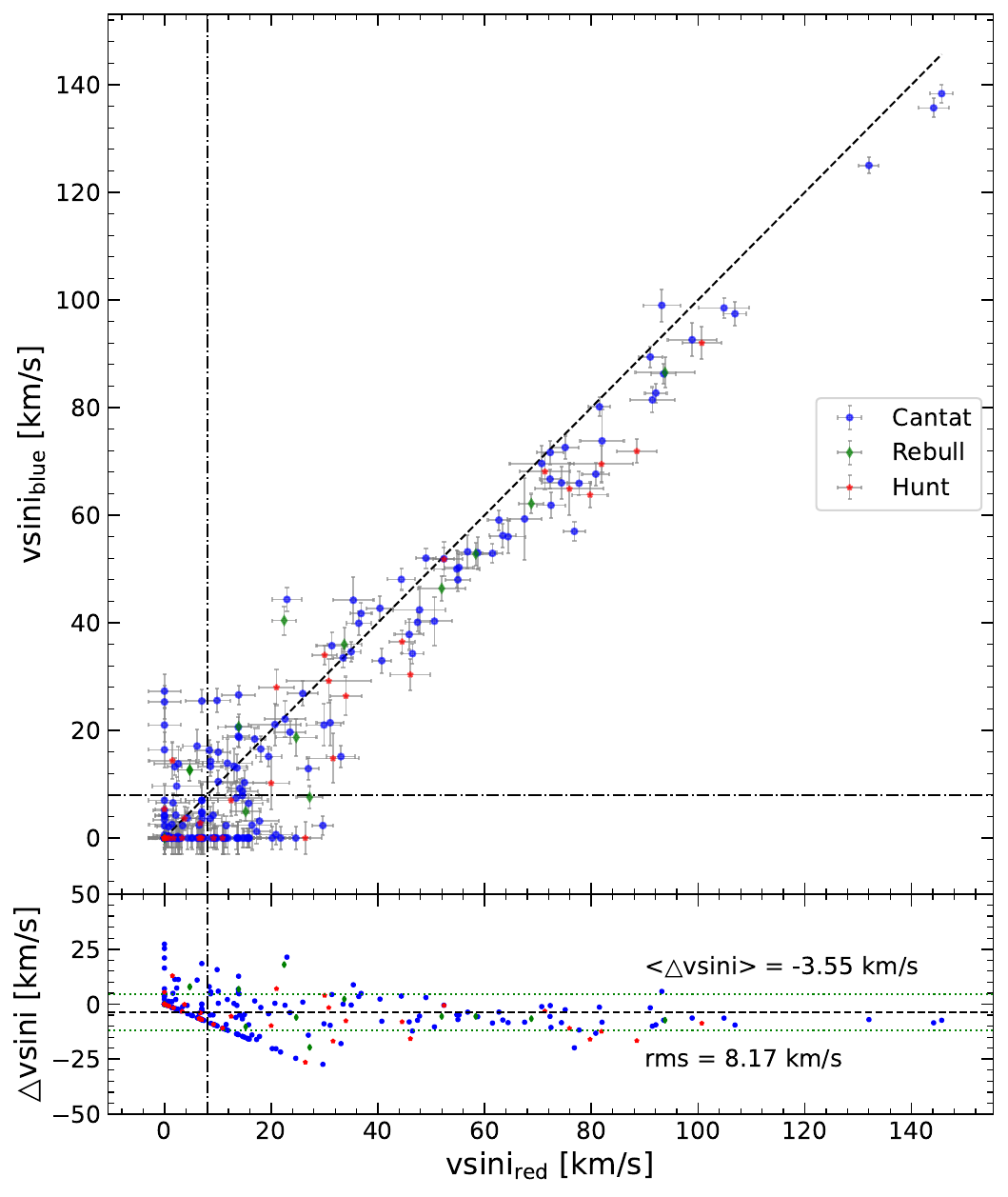}
\caption{Comparison of the \vsini\ values derived from the blue- and red-arm LAMOST-MRS spectra with ROTFIT. The meaning of lines and symbols is the same as in Fig.~\ref{Fig:RV_blue_red}. }
\label{Fig:vsini_blue_red}
\end{figure}

As shown by \citet{Frasca2022} with Monte Carlo simulations, the resolution and sampling of the MRS \lamost\ spectra do not allow one to measure \vsini\ values smaller than 8\,\kms. Whenever we find a smaller value in at least one arm, this must be treated as a non-detection and flagged as an upper limit.
Comparison of the blue-arm with the red-arm \vsini\ shows a fairly good correlation and a low value floor for blue-arm spectra (\vsini$_{\rm blue}$), which extends up to \vsini$_{\rm red}\simeq30$\,\kms\ and  translates into the tilted strip in Fig.~\ref{Fig:vsini_blue_red}. A similar behavior was found by \citet{Frasca2022}. This happens when the red-arm spectra provided a poor constrain to \vsini\ due to the few absorption lines or to the presence of molecular bands. Some objects show the opposite behavior, that is, a range of \vsini\ values extending to about 30\,\kms\ for which we found \vsini$\simeq0$ from the \rotfit\ analysis of red-arm spectra. This is likely due to the low S/N of the blue-arm spectra or some flaws in them. Apart from these issues at low values of \vsini, we note that red values are systematically $\approx$\,5\,\kms\ larger than the blues ones.
We do not know the reason for this discrepancy, which could be related to the different shape of the spectra in the red and blue regions.
However, the weighted average of the values obtained from the blue and red arm gives \vsini\ values that are in good agreement with those derived from APOGEE (see Sect.\,\ref{Subsec:external}). These are the final values of \vsini\ that we adopt in this work and list in Table~\ref{Tab:APs}.

\subsection{Data quality control: comparison with the literature}
\label{Subsec:external}

\begin{figure}[htb]
\includegraphics[width=9cm,viewport= 0 0 480 525]{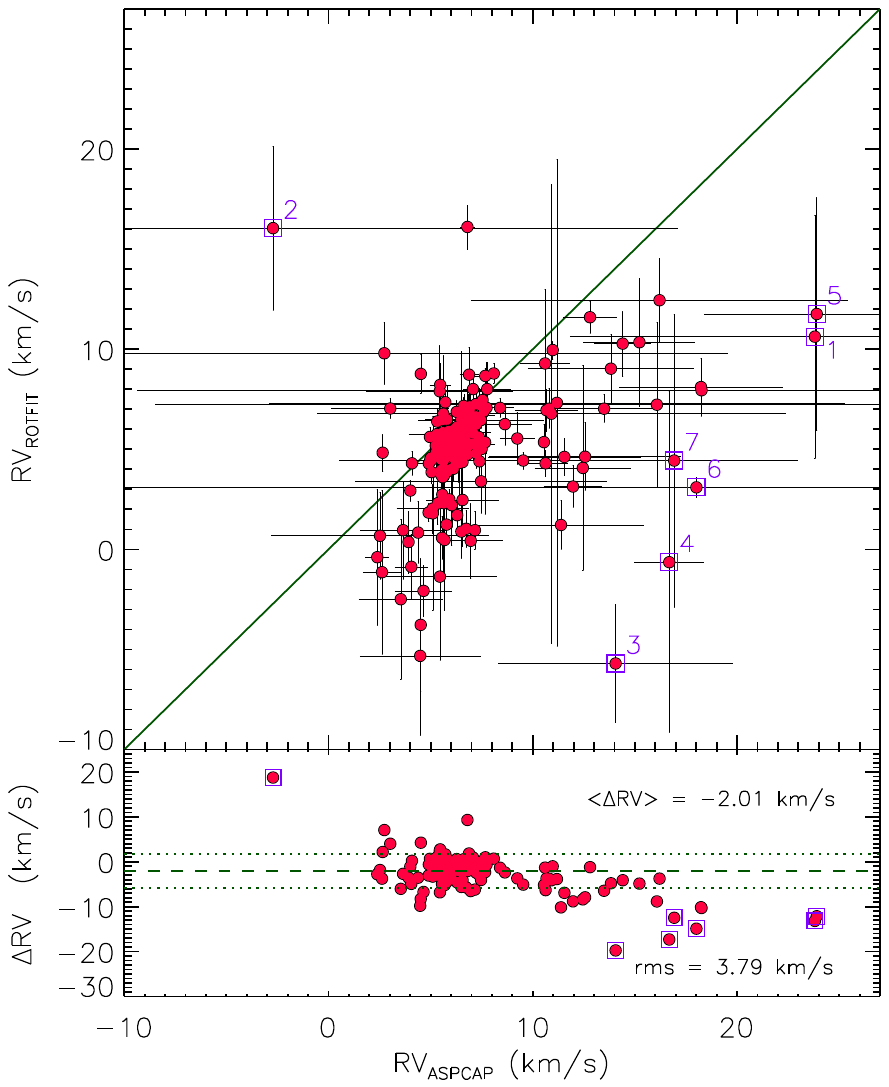}
\caption{Comparison between the RV measured in this work 
and APOGEE DR17 values \citep{Abdurro2022}. 
The one-to-one relation is shown by the full line in the upper box. The RV differences between \rotfit-\lamost\ and APOGEE, $\Delta$RV, are displayed in the lower box and show an average value of $-2.01$\,\kms\ (dashed line) and a standard deviation of 3.79\,\kms\ (dotted lines). 
The purple squares enclose the most discrepant points. 
}
\label{Fig:RV_comp}
\end{figure}

To check the accuracy of our values, we have compared the APs derived with \rotfit\ on the \lamost-MRS spectra with those available in the literature.

\begin{figure}[htb]
\includegraphics[width=9cm,viewport= 30 10 450 280]{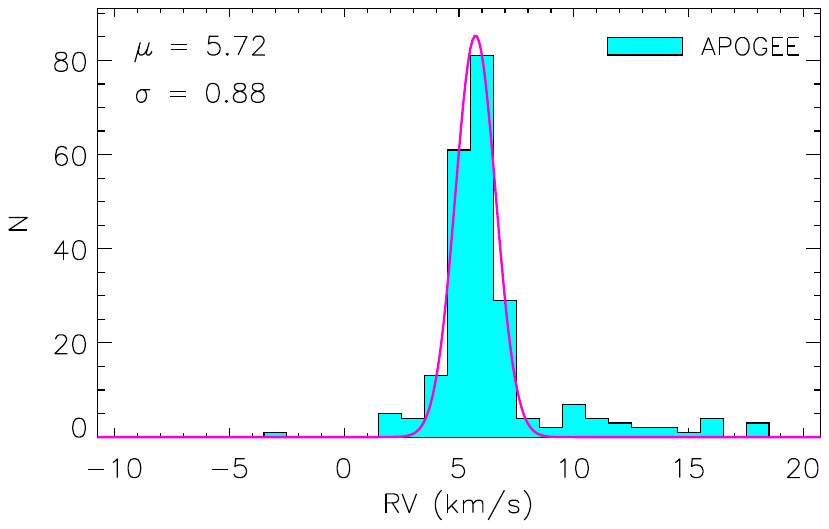}
\caption{RV distribution (cyan histogram) of the Pleiades members observed by APOGEE, as obtained with the average RV values reported in the 17th data release of SDSS \citep{Abdurro2022}. The Gaussian fit is overplotted with a full pink line; the center ($\mu$) and dispersion ($\sigma$) of the Gaussian are also marked. }
\label{Fig:RV_distr_APO}
\end{figure}

As regards the radial velocity, in Fig.~\ref{Fig:RV_comp} we show the comparison of our average RV values, corrected for the systematic offsets, with those measured by APOGEE  with the ASPCAP pipeline and reported in the 17th data release of the Sloan Digital Sky Survey \citep[SDSS APOGEE-2 DR17,][]{Abdurro2022} for the 205 sources in common between \lamost\ and APOGEE.
We note a small negative average offset of $-2.0$\,\kms\  between our corrected RVs and APOGEE and an rms of the data dispersion around the mean of 3.8\,\kms.
If we exclude the seven most discrepant points ($|\Delta{\rm RV}| \ge 3\cdot rms$) the offset becomes $-1.7$\,\kms\ and the rms decreases at 2.7\,\kms. It is worth noticing that these objects are on the tail of the RV distribution, which is peaked at about 5\,\kms\ in both datasets (see Figs.\,\ref{Fig:RV_distr} and \ref{Fig:RV_distr_APO}). Moreover, three of these sources (\#1, \#2, and \#4) have been already flagged as stars with variable 
RV (`RVvar') in Table~\ref{Tab:APs}, according to the $P(\chi^2)<0.05$ criterion.
Details on these discrepant stars are provided in Appendix~\ref{Appendix:discrepant}.

The distribution of the average RVs of the 229 Pleiades members observed by APOGEE, which are contained in the SDSS APOGEE-2 DR17 \citep[][]{Abdurro2022}, is shown in Fig.~\ref{Fig:RV_distr_APO}. The center $\mu=5.72\pm0.02$\,\kms\ and dispersion $\sigma=0.88\pm0.09$\,\kms, where obtained, as for the \rotfit\ RVs measured on \lamost\ spectra, with a Gaussian fit. They again suggest a small negative offset ($\approx-0.7$\,km/s) between \rotfit-\lamost\ and APOGEE and a smaller dispersion for the latter dataset, possibly due to their better accuracy.

\begin{figure*}[htb]
\includegraphics[width=8.1cm]{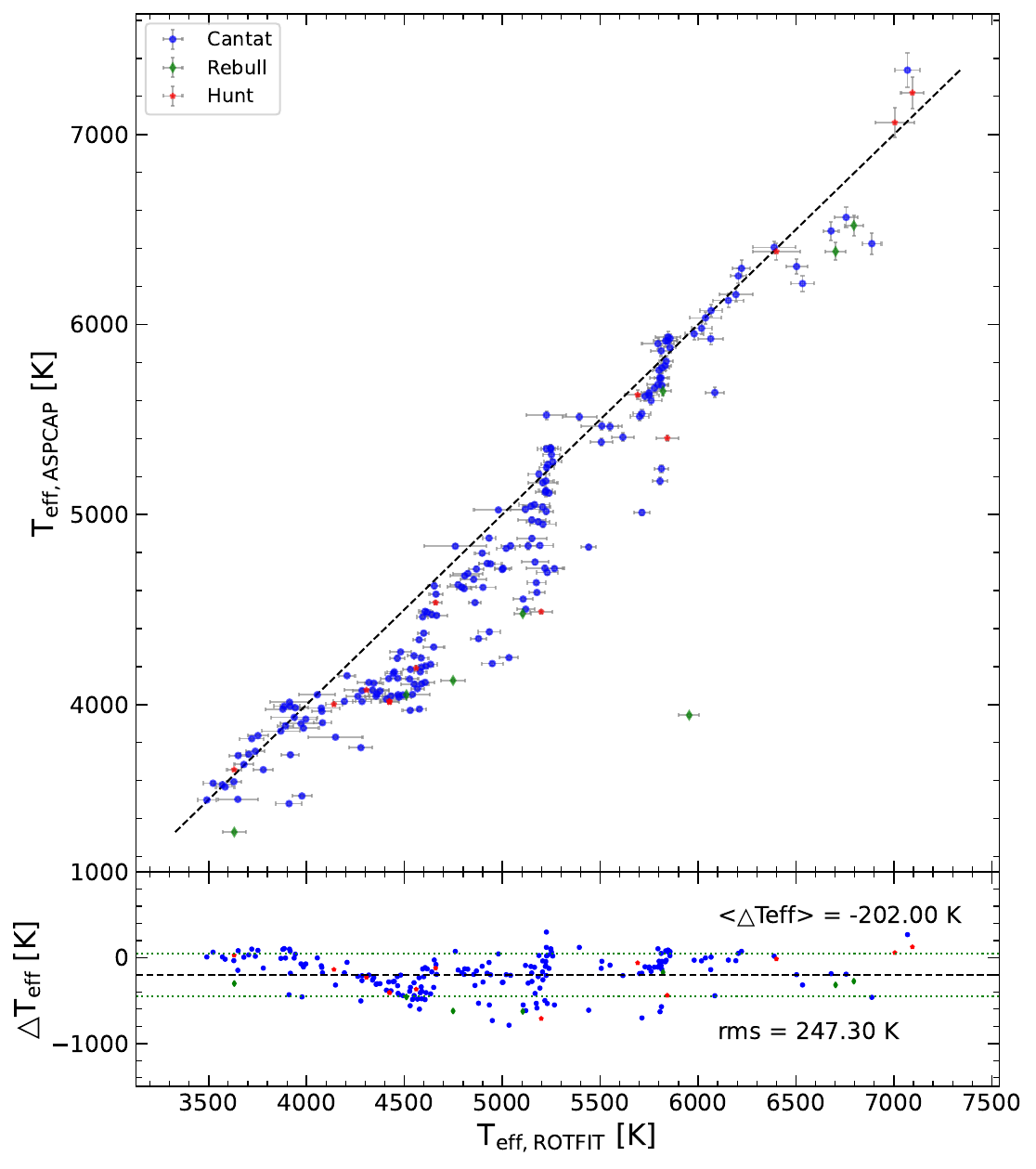}
\hspace{1cm}
\includegraphics[width=8cm]{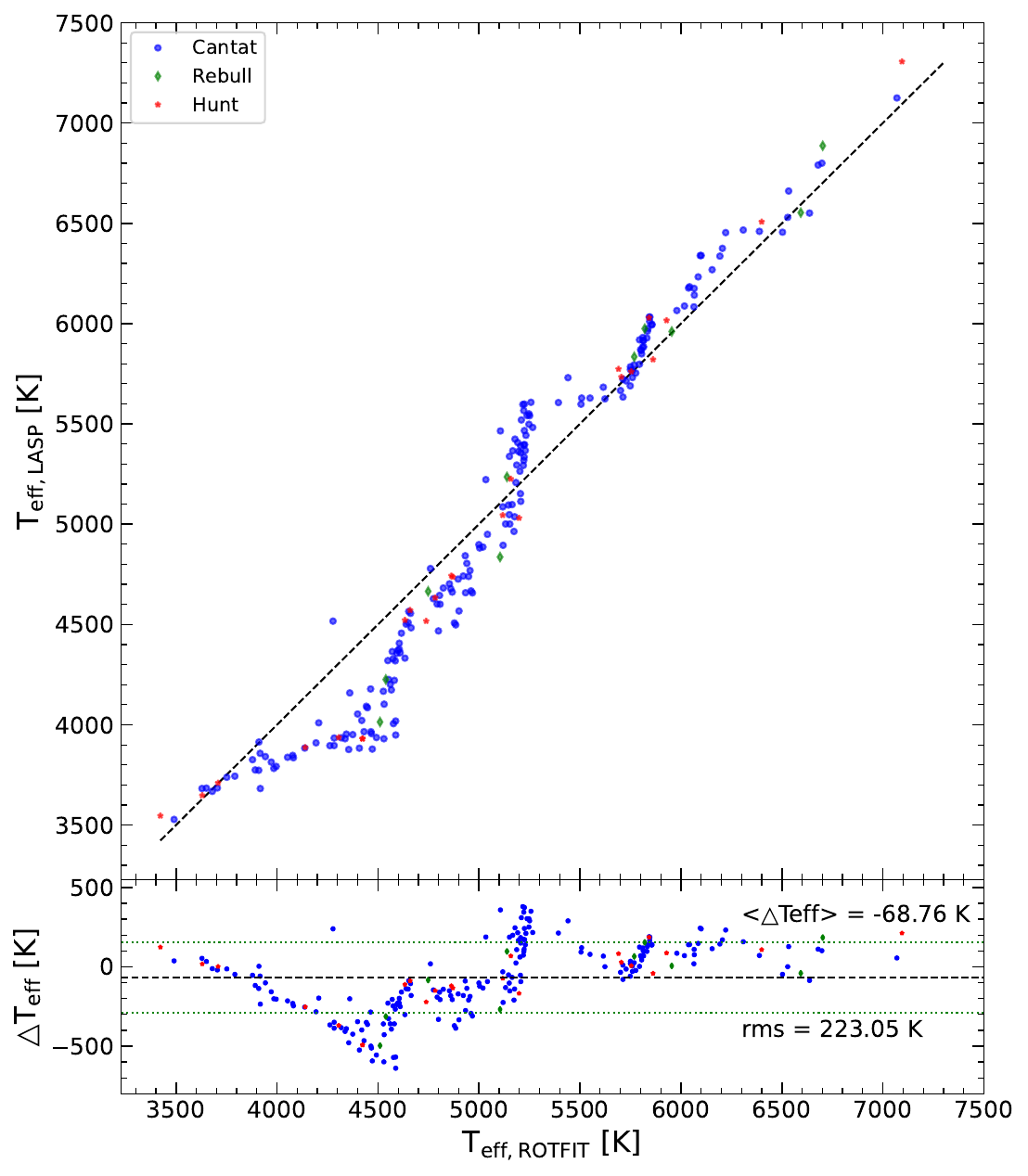}
\caption{Comparison between the \teff\ values measured in this work and those found in the literature. Left panel: \rotfit\ versus APOGEE-2 DR17 Teff values \citep{2020AJ....160..120J}. 
The color of symbols distinguishes the three subsamples as in the previous figures and indicated in the legend. The one-to-one relation is shown by the dashed line. The \teff\ differences between \rotfit\ and APOGEE, $\Delta$\teff, are displayed in the lower box.
Right panel: \rotfit\ versus \lasp\ DR11 v1.0 \teff\ values ({\tt https://www.lamost.org/dr11/v1.0}). The meaning of lines and symbols is as in the left panel.
}
\label{Fig:Teff_comp}
\end{figure*}

The comparison of the \rotfit\ \teff\ with those derived with \lasp\footnote{https://www.lamost.org/dr11/v1.0} and the ASPCAP pipelines is shown in Fig.\,\ref{Fig:Teff_comp}.
One of the most striking feature of these scatter plots is the ``boxy'' shape, which is more evident in the comparison with \lasp\ data.  This pattern was already noted by \citet{Frasca2022}, who interpreted it as a sort of ``pile up'' of the average parameters around those of the best (minimum $\chi^2$) template. This is not surprising, because \rotfit\ does not apply any kind of interpolation or regularization between the parameters of the closest templates.
Despite this pattern, the comparison with APOGEE indicates a scatter of about 247\,K and an overall offset of $\approx-202$\,K, which is mostly due to the stars with \teff$^{ROTFIT}<$\,4500\,K. 
There is only one outstanding outlier in this plot, which is \gaia\ DR3 649494274606140 (= V357\,Tau), with a \teff$\simeq$3950\,K in the APOGEE catalog, for which we find instead 5955\,K. 
The comparison with \lasp\ displays  a smaller offset of $-68$\,K and an rms of $\approx223$\,K that is due to the ``boxy" shape of the scatter plot.  
Interestingly, V357\,Tau does not appear as an outlier in this plot because its LASP temperature of 5907\,K reported by \citet{Wang2021} is nearly identical to our value. We note that this star has a \gaia\ color $G_{\rm BP}-G_{\rm RP}$=1.85\,mag, which is consistent with the APOGEE \teff, but its Re-normalized Unit Weight Error value, RUWE\,=\,14.578 indicates a bad astrometric solution and could be related to the presence of an unresolved optical companion. Therefore, we have discarded this star from the following analysis. The comparison with APOGEE and \lasp\ data suggests that our \teff\ determination is of sufficient quality for our purposes, as also witnessed by the rather smooth distribution of \gaia\ color index versus our \teff\ values (Fig.~\ref{Fig:Gaia_color}).

The comparison of \logg\ values of the present work with those derived with \lasp\ and APOGEE is shown in Fig.\,\ref{Fig:logg_comp}. The values agree well to each other with a small average offset (0.03--0.06 dex) and a small rms scatter of about 0.11\,dex.

The comparison of \vsini\ values measured with \rotfit\ in the present work with those derived with \lasp, APOGEE, and the values reported in \citet{Hartman2010} (which is indeed a compilation of literature data mostly coming from \citet{Quelozetal1998} and \citet{Terndrup2000}) and in \citet{Soderblom1993c} is shown in Fig.\,\ref{Fig:vsini_comp}. The \rotfit\ values agree very well with the APOGEE ones derived with the ASPCAP pipeline, showing only a small residual scatter for the \rotfit\ values less than or close to the \lamost\ upper limit of 8\,\kms. The 
discrepant points at \vsini$^{\rm ASPCAP}=96$\,\kms\ that are enclosed into red squares correspond to the maximum rotational broadening allowed for the ASPCAP pipeline \citep{2020AJ....160..120J}. If we exclude these points, we find an offset of only $\approx2$\,\kms\ with an rms dispersion of $\approx7$\,\kms. 
Another discrepant point is related to the F-type source \gaia-DR3\,66832993955739776 (=HD\,23567) (\#1) for which we find \vsini\,=\,90.1\,\kms\ while ASPCAP reports 50\,\kms. This source is a $\delta$\,Sct-type star for which a \vsini\,=\,98.5\,\kms, in better agreement with our value, is reported by \citet{Solano1997}.
The comparison with the \lasp\ results displays a bad correlation. In particular, the \lasp\ values are systematically higher than ours by about 9\,\kms, on average. Moreover, the minimum \vsini\ in the \lasp\ data is 27\,\kms, which is related to the method, the template grid (minimum \teff=5000\,K), and the velocity steps adopted in the \lasp\ pipeline \citep[e.g.,][]{Zuo2024}. 

The agreement of our results with those reported by \citet{Hartman2010} and \citet{Soderblom1993c} is good, with average differences of about 1\,\kms\ and dispersion of about 9\,\kms, as shown in the bottom panels of Fig.\,\ref{Fig:vsini_comp}.
This strengthens the validity of our \rotfit-based measures and suggests using the (few) \vsini\ values of \lasp\ with caution.

\section{Rotation periods}
\label{Sec:Prot}

As mentioned in Sect.~\ref{Subsec:Obs_photo}, for most (178 out of 283) of the sources  investigated by us  the rotational periods, \prot, were derived by \citet{Rebull2016} from the analysis of \kepler-\ktwo\ photometry. For 89 of the remaining stars we were able to measure \prot\ from \tess\ data in three consecutive quarters, by means of a Lomb-Scargle periodogram analysis \citep{Scargle1982} to which the CLEAN deconvolution algorithm \citep{Roberts1987} was applied. 
To validate the above period analysis we have selected a dozen  stars analyzed by \citet{Rebull2016} with both short and long periods. We have retrieved and analyzed the \tess\ photometry in the three consecutive sectors 42, 43, and 44. The results are shown in Fig.~\ref{fig:TESS_Phot_comp} and are quoted in Table~\ref{Tab:Comp_Prot}. The \prot\ values are in excellent agreement, typically within 1\,\% and always better than 5\,\%, with each other. 
A sample of \tess\ light curves of both short- and long-period variables is shown in Fig.~\ref{fig:TESS_Phot}. Typically, a single peak, corresponding to the rotational period, dominates the periodogram. In some cases, such as TIC\,15902424 and TIC\,427545153, the periodogram shows two close peaks and a beating pattern is apparent in the light curve. This behavior has been observed when two or more spotted areas are located at different latitudes and the star is rotating differentially \citep[e.g.,][and reference therein]{Frasca2011_AA523_A81}. We have reported the values of \prot\ in Table\,\ref{Tab:APs}.

\begin{figure}[htb]
\includegraphics[width=9cm]{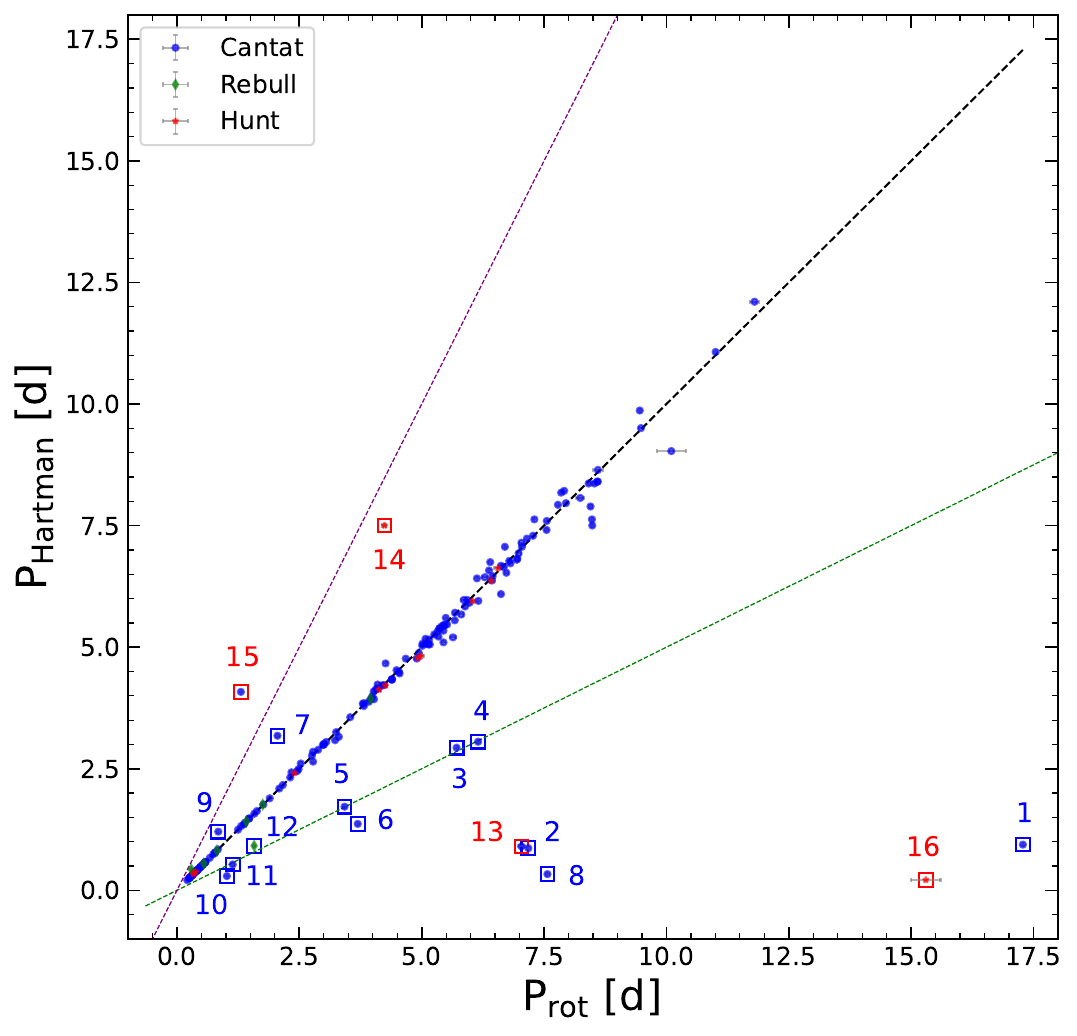}
\caption{Comparison between the \prot\ values adopted in this work (from \ktwo\ or \tess) and those published by \citet{Hartman2010}. 
The color of symbols distinguishes the three subsamples as 
indicated in the legend. The one-to-one relation is shown by the black dashed line. The green and purple dotted lines with slopes 0.5 and 2, respectively, are also shown.
The most discrepant stars are enclosed into squares in blue \citep[for \prot\ reported in][]{Rebull2016} and red (for \prot\ values derived in this work from \tess\ data). 
}
\label{Fig:Prot_comp}
\end{figure}

We have compared these \prot\ values with those of \citet{Hartman2010}, which were derived from ground-based photometry collected at the Hungarian-made Automated Telescope Network (HATNet) or retrieved from the literature in a few cases.
The results are shown in Fig.~\ref{Fig:Prot_comp} along with the one-to-one relation (black dashed line).  
Most of the 187 targets in common with \citet{Hartman2010} display a very good agreement of \prot\ values and lie on the bisector of the plot. Among the 16 most discrepant objects marked in Fig.~\ref{Fig:Prot_comp} and listed in Table~\ref{Tab:Discrepant_Prot}, 12 have 
\prot\ measured by \citet{Rebull2016}, while for the last four \prot\ has been determined in the present work with the \tess\ photometry.
In order to explain the differences observed, in Fig.~\ref{Fig:Prot_comp} are also plotted the 2$\times$\prot\ relations with slopes 0.5 (green dotted line) and 2 (purple dotted line), respectively. Objects near the green dotted line are those for which \citet{Hartman2010} derives a rotation period that is half of ours. This could be due to the temporal sampling and photometric precision of the ground-based data and to the presence of spots of similar size at nearly opposite longitudes during the HATNet observations mimicking a light curve with a single spot and half the period. This is clearly the case for  \#3, \#4, \#5, \#11, and \#12. In the opposite case, for data near the line with slope\,=\,2, it could be that the duration of the \ktwo\ or \tess\ observations is insufficient to correctly determine the rotation period, but this is more likely to happen with rotators slower than a fortnight.

Furthermore, to try to understand this discrepancy, we have retrieved and analyzed the \tess\ photometry in sectors 42, 43, and 44 of the first 12 stars, in the same way we did for the sources with no data in \citet{Rebull2016} and found the same period within the errors, suggesting that Rebull's (and our) determination is very likely the correct value. We found only two exceptions, namely \#8 for which we find \prot\,$\simeq$\,13.5\,d instead of 7.56\,d measured by \citet{Rebull2016}, and \#9 for which we measure \prot\,$\simeq\,1.204\pm0.005$\,d instead of 0.845\,d. For the first object, adopting our \prot\ determination increases the discrepancy, considering the very small period of 0.333\,d reported by \citet{Hartman2010}. Moreover, we do not find any indication of high-frequency peaks in our periodogram; this, along with the low \vsini\ measured in our \lamost\ spectra points to a long rotation period. For \#9 our \prot\ determination is nearly identical to the \citet{Hartman2010} period (1.207\,d), then resolving the discrepancy.

Regarding the remaining four stars, \#13 to \#16, as mentioned before, their periods are only measured by us from the \tess\ data.
For \#13 \citet{Hartman2010} found a period of 0.904\,d while ours is 7.04\,d, which is more consistent with the upper limit \vsini$\leq$8\,\kms\ or the value of 5\,\kms\ measured by \citet{Quelozetal1998}. Indeed, such a low \vsini\ would imply an inclination $i<10\degr$ with \prot=0.904\,d, which is too small to produce a significant variation for ground-based observations. Moreover, no peak around 0.90\,d is visible in our periodogram. For \#14 \citet{Hartman2010} found a period of 7.50\,d, almost twice our determination (4.24\,d). However, a second peak with a slightly lower amplitude in our periodogram is found at \prot=7.60\,d, which is close to the \citet{Hartman2010} determination. 
For \#15 \citet{Hartman2010} found a period of 4.082\,d, while our determination is 1.307\,d without any indication of significant peaks at longer periods. This \prot\ value agrees with our determination of \vsini=39.2\,\kms, which excludes a longer rotation period. 
Interestingly, the \prot\ determination of 4.082\,d quoted by \citet{Hartman2010} is an uncertain value reported by \citet{Marilli1997}  that is likely a wrong determination.

The last discrepant source, \#16 (=\,TIC\,440686834 = \gaia\,DR3 64030785494725632), is a very puzzling object. Indeed, we find a large-amplitude variation with a period of about 15.3 days (see also Fig.~\ref{fig:TESS_Phot}). In the high-frequency domain of the periodogram there is also a small amplitude ($\sim$ 30 times lower than the former) peak a \prot\,$\simeq$\,0.215\,d, which is indicated with a blue arrow in Fig.~\ref{fig:TESS_Phot} and corresponds to the rotation period reported by \citet{Hartman2010}. This value of \prot\ is also consistent with the \vsini\,=\,123.6\,\kms\ measured by us. 
This object was classified by us as a possible SB2 on the basis of the appearance of the spectrum and the shape of
the CCF peak that displays a broad and a narrow component (see Fig.~\ref{Fig:64030785494725632}). It is not clear if it is a physical binary or an optical unresolved double. It is interesting to note that there is no bright source within 21\arcsec\ in the \gaia\ DR3, but the Re-normalized Unit Weight Error value (RUWE) is 2.386. The RUWE is expected to be close to 1.0 when a single-star model fits the astrometric observations adequately. A
value noticeably higher than 1.0, like in this case, may suggests that
the source is either not a single star or presents challenges for
the astrometric solution \citep{Castro-Ginard2024}.
Therefore, we consider 0.215\,d as the rotation period of the star with the broader lines. The \prot\,$\simeq$\,15.3\,d could be instead related to the narrow-lined component.

\section{Chromospheric emission and lithium abundance}
\label{Sec:chrom_lithium}

\begin{figure}[th]
\begin{center}
\includegraphics[width=9.5cm,viewport= 10 0 540 540]{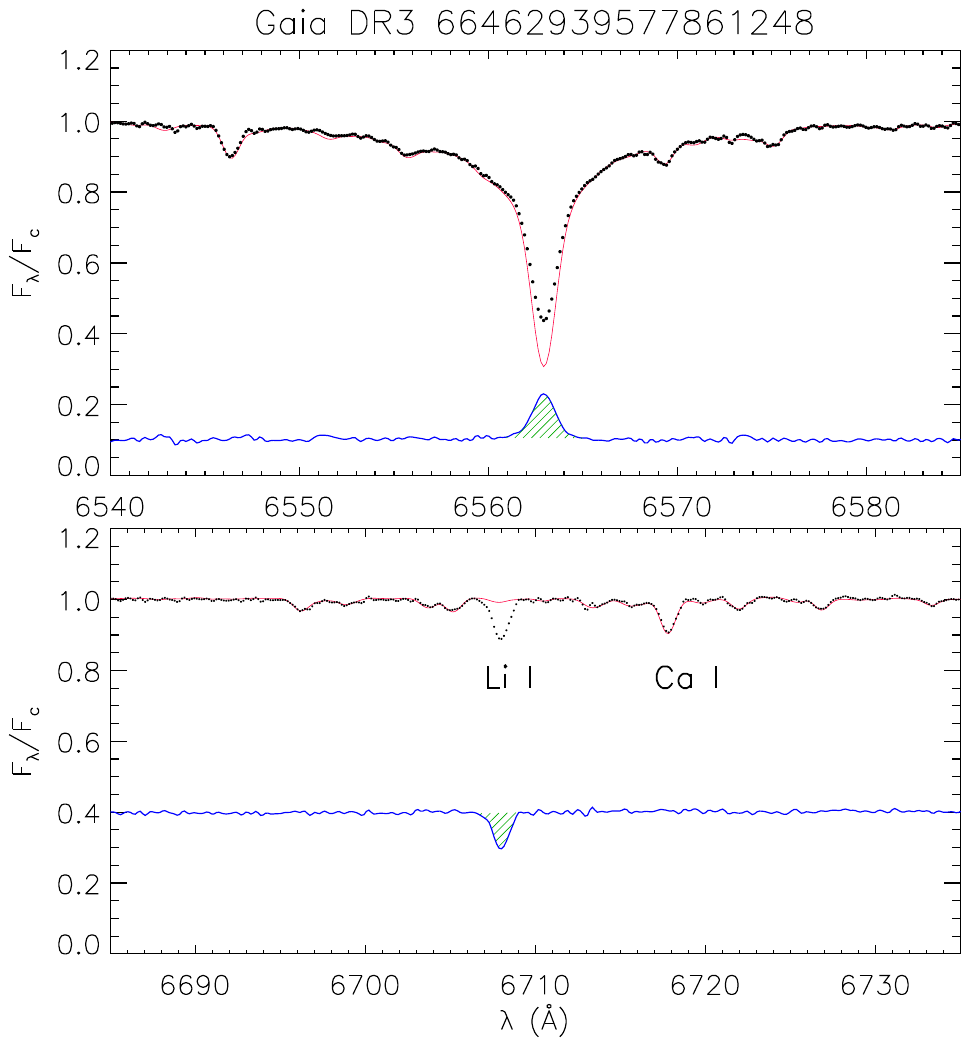}
\vspace{-0.5cm}
\caption{Example of the subtraction of the best non-active, lithium-poor template (red line) from the spectrum of the F9-type star \gaia-DR3\,66462939577861248 = TIC 35205639 = HII 2786 (black dots) in the H$\alpha$ (top panel) and \ion{Li}{i}\,$\lambda$6708\,\AA\ (bottom panel) spectral regions. The difference spectrum (blue line), whose continuum has been arbitrarily shifted for clarity,} reveals the chromospheric emission in the H$\alpha$ core and emphasizes the lithium line, removing the blended photospheric lines. The green hatched areas represent the excess H$\alpha$ emission and \ion{Li}{i} absorption  that were  integrated to obtain \Whalpha\ and \WLi, respectively.
\label{Fig:Ha_Li}
\end{center}
\end{figure}

For stars belonging to young OCs both the chromospheric emission (traced by Balmer H$\alpha$ in the \lamost\ MRS spectra) and lithium absorption are age-dependent parameters \citep[see, e.g.,][and references therein]{Jeffries2014, Frasca2018}.
As chromospheric emission in the \halpha\ can only result in a small to moderate filling of the line core, depending on the activity level and on the photospheric flux of the star, the removal of the photospheric spectrum is mandatory.
To this end, we have subtracted the non-active, lithium-poor template that best matches the final APs from each \lamost\ red-arm spectrum. 
This template has been aligned to the target RV, rotationally broadened by the convolution with a rotational profile with the \vsini\ of the target and resampled on its spectral points. 
The ``emission'' \halpha\ equivalent width, \Whalpha, was measured by integrating the residual emission profile (see Fig.~\ref{Fig:Ha_Li}, top panel). We excluded the SB2s from this analysis and kept only the values of \Whalpha\ significantly larger than zero, i.e. those larger or equal to their respective errors (256 sources). For the remaining single-lined sources, we adopted the error as the upper limit of the measure. These data are reported in Table~\ref{Tab:Halpha}, where we list the weighted average of the values measured in the individual spectra for stars with multiple observations.  

The equivalent width of the \ion{Li}{i} $\lambda$6708\,\AA\ absorption line (\WLi) was also measured in the subtracted spectra where the blends with nearby photospheric lines have been removed (see Fig.~\ref{Fig:Ha_Li}, bottom panel). This allows us to get a better measure of \WLi\ and a reliable estimate of its error, 
which  was calculated as $\sigma_{W_{\rm Li}} = D\cdot\sqrt{w\Delta\lambda}$, where $D$ is the average dispersion of the flux values in the residual spectrum on the two sides of the line ($D\simeq \frac{1}{S/N}$), $w$ is the integration width in wavelength units and $\Delta\lambda$ (=0.15\,\AA) is the pixel size in wavelength units. This expression is similar to the formula proposed by \citet{Cayrel1988}.
We were able to detect the lithium line (\WLi\ larger than the error in at least one spectrum) for 224 objects, while for 52 of them we could only determine an upper limit. We did not measure \WLi\ for the SB2 systems. For the stars with time-series spectra we calculated the weighted average of the individual values of \WLi\ and took the weighted standard deviation or the standard error of the weighted mean (whenever greater than the former) as the error estimate. For the objects with only non-detections in all their spectra we took the lowest upper limit. These values are quoted in Table \ref{Tab:lithium}.

The equivalent width of a chromospheric line is not the best diagnostic of magnetic activity, and more accurate indicators of chromospheric activity are the line flux in units of stellar surface, $F_{\rm H\alpha}$, and the ratio between the line luminosity and bolometric luminosity, \rha, which have been evaluated according to Eq.\,2 and 3 of \citet{Frasca2022}. These values are also reported in Table\,\ref{Tab:Halpha}.

\begin{figure*}[htb]
\includegraphics[width=9.2cm,viewport= 20 10 440 400]{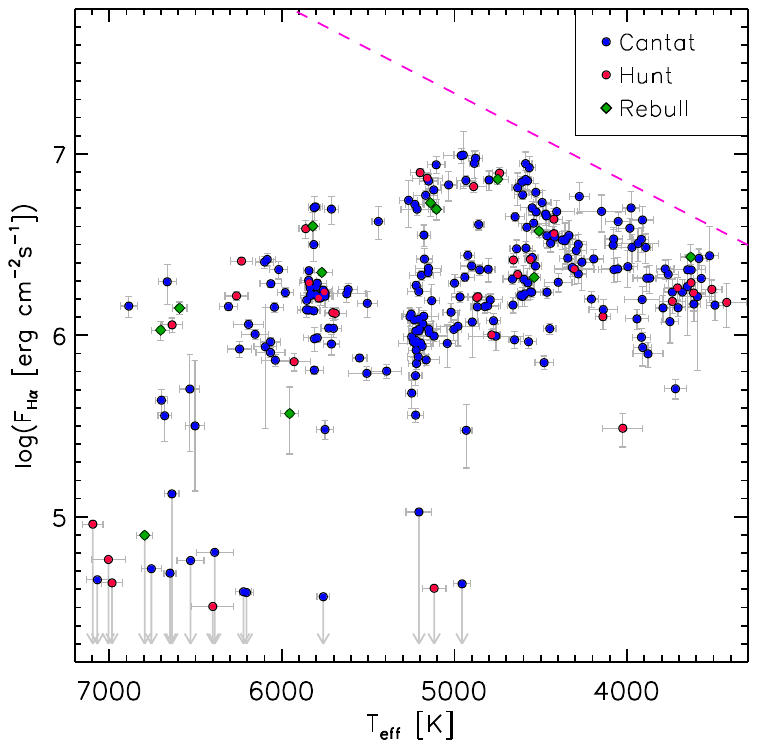}   
\includegraphics[width=9.2cm,viewport= 30 10 440 400]{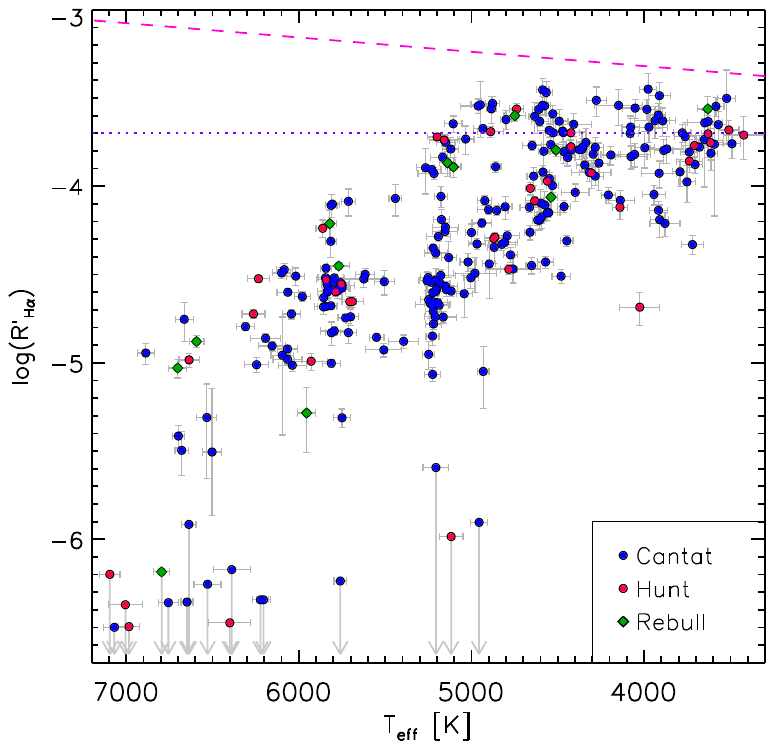}    
\caption{Activity indicators. Left panel: \halpha\ flux versus \teff. Right panel: \rha\ versus \teff. The three subsamples are distinguished with different symbols, as indicated in the legend. Upper limits are indicated by downward arrows. The straight dashed line in each panel is the boundary between chromospheric emission (below) and accretion as derived by \cite{Frasca2015}.
The dotted line in the \rha\ plot marks the saturation level $\log($\rha$)=-3.7$ estimated by \citet{Fang2018}. }  
\label{Fig:indicator}
\end{figure*}

The \halpha\ flux and \rha\ are plotted as a function of \teff\ in Fig.~\ref{Fig:indicator}. This plot also shows the dividing line
between chromospherically active stars and objects still undergoing mass accretion, which was empirically determined by \citet{Frasca2015}. As discussed by these authors, this boundary is close to the saturated chromospheric activity observed for main-sequence stars in young OCs (including the Pleiades), which has been found to be $\log$\rha\,$\simeq-3.3$ by \citet{Barrado2003}. Other authors, instead, estimate a lower value of $\log$\rha\,$\simeq-3.7$ for the saturation level \citep{Fang2018}.
The \rha\ values measured by us place all the Pleiades members in the region of chromospherically active sources as also seen in the \halpha\ flux diagram. This is what is expected for a cluster of $\approx$\,100\,Myr, for which  the  magnetospheric accretion from the circumstellar disks ended long ago. We note that the objects with the strongest activity, close to the saturation limit, are found among the coolest sources (\teff$<$\,5000\,K), as also found in previous works \citep[e.g.][]{Frasca2015, Fang2018}. 
An average trend of decreasing chromospheric activity with the increase of \teff\ is also apparent in Fig.\,\ref{Fig:indicator}, where the most discrepant objects are those with upper limits in \Whalpha. Most of them are warm stars (\teff$>6200$\,K) for which a small H$\alpha$ excess emission is hard to detect against the strong photospheric flux. However, four objects with upper limits in \rha\ are colder than 6000\,K. They are, in decreasing \teff\ order, Gaia\,DR3 64770241424108032, Gaia\,DR3 64923279699744256, Gaia\,DR3 121349735399525632, and Gaia\,DR3 63916431989200256. With the exception of the penultimate one, these stars have a high membership probability ($P_{\rm memb}=1.0$) according to \citet{Cantat2018} and a lithium abundance compatible with the cluster isochrone. However, they are all slow rotators (\prot$>$6 d), thus justifying their low chromospheric activity.

\renewcommand{\tabcolsep}{0.1cm}

\begin{table*}[ht]
  \caption{Chromospheric activity indicators: net H$\alpha$ equivalent width (\Whalpha), line flux (\fha), and luminosity ratio (\rha\,=\,$L_{\rm H\alpha}/L_{\rm bol}$).}
\begin{center}
\begin{tabular}{ccccccrcrcrcc}
\hline\hline
\noalign{\smallskip}
Designation &  \gaia-DR3  & \teff  & err & \logg  & err & \Whalpha & err  &  \fha & err   &  $\log$(\rha)  &  err    & Sub$^a$ \\  
   &  &   \multicolumn{2}{c}{(K)} & \multicolumn{2}{c}{(dex)} & \multicolumn{2}{c}{(\AA)} & \multicolumn{2}{c}{(\erg)} & \multicolumn{2}{c}{(dex)} & \\  
\noalign{\smallskip}
\hline
\noalign{\smallskip}

 J035427.83+212323.0 & 51742471745296768 & 5624 & 111 & 4.40 & 0.11 &        0.25 & 0.02 & 1.693e+06 & 1.709e+05 & -4.53 & 0.04 & C \\
 J035922.89+223416.3 & 53335045618385536 & 4360 &  42 & 4.66 & 0.04 &        1.71 & 0.42 & 3.318e+06 & 8.453e+05 & -3.79 & 0.11 & C \\
 J033305.82+220803.3 & 61554650949438208 & 5196 &  58 & 4.54 & 0.08 &        0.19 & 0.01 & 9.048e+05 & 5.268e+04 & -4.66 & 0.03 & C \\
 J035142.08+214006.0 & 63730309584280960 & 5222 &  45 & 4.46 & 0.11 &        0.17 & 0.01 & 8.272e+05 & 6.646e+04 & -4.71 & 0.03 & C \\
 J035434.79+215302.0 & 63795455648004608 & 3918 &  44 & 4.65 & 0.05 &        3.58 & 0.50 & 3.382e+06 & 5.327e+05 & -3.60 & 0.07 & C \\
 J035123.87+220648.1 & 63849572235829248 & 4640 &  38 & 4.62 & 0.05 &        1.07 & 0.12 & 2.998e+06 & 3.523e+05 & -3.94 & 0.05 & C \\
 J034734.16+214448.5 & 63916431989200256 & 4956 &  48 & 4.56 & 0.05 & $\leq$ 0.01 & 0.01 & $\leq$ 4.266e+04 & \dots & $\leq$ -5.90 & \dots & C \\
 J034935.88+220905.1 & 63958801843006208 & 4052 &  74 & 4.68 & 0.11 &        3.58 & 0.02 & 4.238e+06 & 5.209e+05 & -3.56 & 0.05 & C \\
 J035022.91+221118.0 & 64040818540162304 & 4193 &  57 & 4.66 & 0.04 &        1.79 & 0.14 & 2.635e+06 & 3.229e+05 & -3.82 & 0.05 & C \\
 J035234.46+223007.7 & 64073597730589952 & 3762 &  61 & 4.68 & 0.05 &        3.05 & 0.52 & 2.168e+06 & 4.474e+05 & -3.72 & 0.09 & C \\
\noalign{\smallskip}
\hline
\end{tabular}
\end{center}
{\bf Notes.} The full Table is available at the CDS.\\ $^{(a)}$ Subsample to which the target belongs as in Table\,\ref{Tab:APs}.
\label{Tab:Halpha}
\end{table*}

\begin{figure}[htb]
\includegraphics[width=7.7cm,viewport= 30 10 450 400]{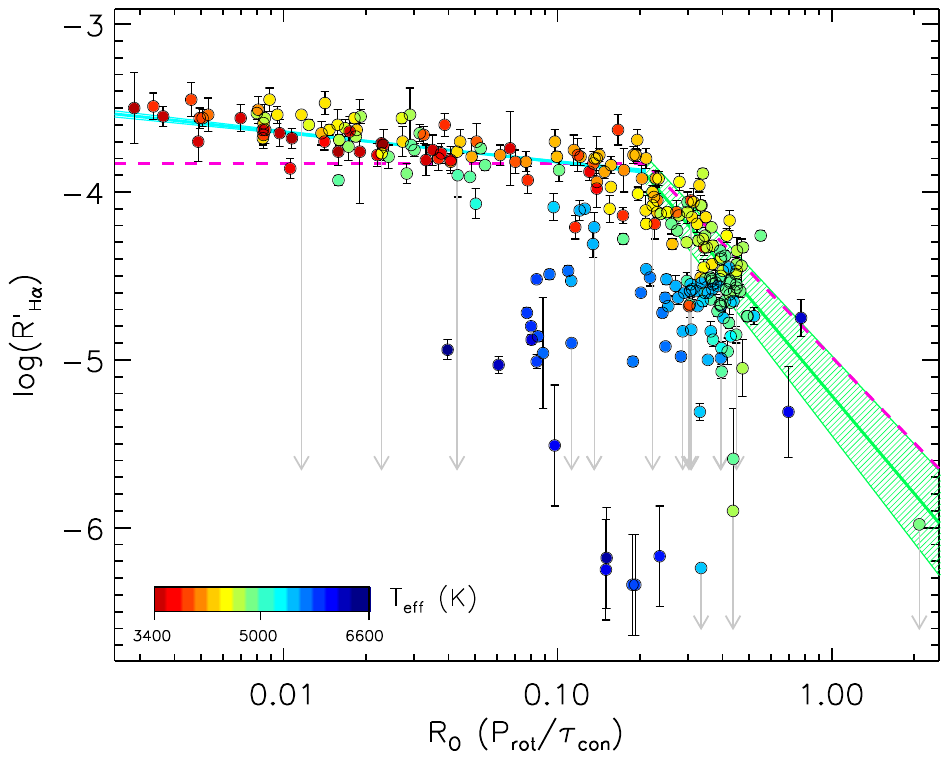}   
\caption{\rha\ versus the Rossby number $R_{\rm O}$. The
symbols are color coded by \teff. Upper limits are highlighted with downward gray arrows. The magenta dashed line denotes the relation of \citet{Newton2017}, while the cyan full line (slope = $-0.18\pm0.02$) is our best fit to the cold stars (\teff$\leq 4500$\,K) in the saturated activity domain ($R_{\rm O}<0.21$). The power law with slope $-1.9\pm0.2$ we found for the stars with $R_{\rm O}\ge0.21$ is shown with a green line. The 1$\sigma$ errors are indicated by the hatched regions.} 
\label{Fig:Rha_Rossby}
\end{figure}

For 257 sources with a measure of \Whalpha\ or an upper limit we have the rotation periods (Sect.~\ref{Sec:Prot}).
This allowed us to investigate the dependence of chromospheric activity on parameters related to the efficiency of the dynamo action. 
We found a correlation between \fha\  and \prot\ with a Pearson's coefficient $\rho=-0.49$.
Another important parameter expressing the efficiency of the dynamo in convective stellar interiors is the Rossby number, $R_{\rm O}$, which is defined as the ratio of \prot\ and the convective turnover time, $\tau_{\rm con}$. The latter is not a directly measurable quantity, but can be derived from theoretical models for MS stars or from calibrations as a function of temperature or color indices. We have used the empirical relation proposed by \citet[][Eq.\,10]{Wright2011} as a function of $V-K_S$. The de-reddened color index $(V-K_S)_0$ is provided by \citet{Rebull2016} for the sources in their catalog. For the additional sources with periods determined by us, we have taken the $V$ and $K_S$ magnitudes from the APASS \citep{Henden2018} and 2MASS \citep{Cutri2006} catalogs, respectively, and corrected the color index for the excess $E(V-K_S)=0.11$\,mag as in \citet{Rebull2016}. The correlation between activity indicators and $R_{\rm O}$ is even better, with a coefficient $\rho=-0.54$ for \fha\ and $-0.58$ for \rha. We have plotted the values of \rha\ as a function of the Rossby number in Fig.~\ref{Fig:Rha_Rossby}, distinguishing our targets by a \teff-dependent color code. The relation proposed by \citet{Newton2017} for nearby M dwarfs is overlapped with a dashed line. They fitted a canonical activity-rotation relation to their data, where the saturation level, for $R_{\rm O}\leq 0.21$, was found to be $\log$\rha\,=\,$-3.83$. For larger values of $R_{\rm O}$ they found a power-law decay with an exponent $\beta=-1.7$. A linear regression to our data for cold stars (\teff$\leq 4500$\,K) with $R_{\rm O}>0.21$ also suggests a power-law decay with $\beta=-1.94\pm0.19$ (green line in Fig.~\ref{Fig:Rha_Rossby}), which agrees, within the errors, with the results of \citet{Newton2017}. However, for rapid rotators
we find a higher saturation level, as already mentioned, with a slowly increasing trend towards the shortest values of $R_{\rm O}$. The fit of the data of cold stars (\teff$\leq 4500$\,K) with $R_{\rm O}\leq 0.21$ gives rise to a slope of $-0.18\pm0.02$ (cyan line in Fig.~\ref{Fig:Rha_Rossby}).
A similar result was also found by \citet{Fang2018} by analyzing \lamost-LRS data of young clusters including the Pleiades, where they found that a power law with slope $-0.2$  fitted the low $R_{\rm O}$ domain better than a constant saturated value.  This behavior has been also  noticed by some researchers who used other activity diagnostics, such as the X-ray luminosity. For instance, \citet{Reiners2014} found a slow increase of the coronal activity in the saturated regime with a law $L_{\rm X}/L_{\rm bol}\propto R_{\rm O}^{-0.16}$. 
As already apparent in Fig.~\ref{Fig:indicator}, it is once again shown that the hottest stars (\teff$\geq$\,6000\,K) are those with the lowest activity level.

\begin{figure}[htb]
\includegraphics[width=7.5cm,viewport= 10 10 500 550]{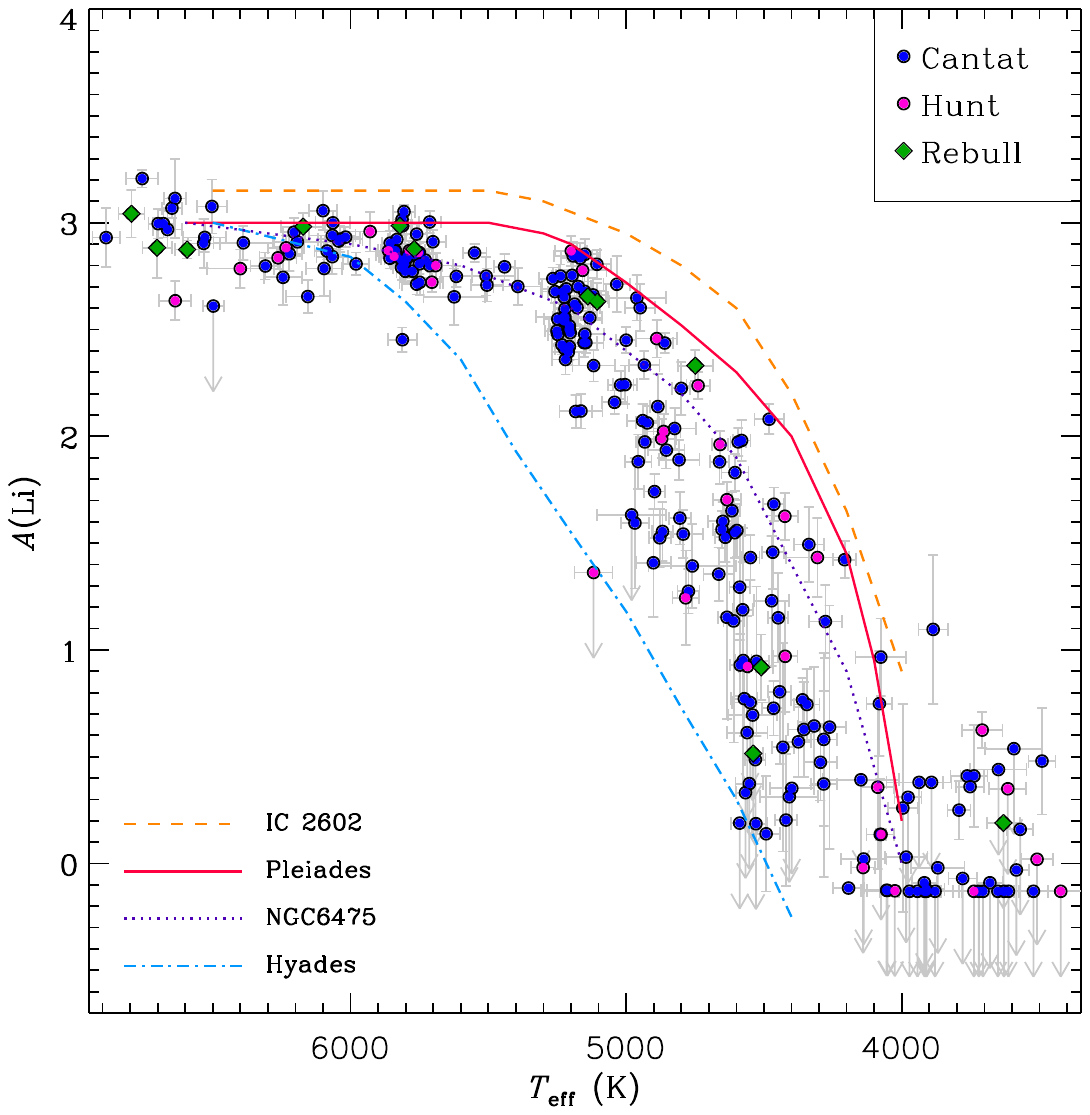}           
\caption{Lithium abundance of the Pleiades members with \lamost-MRS spectra as a function of \teff. Upper limits are highlighted with downward arrows.
The upper envelopes of $A$(Li) for the IC~2602 (age\,$\approx$\,30 Myr), Pleiades,  NGC\,6475 ($\approx$\,300 Myr), and Hyades ($\approx$\,650 Myr) clusters adapted from \citet{Sestito2005} are overplotted. 
}
\label{Fig:NLi}
\end{figure}

Lithium is a fragile element that is burned in stellar interiors at temperature as low as 2.5$\times10^6$\,K. It is progressively depleted 
from the stellar atmosphere in a way depending on the internal structure (i.e., stellar mass) when it reaches layers with the Li-burning temperature.  
Thus, its abundance can be used as an age proxy for stars cooler than about 6500\,K. Theoretical or empirical (based on coeval star groups) isochrones 
are normally used to infer stellar ages \citep[e.g.,][]{Jeffries2014, Frasca2018}. To this end, it is very important to determine as accurately as possible the age of the OCs used as a reference. In this sense, the Pleiades is one of the most widely used clusters and its age has been a long debated topic (which will be discussed in more depth in Sect.~\ref{Sec:age}).
We derived the lithium abundance, $A$(Li), from our values of \teff, \logg, and $W_{\rm Li}$ by interpolating the curves of growth of \citet{Lind2009}, which span the \teff\ range 4000--8000\,K and \logg\ from 1.0 to 5.0 and include non-LTE corrections. The errors of $A$(Li) have been calculated by propagating the \teff\ and $W_{\rm Li}$ errors.
These abundances are also listed  in Table~\ref{Tab:lithium}.
A plot of $A$(Li) versus \teff\ is shown in Fig.~\ref{Fig:NLi} along with the upper envelopes of the $A$(Li) distributions for young OCs shown by \citet{Sestito2005}. Our data are correctly located between the upper envelopes corresponding to 300 and 100\, Myr.

\renewcommand{\tabcolsep}{0.1cm}

\begin{table*}[ht]
  \caption{Lithium equivalent widths (\WLi) and abundances ($A$(Li)).}
\begin{center}
\begin{tabular}{ccccccccrcrcc}
\hline\hline
\noalign{\smallskip}
 Designation &  \gaia-DR3  &RA & DEC & \teff  & err & \logg  & err & \WLi & err &  $A$(Li)  &  err    & Sub$^a$ \\  
   &   &(J2000)  &(J2000) & \multicolumn{2}{c}{(K)} & \multicolumn{2}{c}{(dex)} & \multicolumn{2}{c}{(m\AA)} & \multicolumn{2}{c}{(dex)} & \\  
\noalign{\smallskip}
\hline
\noalign{\smallskip}
 J035427.83+212323.0 &  51742471745296768 &  58.615971 &  21.389748  &  5624 &  111  &  4.40 & 0.11  &        115  &  11 & 	   2.65 & 0.13  & C \\
 J035922.89+223416.3 &  53335045618385536 &  59.845396 &  22.571372  &  4360 &   42  &  4.66 & 0.04  &         29  &   3 & 	   0.77 & 0.10  & C \\
 J033305.82+220803.3 &  61554650949438208 &  53.274270 &  22.134250  &  5196 &   58  &  4.54 & 0.08  &        228  &   4 & 	   2.75 & 0.06  & C \\
 J035142.09+214006.0 &  63730309584280960 &  57.925378 &  21.668355  &  5222 &   45  &  4.46 & 0.11  &        164  &  15 & 	   2.56 & 0.09  & C \\
 J035434.79+215302.0 &  63795455648004608 &  58.644969 &  21.883909  &  3918 &   44  &  4.65 & 0.05  & $\leq$  14  &  14 & $\leq$ -0.09 & \dots & C \\
 J035123.87+220648.1 &  63849572235829248 &  57.849460 &  22.113383  &  4640 &   38  &  4.62 & 0.05  &         68  &   9 & 	   1.53 & 0.11  & C \\
 J034734.16+214448.5 &  63916431989200256 &  56.892334 &  21.746810  &  4956 &   48  &  4.56 & 0.05  &         70  &  11 & 	   1.88 & 0.13  & C \\
 J034935.88+220905.1 &  63958801843006208 &  57.399503 &  22.151421  &  4052 &   74  &  4.68 & 0.11  & $\leq$   4  &   4 & $\leq$ -0.13 & \dots & C \\
 J035022.91+221118.0 &  64040818540162304 &  57.595461 &  22.188337  &  4193 &   57  &  4.66 & 0.04  &          4  &   2 & 	  -0.11 & 0.00  & C \\
 J035234.46+223007.7 &  64073597730589952 &  58.143603 &  22.502144  &  3762 &   61  &  4.68 & 0.05  &         41  &   2 & 	   0.41 & 0.02  & C \\

\noalign{\smallskip}
\hline
\end{tabular}
\end{center}
{\bf Notes.} The full Table is available at the CDS.\\ $^{(a)}$ Subsample to which the target belongs as in Table~\ref{Tab:APs}.
\label{Tab:lithium}
\end{table*}

\section{Flares}
\label{Sec:Flares}

The multi-epoch \lamost\ spectra allow us to study the variation of activity on different timescales. The data cadence is normally too scarce to properly sample variations of \Whalpha\ produced by rotational modulation of chromospheric active regions, but it is helpful to characterize the variation range of the investigated sources and to detect flare events.
The potential of these times series data will be exploited in forthcoming works.
We only would like to mention three stars for which remarkable flares were detected.

\begin{figure}
\begin{center}
\includegraphics[width=9.3cm,viewport= 10 10 550 380]{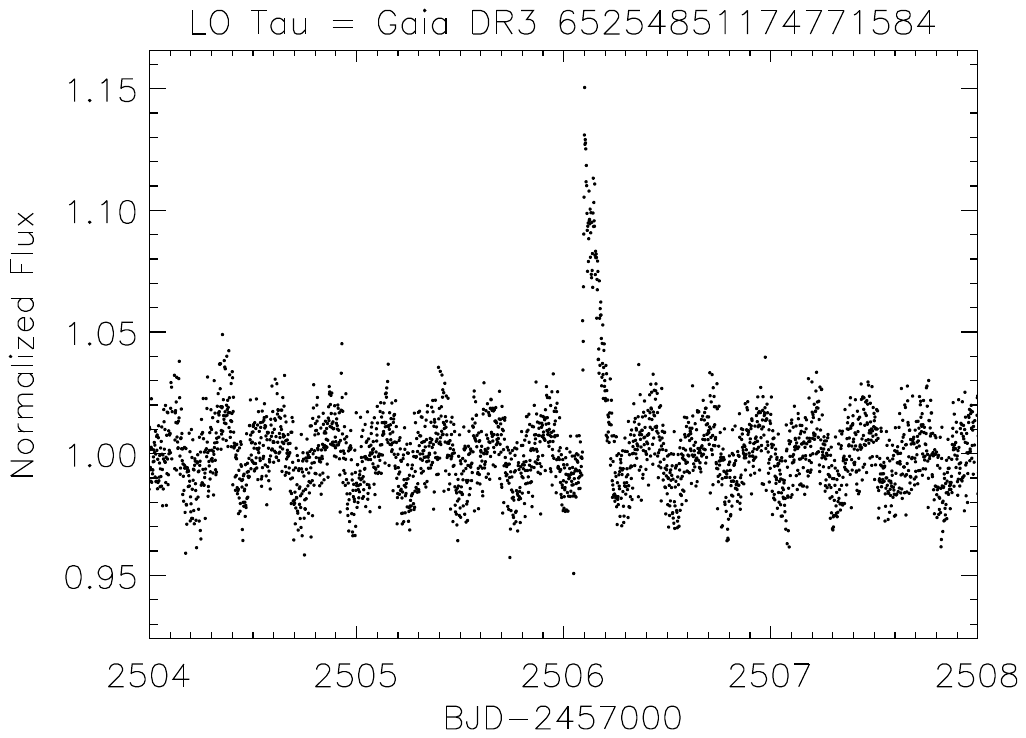}
\caption{A portion of a \tess\ light curve of LO\,Tau (= TIC\,258067348) that displays a strong flare standing out above the rotational modulation.
}
\label{fig:Flare_TESS}
\end{center}
\end{figure}

The first case is \gaia-DR3\,65254851174771584 (= TIC\,258067348 = LO\,Tau). It is classified as an ``eruptive variable'' in Simbad, where a spectral type M2.9, based on APOGEE data \citep{Birky2020}, is reported. This is in very good agreement with our M2-type classification. The rapid rotation of this star is witnessed by the large value of \vsini\,=\,68\,\kms, which agrees with the value of 80\,\kms\ reported in the APOGEE DR17 catalog \citep{Abdurro2022} and with the rotational period \prot\,=\,0.2587\,d reported by \citet{Rebull2016}. Several strong white-light flares are visible in the \ktwo\ and \tess\ light curves of this source (see Fig.~\ref{fig:Flare_TESS} for an example).

\begin{figure}
\begin{center}
\includegraphics[width=9.5cm,viewport= 0 10 430 600]{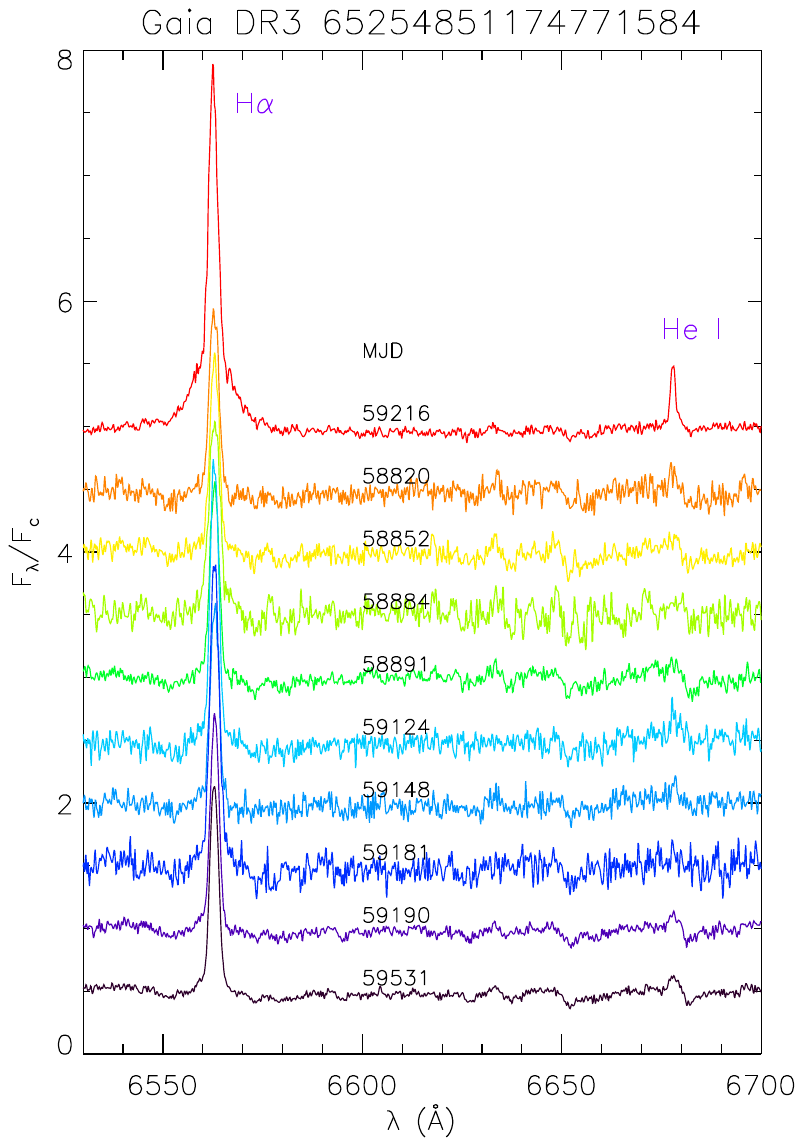}
\caption{\lamost\ MRS photospheric-subtracted spectra of LO\,Tau in the spectral region containing the H$\alpha$ and \ion{He}{i}\,$\lambda$6678\,\AA\ line. The spectra have been sorted in time order from top to bottom (except for the uppermost one) and have been vertically shifted for clarity. The modified Julian day (MJD) is written next to the spectrum. The uppermost spectrum plotted with a red line displays a very strong and broad H$\alpha$ profile as well as the 
\ion{He}{i} line in emission, which is indicative of a flare event.}
\label{fig:Flare_LOTau}
\end{center}
\end{figure}

In Fig.~\ref{fig:Flare_LOTau} we show the \lamost\ MRS spectra of LO\,Tau taken from November 2019 to December 2021. The strong and broad H$\alpha$ profile with emission wings extending up to $\approx\pm500$\,\kms\ observed at MJD\,=\,59216 is apparent. Another distinctive feature of this spectrum is the \ion{He}{i}\,$\lambda$6678\,\AA\ emission line, which has never been observed during the quiescent phase.
Emission in \ion{He}{i} lines, notably the D$_3\,\lambda$5876\AA\ line, has been reported during strong flares of RS~CVn systems \citep[e.g.,][]{Montes1997,Garcia-Alvarez2003,Frasca2008,Cao2019} as well as in the strongest solar flares \citep[e.g.,][]{Yakovkin2024}.
Unfortunately, our \lamost\ spectra are not taken during \ktwo\ or \tess\ observations, and therefore no high-precision contemporaneous photometry is available.

Another strong flare has been clearly detected on \gaia-DR3\,66555573432261376 (= TIC\,640641946 = V371\,Tau).
 V371\,Tau has nearly the same spectral type as LO\,Tau \citep[M2e,][]{Prosser1991} but a lower \vsini\,=\,5.9\,\kms\ \citep{Abdurro2022}. The lower rotation rate is confirmed by the longer \prot\,=\,4.34\,d we derived from the \tess\ light curves. From the analysis of the \lamost\ MRS spectra, we found the same spectral type (M2V) but a different value of \vsini\,=\,33.9\,\kms; we also noticed RV variations with a $P(\chi^2)=0.006$ that make this star an SB1 candidate. Indeed, it is a visual binary composed of two stars with $G=14.5$\,mag and $G=15.3$\,mag separated by 2\arcsec, whose light entered the \lamost\ fiber. The RV variation and the different \vsini\ values can be the result of the binarity.
Although not explicitly mentioned in Simbad, this star displays frequent white-light flares in the space-based photometry. An example of a portion of a \tess\ light curve with two flares is shown in Fig.\,\ref{fig:Flare_TESS2}.
The spectrum acquired on MJD\,=\,59216 displays a strong and broad H$\alpha$ profile and \ion{He}{i} emission, which are not observed in the quiescent phase
 (Fig.\,\ref{fig:Flare_V371Tau}).

The third object displaying a spectrum with flare characteristics is \gaia-DR3\,66739982146803456 (= TIC\,35155775 = V343\,Tau = HII 1785). It is classified as a young stellar object in Simbad, where an M1.4 spectral type \citep{Birky2020}
and a \vsini\,=\,7.2\,\kms\ \citep{Abdurro2022} are reported. We find a slightly earlier spectral type (K6V) and a \vsini\,$<$\,8\,\kms. The spectrum observed at MJD\,=\,59531 (Fig.\,\ref{fig:Flare_V343Tau}) is typical of a flare event.

\section{The age of the Pleiades}
\label{Sec:age}

The Pleiades, along with $\alpha$ Per and Hyades, is one of the best known OCs. It has been long used as benchmark to probe stellar evolution and calibrate the theoretical models due to its proximity and low foreground extinction.

However, the accepted (canonical) age for the Pleiades has changed by a factor of two in recent decades. To calculate the age of a cluster, the most widely used procedure is the isochrone-fitting method, in which theoretical models of different ages (isochrones) are superimposed on the CMD of the cluster until the isochrone that best reproduces the arrangement of the cluster members in the diagram is found. In this case, the position of giants (not present in the Pleiades) and stars in the upper MS, close to the turn-off point, are essential to make a good fit and, therefore, find a precise age. In the latter, the treatment of the convective mixing after exhausting the hydrogen in their cores plays a fundamental role in their (observable) properties. In the 1970s and 1980s, classical models (aside from using a different chemical composition) did not include the overshooting as most modern models do. The resulting consequence is that the first age estimates for the Pleiades were $\leq$\,80\,Myr \citep{Hazlehurst70, Simpson70, Mermilliod81,Steele93}. On the contrary, nowadays, works devoted to it find older ages, about 110--160\,Myr \citep{Mazzei89, Gossage18}. However, even in current studies using an automated approach on a large cluster sample and taking $Gaia$ distances into account \citep{Kharchenko13,Cantat2020,Dias21,Hunt2024,Cavallo24} the age spread still persists, giving values in the range $\approx$\,80--140\,Myr.

The determination of the age of young clusters is affected not only by the lack of giant or turn-off stars, but also by the radius inflation caused the strong magnetic activity in their late-type members \citep[e.g.,][]{Jackson2009, Jackson2014b}. The average radii of M-type dwarfs in the Pleiades have been found to be on average $\sim$\,14\,\% larger than inactive field M dwarfs \citep{Jackson2018}. Similar results have been found for other young MS or PMS clusters \citep{Jackson2016}. If not properly taken into account by the evolutionary models, the presence of starspots and magnetic fields, which make the low-mass stars redder and brighter, would lead to an age underestimation \citep[e.g.,][]{Jeffries2023a}. 

On the other hand, an alternative to the isochrone-fitting method is the Li depletion boundary (LDB). This tool has proven to be very useful in determining the age of young OCs \citep[$\approx$\,20--200\,Myr, see e.g.][]{Stauffer1998,BN04}. Indeed, this technique, according to \citet{Soderblom14}, would have the advantage of being model-independent, which is not a closed debate \citep{Gossage18}. In any case, the results derived from the LDB for the Pleiades (using both MS stars and brown dwarfs) show a high level of agreement around 120\,Myr \citep{Stauffer1998,Martin01,Dahm15}. This age is currently the accepted canonical age for the Pleiades and is fully compatible with that obtained from isochronal analysis, whose stellar models include moderate rates of convective mixing and rotation.

Taking the results presented in Sect.~\ref{Sec:APs} into account we estimated the cluster age from the fitting of empirical lithium isochrones. In this way, as done previously \citep[see e.g.][]{Radcliffe,M39}, we used the code \eagles\ \citep{Jeffries2023}. This code fits  Li-depletion isochrones to the values of \teff\ (in the range 3000--6500\,K) and $W_{\rm Li}$ of a coeval stellar group. We note that \eagles\ works better for clusters with a wide \teff\ distribution of members. This is indeed the case for our \lamost\ \WLi\ measurements. For our purposes, we have selected the best measurements, namely those with \WLi\ errors less than 15\,m\AA\ (in total 144 objects), among the stars comprised in the working temperature range of the code. The results of the application of this code to our data are shown in Fig.~\ref{fig:EAGLES}. We find an age of about 118$\pm$\,6\,Myr, considering only the detections; if we include the upper limits, the age remains basically unchanged  (122$\pm$\,6\,Myr). This age value is in line with the most recent determinations. We note that this result is very precise thanks to the large sample used, which covers a large part of the temperatures interval {used for the isochrone fitting, as mentioned above. The opposite occurred in one of our previous works \citep{Radcliffe}, where we used a sample of three stars with similar temperatures only, resulting in a poor age estimate of $75 \pm 36$\,Myr.

\begin{figure}
\begin{center}
\includegraphics[width=9.5cm]{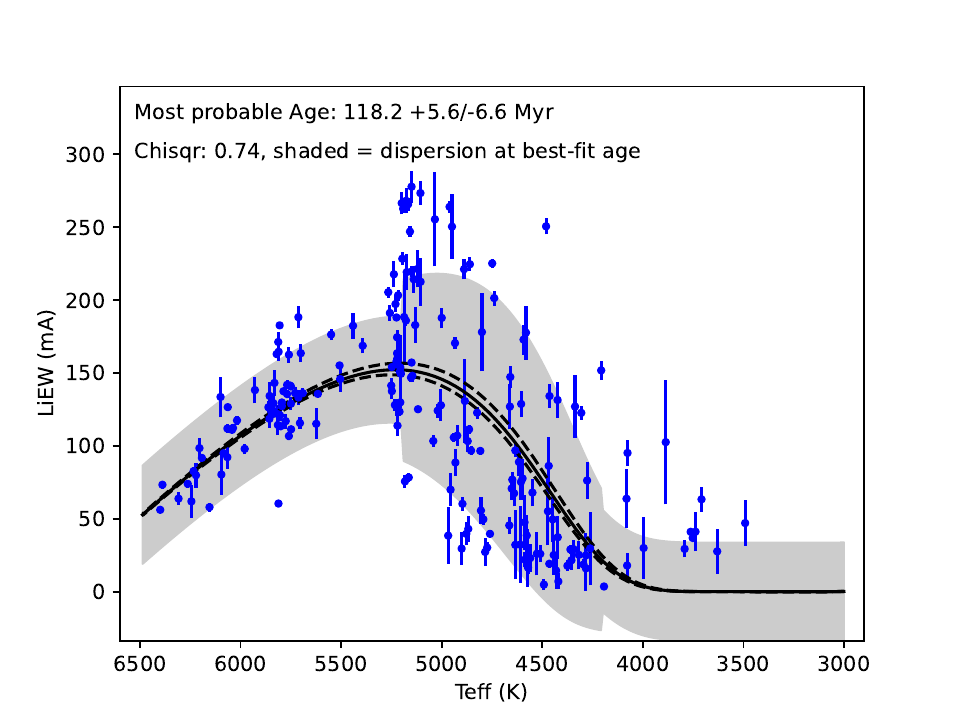}    
\caption{Fit to the lithium depletion pattern of the Pleiades \WLi\ measured in this work, obtained with the \eagles\ code \citep{Jeffries2023}.
}
\label{fig:EAGLES}
\end{center}
\end{figure}

\section{Spectroscopic binaries}
\label{Sec:SB}

The RV precision of \lamost\ MRS is sufficient to detect binary systems and to  derive their orbital elements and the physical parameters of the system components from the analysis of their RV curves \citep[see, e.g.,][]{Pan2020,Wang2021}. 
As mentioned in Sect.\,\ref{Sec:Analysis}, in some spectra 
corresponding to ten sources we noted two peaks in the CCF that are significant in comparison with the CCF noise, which enabled us to classify these objects as SB2s. No stellar parameters are reported in Table~\ref{Tab:APs} for these objects.
In some cases the binary nature is questionable because the CCF displays only a distorted peak, which can either be due to an unresolved companion or to starspots. We have flagged these sources as `SB2?' in Table~\ref{Tab:APs}. 
Whenever an  SB2 system is observed near the conjunctions, only one peak is visible in the CCF and the RV, measured with a single Gaussian fit, has to be considered as a ``blended'' value of the RVs of the two components, which is close to the velocity of the center of mass of the system.

In addition, we have objects with a single CCF peak that display a clear RV variation (SB1s). For these systems, for which a single component is visible in the spectrum, we report the APs. 
Some of these binaries were already known and their RVs and, in some cases, the orbital solutions were reported in the literature \citep[e.g.,][]{Mermilliod1992,Torres2021}.
In the following, we show some examples of RV curves of well-studied systems, some of which, when treated as single stars, are among the most discrepant objects compared to the APOGEE RVs (Fig~\ref{Fig:RV_comp}). Our main aim is to show the agreement of our RV data with the RV curves in the literature. We also show cases of poorly studied or unknown system for which we could build an RV curve and made a preliminary orbital solution.

The first case is HII\,571, discovered as an SB1 by \citet{Mermilliod1992}, who also reported an orbital period of 15.9 days and provided an orbital solution. New data and a recent orbital solution were presented by \citet{Torres2021}. The RV curve of HII\,571 based on these data, three APOGEE epochs \citep{Abdurro2022}, and one \lamost-\rotfit\ RV is presented in the left panel of Fig.~\ref{Fig:RV_HII571}, where the orbital solution of \citet{Torres2021} is also displayed. 
Two other SB1s discovered by \citet{Mermilliod1992} are HII\,2172 ($P_{\rm orb}\simeq 30.2$\,d) and HII\,2406 ($P_{\rm orb}\simeq 33.0$\,d), whose RV curves are also displayed in the middle and right panel of Fig.~\ref{Fig:RV_HII571}, respectively, along with the literature data and the solutions of \citet{Torres2021}. We note that \citet{Torres2021} were able to detect the secondary CCF peak in their TRES spectra pf HII\,2406, making this an SB2 system, but this component was not visible in the \lamost\ spectra analyzed by us, so that we have only shown the RV curve of the much brighter primary star. The agreement of our \lamost-MRS velocities derived with \rotfit\ with the literature data is apparent.

\begin{figure*}[th]
\begin{center}
\includegraphics[width=6cm,viewport= 10 0 440 400]{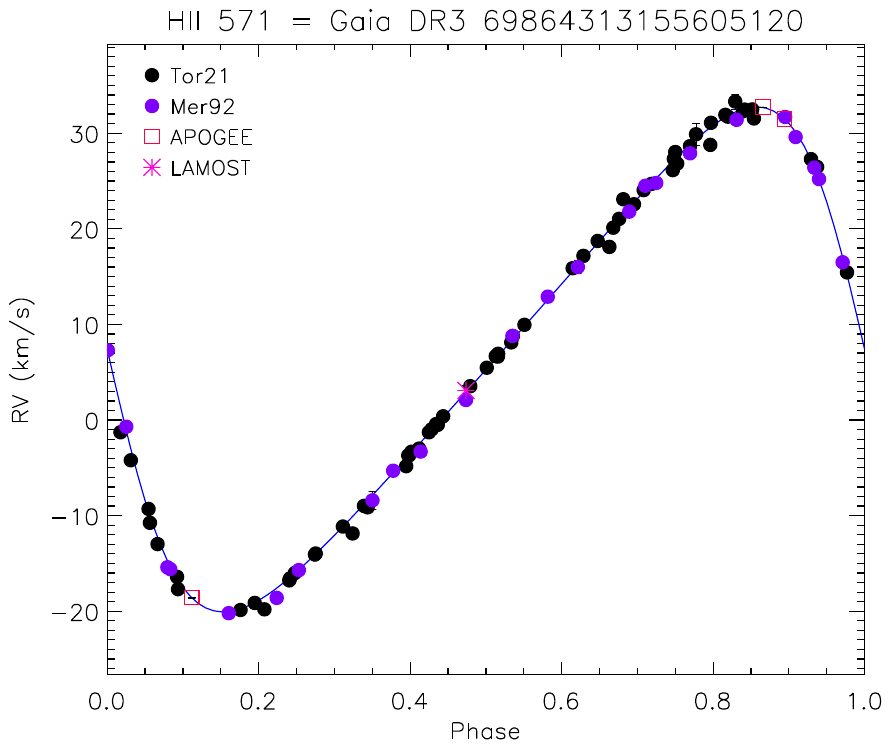}           
\includegraphics[width=6cm,viewport= 10 0 440 400]{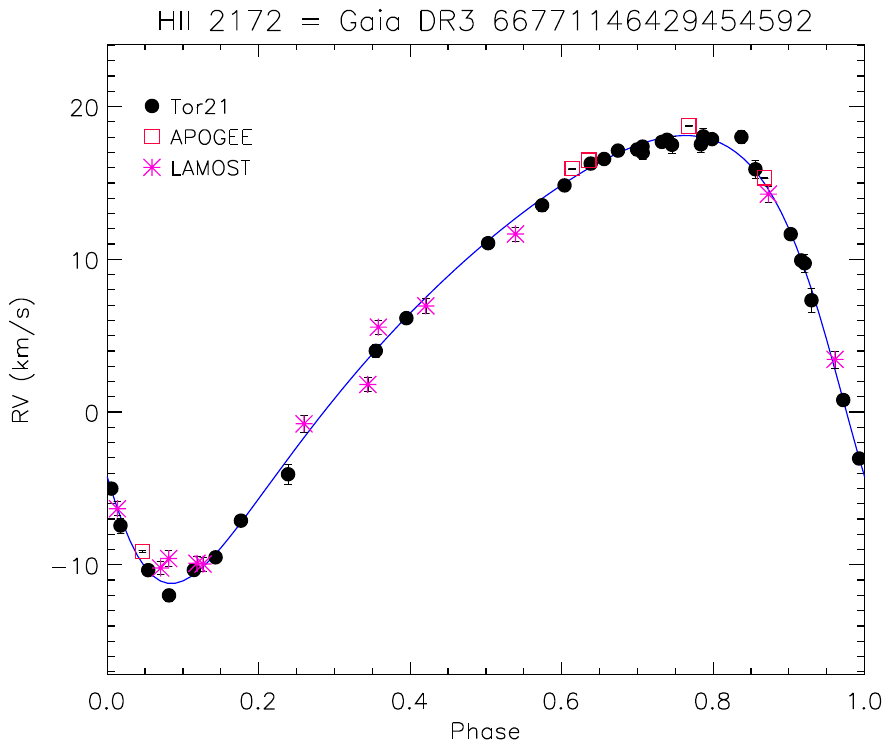}           
\includegraphics[width=6cm,viewport= 10 0 440 400]{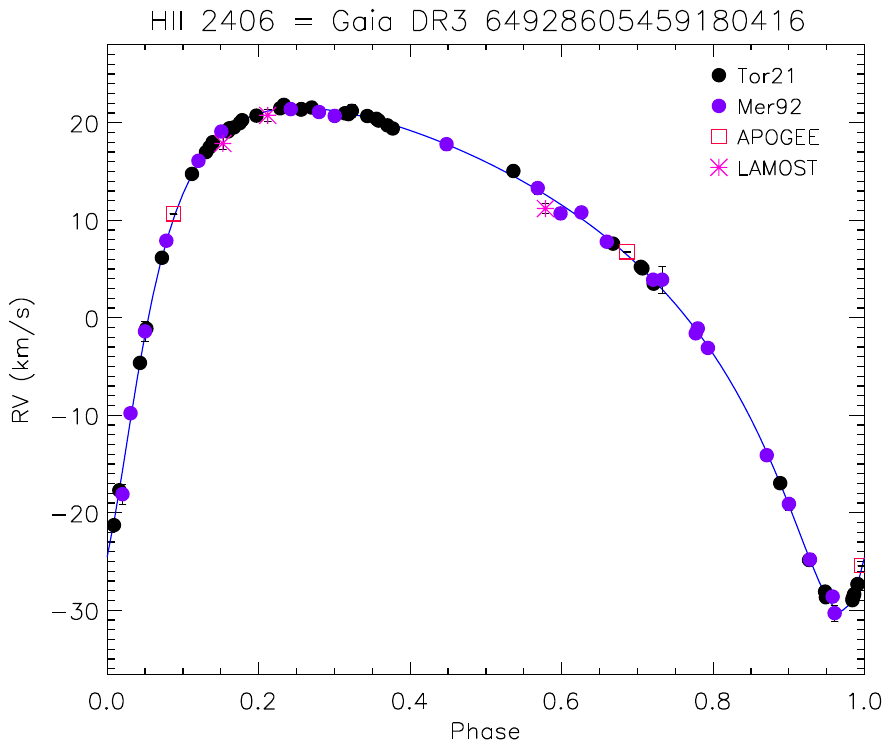}           
\caption{Radial velocity curves of three single-lined binaries: HII\,517 (left panel), HII\,2172 (=HD\,282965; middle panel), and HII\,2406 (right panel). The RV data of \citet{Mermilliod1992} and \citet{Torres2021} are shown by purple and black dots, respectively, while the APOGEE DR17 data \citep{Abdurro2022} and our \lamost\ RVs are plotted with red squares and asterisks, respectively. The orbital solutions are superimposed with a full blue line.
}
\label{Fig:RV_HII571}
\end{center}
\end{figure*}

Another interesting case is the short-period ($P_{\rm orb}\simeq 5.69$\,d) double-lined binary DH\,794 discovered by \citet{Torres2021}, who found a slightly eccentric orbit ($e\simeq0.013\pm0.002$). 
We could measure with \rotfit\ the RVs of both components of this system in three \lamost\ MRS spectra, while in the remaining three only the peak of the primary component could be identified. In addition, there are four 
APOGEE spectra taken with the 2.5-m telescope of the SDSS at the Apache Point Observatory. 
Since the radial velocities measured on these spectra and reported in the SDSS DR17 \citep{Abdurro2022} were derived by assuming the object 
as a single star, they are related to the primary (brighter) component only or are a ``blended'' measure taken when the two components are not well separated in wavelength. We therefore downloaded the APOGEE spectra\footnote{Available at \tt{http://skyserver.sdss.org/dr17/}} and derived the velocities of both components by cross-correlating the continuum-normalized APOGEE spectra with a template made with a BT-Settl synthetic spectrum \citep{Allard2012}, with [Fe/H]=0, \logg=4.0 and  \teff\ in the range 3800--4800\,K, similar to what we did in \citet{Frasca2021} for a newly discovered SB2 system in Orion. 
Our \lamost-\rotfit\ RV measures and those we measured on the four DR17 APOGEE data are reported in Table\,\ref{Tab:RV_DH794} and show a very good agreement with the \cite{Torres2021} data and orbital solution (Fig.\,\ref{Fig:RV_DH794}).

The same holds true for HII\,761, which was discovered as an SB1 system by \citet{Mermilliod1992} and subsequently classified as an SB2 by \citet{Torres2021} who were able to measure the RV of the secondary component, although with a lower accuracy than the primary, in their spectra. We were also able to detect the CCF peak of the secondary component in the APOGEE spectra and in most of the \lamost-MRS ones and report these data in Table\,\ref{Tab:RV_HII761}. Our RV measures follow closely the circular orbital solution of \citet{Torres2021}  (middle panel of Fig.\,\ref{Fig:RV_DH794}).
A third SB2 system, discovered by \citet{Torres2021}, is HCG\,495. The RV measured by us on three APOGEE and one \lamost\ MRS spectra, reported in Table\,\ref{Tab:RV_HCG495} and displayed in the right panel of Fig.~\ref{Fig:RV_DH794}, allowed us to refine the orbital solution proposed by \citet{Torres2021} and the new orbital elements are listed in Table~\ref{Tab:Param_HCG495}.

\begin{table}[ht]
\caption{Orbital parameters of HCG\,495.}
\begin{tabular}{lr}
\hline
\hline
\noalign{\smallskip}
Parameter &  Value   \\ 
\hline
\noalign{\smallskip}
 HJD0$^a$      & 56967.12$\pm$0.05  \\
 $P_{\rm orb}$ ($d$) & 8.5797$\pm$0.0005    \\
 $e$            &  0. 147$\pm$0.005  \\
 $\omega$ (\degree) &  278.7$\pm$1.3   \\
 $\gamma$ (\kms)    &  $6.06\pm$0.20 \\
 $K_1$ (\kms)       &  52.43$\pm$0.35  \\
 $K_2$ (\kms)       &  54.14$\pm$0.43  \\
 $M_1\sin^3i$ ($M_{\sun}$) & 0.530$\pm$0.009 \\
 $M_2\sin^3i$ ($M_{\sun}$) & 0.513$\pm$0.008 \\
 $M_2$/$M_1$  &  0.97$\pm$0.01  \\
 $a\sin i$ ($R_{\sun}$) & 17.87$\pm$0.09  \\
\noalign{\smallskip}
\hline \\
\end{tabular}
\label{Tab:Param_HCG495}
\begin{list}{}{}
\item[$^{(a)}$] Heliocentric Julian Date (HJD-2,400,000) of the 1st conjunction (primary behind).
\end{list}
\end{table}

\begin{figure*}[th]
\begin{center}
\includegraphics[width=6cm,viewport= 10 0 440 400]{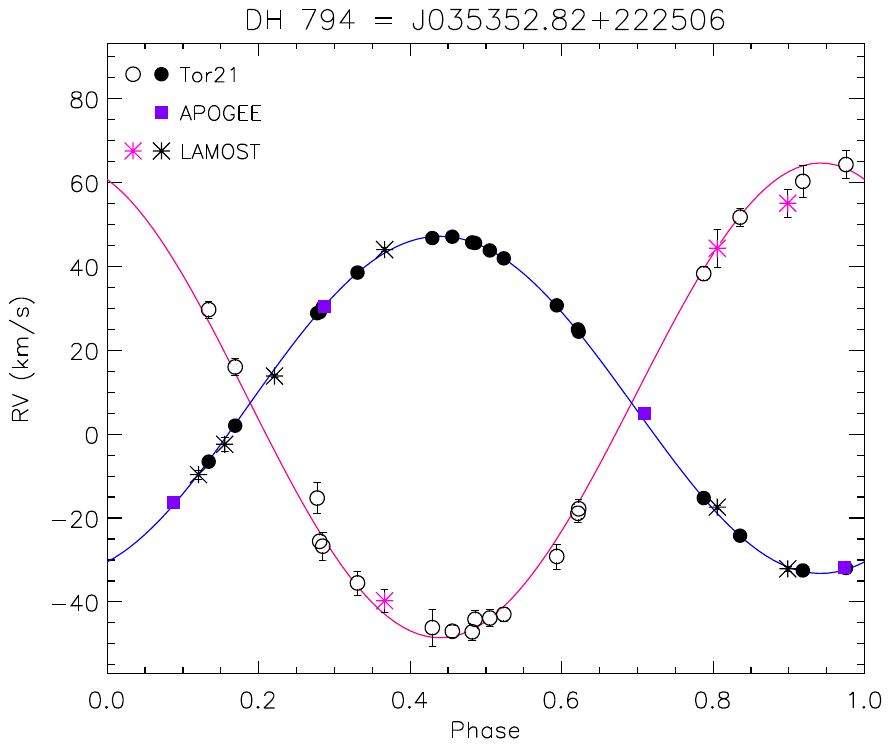}           
\includegraphics[width=6cm,viewport= 10 0 440 400]{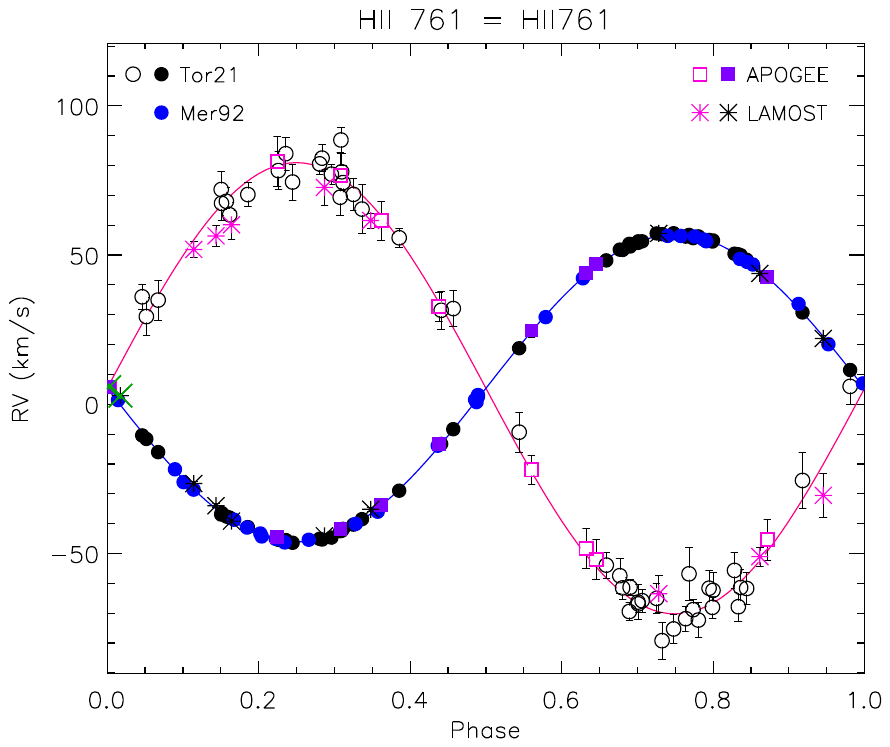}           
\includegraphics[width=6cm,viewport= 10 0 440 400]{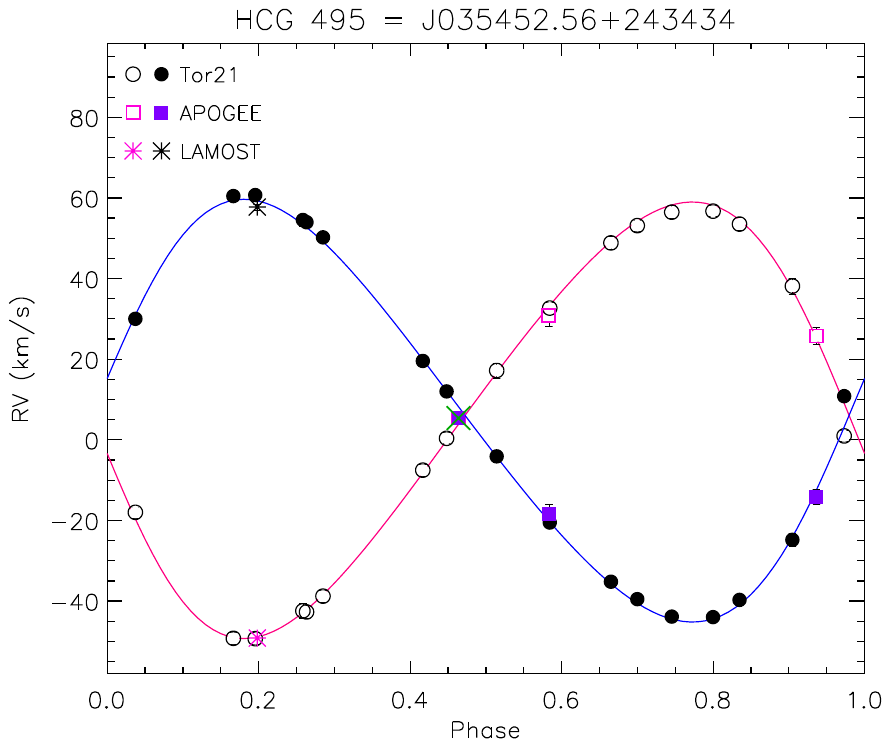}           
\caption{Radial velocity curves of three SB2 binaries: DH\,794 (left panel), HII\,761 (middle panel), and HCG\,495 (right panel). The RV data of \citet{Torres2021} are shown by black dots and open circles for the primary and secondary component, respectively. The RVs of the primary and secondary components measured by us from the APOGEE DR17 spectra are plotted with purple filled and magenta open squares, respectively. Our \lamost\ RVs are plotted with black and magenta asterisks for the primary and secondary component, respectively. The RV values measured near the conjunctions, when only one CCF peak is visible, are marked with green crosses.  The orbital solution of \citet{Torres2021} is superimposed with a full blue and red line for the primary and secondary component, respectively.
}
\label{Fig:RV_DH794}
\end{center}
\end{figure*}

\begin{table}[ht]
\caption{Heliocentric RVs of QQ\,Tau (= HII\,1286).}
\begin{tabular}{lrcrcl}
\hline
\hline
\noalign{\smallskip}
 HJD     &  RV$_{\rm 1}$   &  $\sigma_{\rm RV_1}$ &  RV$_{\rm 2}$   &  $\sigma_{\rm RV_2}$ & Instrument \\
(2\,400\,000+) &  \multicolumn{2}{c}{(\kms)} & \multicolumn{2}{c}{(\kms)} & \\
 \hline
\noalign{\smallskip}
55847.869  &  $-$30.30  &  2.21  &  43.61  &  2.75  &  APOGEE \\
55851.863  &      4.01  &  1.21  &   4.01  &  1.21  &  APOGEE \\
55854.888  &  $-$39.81  &  2.12  &  52.66  &  2.91  &  APOGEE \\
57408.659  &  $-$40.60  &  2.18  &  54.15  &  2.99  &  APOGEE \\
57649.914  &  $-$39.28  &  1.89  &  53.26  &  2.82  &  APOGEE \\
57652.945  &  $-$10.33  &  1.47  &  22.79  &  1.82  &  APOGEE \\
58802.2417 & $-$40.34  &  2.26  &     47.26  &   2.60  & LAMOST  \\ 
58820.1521 &  $-$0.30  &  0.83  &   $-$0.30  &   0.83  & LAMOST  \\ 
58830.1414 &  $-$0.37  &  0.97  &   $-$0.37  &   0.97  & LAMOST  \\ 
58836.0329 &    27.85  &  4.07  &  $-$29.47  &   6.29  & LAMOST  \\ 
58852.0889 &  $-$7.86  &  8.66  &     28.98  &  13.90  & LAMOST  \\ 
58883.9938 & $-$26.56  &  1.95  &     35.60  &   2.74  & LAMOST  \\ 
58890.9855 & $-$43.56  &  1.42  &     50.17  &   1.91  & LAMOST  \\ 
59124.3258 &     0.80  &  0.68  &      0.80  &   0.68  & LAMOST  \\ 
59148.2495 &    44.04  &  1.70  &  $-$44.84  &   2.40  & LAMOST  \\ 
59181.1202 & $-$15.31  &  9.89  &     19.79  &  11.76  & LAMOST  \\ 
59190.1513 &    47.21  &  1.20  &  $-$47.93  &   1.57  & LAMOST  \\  
59216.0946 & $-$46.11  &  1.59  &     48.37  &   2.19  & LAMOST  \\ 
59531.1785 & $-$46.12  &  1.41  &     48.16  &   1.96  & LAMOST  \\ 
\noalign{\smallskip}
\hline \\
\end{tabular}
{\bf Notes.} The RV of the single CCF peak observed near the conjunctions has been assigned to both components. These values have not been considered for the fit of the RV curve.
\label{Tab:RV_QQTau}
\end{table}

We discovered QQ\,Tau (= HII\,1286) as a new SB2 system. It is classified as an ``eruptive variable'' (likely related to the strong flare activity detected by space photometry) in the Simbad database, with no indication of binarity.
We have augmented the 13 RV values derived with \rotfit\ from \lamost\ MRS spectra taken from 14/11/2019 to 12/11/2021 with RV measurements based on six APOGEE spectra taken from 13/10/2011 to 21/09/2016. 
All these RVs are reported in Table\,\ref{Tab:RV_QQTau}. 
We applied the Lomb-Scargle periodogram \citep{Scargle1982} with the CLEAN deconvolution algorithm \citep{Roberts1987} to the RV data of the primary and secondary component to search for the orbital period of the system, which turned out to be 2.462\,days, in close agreement with the one we derived from the periodic analysis of the \tess\ photometry.
Given the short period and the small number of RV points with relatively large errors we have searched for a circular orbit as a preliminary solution. We note that the solution improves if we apply a  shift of +5 \kms\ to the \lamost-MRS RVs.
The RV curve and solution are shown in Fig.\,\ref{Fig:RV_QQTau} and the orbital parameters are listed in Table\,\ref{Tab:Param_QQTau}. We remark that this is a preliminary solution and new  medium- or high-resolution spectra are highly desirable to improve it. 

\begin{figure}[th]
\begin{center}
\includegraphics[width=8.7cm,viewport= 10 0 460 410]{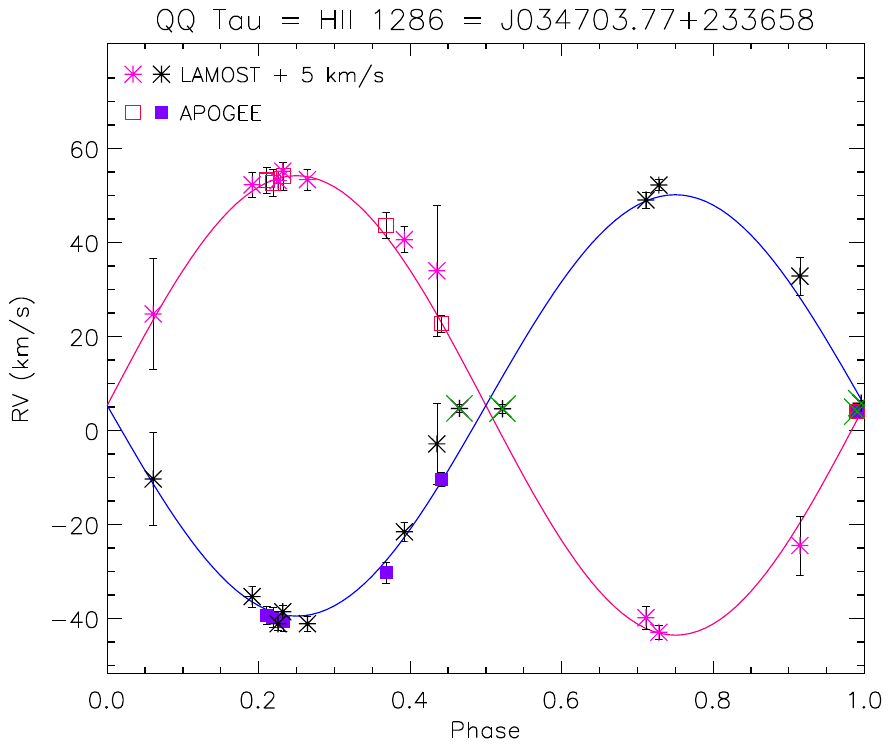}           
\caption{Radial velocity curve of the newly discovered SB2 binary QQ\,Tau (= HII\,1286). The RVs measured by us from the APOGEE spectra are marked with purple filled  and red open squares, respectively. Our \lamost\ MRS RVs are shown with black and magenta asterisks for the primary and secondary component, respectively. The RVs taken near the conjunctions, when only one peak is visible, are marked with large green X symbols. The RV solution for a circular orbit is represented by the full blue and red line for the primary and secondary component, respectively. 
}
\label{Fig:RV_QQTau}
\end{center}
\end{figure}

\begin{table}[ht]
\caption{Orbital parameters of QQ~Tau (= HII\,1216).}
\begin{tabular}{lr}
\hline
\hline
\noalign{\smallskip}
Parameter &  Value   \\ 
\hline
\noalign{\smallskip}
 HJD0$^a$      &  55844.50$\pm$0.01  \\
 $P_{\rm orb}$ ($d$) & 2.46234$\pm$0.00008    \\
 $e$            &  0.0  \\
 $\gamma$ (\kms)    &  $5.33\pm$0.34 \\
 $K_1$ (\kms)       &  44.82$\pm$0.55  \\
 $K_2$ (\kms)       &  46.86$\pm$0.73  \\
 $M_1\sin^3i$ ($M_{\sun}$) & 0.100$\pm$0.003 \\
 $M_2\sin^3i$ ($M_{\sun}$) & 0.096$\pm$0.003 \\
 $M_2$/$M_1$  &  0.96$\pm$0.02  \\
 $a\sin i$ ($R_{\sun}$) & 4.46$\pm$0.06  \\
\noalign{\smallskip}
\hline \\
\end{tabular}
\label{Tab:Param_QQTau}
\begin{list}{}{}
\item[$^{(a)}$] Heliocentric Julian Date (HJD-2,400,000) of the 1st conjunction (primary behind).
\end{list}
\end{table}

\section{Summary}
\label{Sec:Concl}

We have presented the results of the analysis, performed with the code \rotfit, of 1581 \lamost\ medium-resolution spectra of cool (FGKM-type) stars that are candidate members of the Pleiades open cluster. 
We were able to determine the APs (\teff, \logg, and \feh), the radial velocity (RV), and the projected rotational velocity (\vsini) for most of these spectra corresponding to 283 different stars.

Due to systematic offsets in different runs of \lamost, we corrected the RVs measured in the blue and red arms for all the target spectra
using spectra of RV standard stars present in the same plates as our Pleiades samples. We discussed the RV distribution and spotted the stars with variable RV in our catalog. 
The average uncertainties for RV, \teff, \logg, and \feh\ are about 2.1 \kms, 277 K, 0.15 dex, and 0.09 dex, respectively.  
Typical \vsini\ uncertainties are 5--6 \kms, but we could only measure values larger than 8\,\kms\ due to the MRS resolution and sampling. We validated the accuracy of our APs by comparing them to values from APOGEE DR17 and other sources in the literature, which shows an overall good agreement, confirming the reliability of the methodology.
The RV distribution of the cluster members peaks at about 5.0\,\kms\ with a dispersion, measured by the $\sigma$ of the fitted Gaussian, of 1.4\,\kms, in line with other determinations. The mean metallicity of the cluster was found to be [Fe/H]\,=\,$-0.03\pm0.06$.

By subtracting the best-matching non-active, lithium-poor template spectra from each \lamost\ red-arm spectrum, we measured the equivalent widths of the Balmer H$\alpha$ core emission and \ion{Li}{i} $\lambda$6708\,\AA\ absorption lines. This allowed us to investigate the chromospheric emission and lithium absorption for these Pleiades members. All our targets show a chromospheric activity level compatible with the young cluster age and most of the coolest (\teff$<$5000\,K) members display a saturated activity level. 
For three sources with multi-epoch data we have acquired spectra during flares, which are characterized by strong and broad H$\alpha$ profiles and the presence of the \ion{He}{i}\,$\lambda$6678\AA\ emission line.
From the lithium abundance of the late-type members  (\teff$<$6500\,K) we provide a robust determination of the cluster age, $118\pm6$\,Myr, supporting the current accepted age for the Pleiades.

We also derived rotational periods (\prot) for 89 stars lacking \prot\ values in \citet{Rebull2016} using \tess\ data. We cross-matched our targets with those in \citet{Hartman2010} to compare \prot\ values. Among 187 targets in common, most showed excellent agreement, with 16 exhibiting significant discrepancies. For 12 of these having Rebull \prot\ values, we calculated their \tess\ \prot\ and finally confirmed that Rebull's (and our) measurements are most likely correct. Thanks to the values of \prot\ we could investigate the dependence of activity on the period or the Rossby number ($R_{\rm O}=P_{\rm rot}/\tau_{\rm con}$). Interestingly, we found that the stars in the saturated activity regime ($R_{\rm O}\lesssim0.21$) do not show a flat distribution of \rha\ but rather a slight enhancement with the decreasing \prot\ or Rossby number. 

Moreover, thanks to their RV variation in these time series data, we have found 39 candidates to single-lined spectroscopic systems (SB1s) and ten double-lined binaries (SB2s) that were identified based on the double-peak shape of the cross-correlation function (CCF). Many of them were already known as binaries from the literature, but some are not. In particular, we discovered the object QQ\,Tau to be an SB2 system for which we also provide a preliminary orbital solution of the RV curve. We also presented RV curves for some already known SB1s and SB2s in which the addition of our and APOGEE RVs allowed us to improve the solutions.

This work not only strengthens our understanding of the relationship between rotation and magnetic activity in young stars, but also highlights the importance of high-quality spectroscopic data for exploring key questions in stellar astrophysics.
Our results, in addition to displaying a robust dataset for future research, provide a benchmark for theoretical models of stellar activity and evolution in young clusters, offering new insights into the complex interplay between rotation, magnetic activity, and lithium abundance.

\section{Data availability}
\label{Sec:Availability}

Tables~\ref{Tab:APs}, \ref{Tab:Halpha}, \ref{Tab:lithium}, and \ref{Tab:RV_data} are only available at the CDS via anonymous ftp to {\tt cdsarc.u-strasbg.fr (130.79.128.5)} or via {\tt http://cdsarc.u-strasbg.fr/viz-bin/qcat?J/A+A/?/?}. 

\begin{acknowledgements}
We are grateful to the anonymous referee for a careful reading of the manuscript and very useful suggestions that helped us to improve our work.
Based on observations collected with the Large Sky Area Multi-Object Fiber Spectroscopic Telescope (\lamost) located at the Xinglong observatory, China.
Guoshoujing Telescope (the Large Sky Area Multi-Object Fibre Spectroscopic Telescope \lamost) is a National Major Scientific Project built by the Chinese Academy of Sciences. Funding for the project has been provided by the National Development and Reform Commission. \lamost\ is operated and managed by the National Astronomical Observatories, Chinese Academy of Sciences.
Support from the Italian {\it Ministero dell'Universit\`a e della Ricerca} (MUR) is also acknowledged. 
AF acknowledges funding from the Large-Grant INAF YODA (YSOs Outflow, Disks and Accretion). 
JZ and JNF acknowledge the support of National Natural Science Foundation of China (NSFC) through the Grants 12090040, 12090042 and 12427804, and the science research grants from the China Manned Space Project.
This research made use of SIMBAD and VIZIER databases, operated at the CDS, Strasbourg, France.
This publication makes use of data products from the Two Micron All Sky Survey, which is a joint project of the University of Massachusetts and the Infrared Processing and Analysis Center/California Institute of Technology, funded by the National Aeronautics and Space Administration and the National Science Foundation.
This publication makes use of data products from the Wide-field Infrared Survey Explorer, which is a joint project of the University of California, Los Angeles, and the Jet Propulsion Laboratory/California Institute of Technology, funded by the National Aeronautics and Space Administration.
\end{acknowledgements}

\bibliographystyle{aa}
\bibliography{Pleiades_lamost.bib}

\newpage
\eject

\begin{appendix}

\section{Additional tables and figures}
\label{Appendix:TabFig}

\begin{figure}[htb]
\includegraphics[width=8cm]{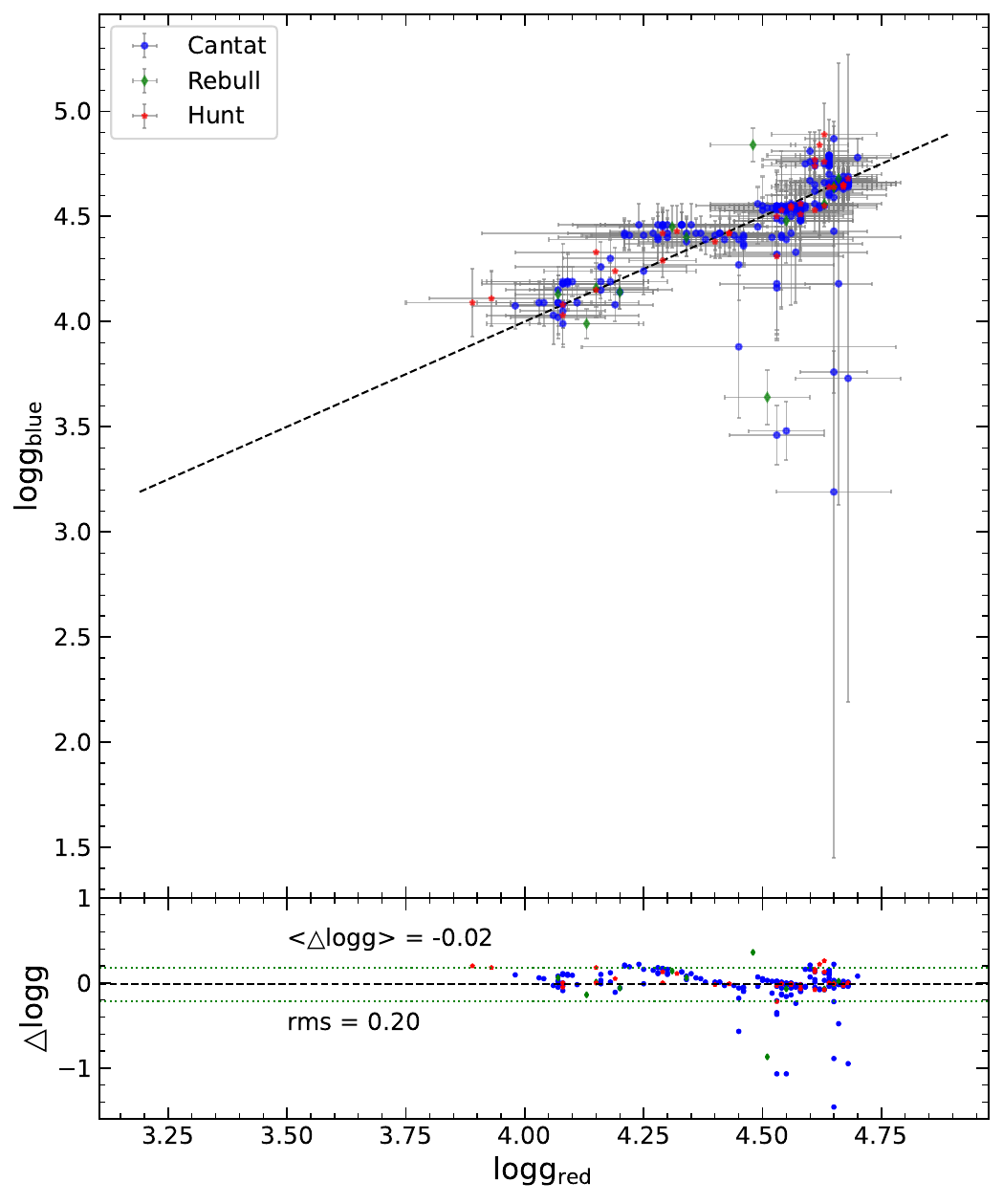}
\caption{Comparison of the \logg\ values derived from the blue- and red-arm LAMOST-MRS spectra with ROTFIT. The meaning of lines and symbols is the same as in Fig.~\ref{Fig:RV_blue_red}. Note the few discrepant points at \logg$_{\rm red}>4.5$ dex.}
\label{Fig:logg_blue_red}
\end{figure}

\begin{figure}[htb]
\includegraphics[width=8cm]{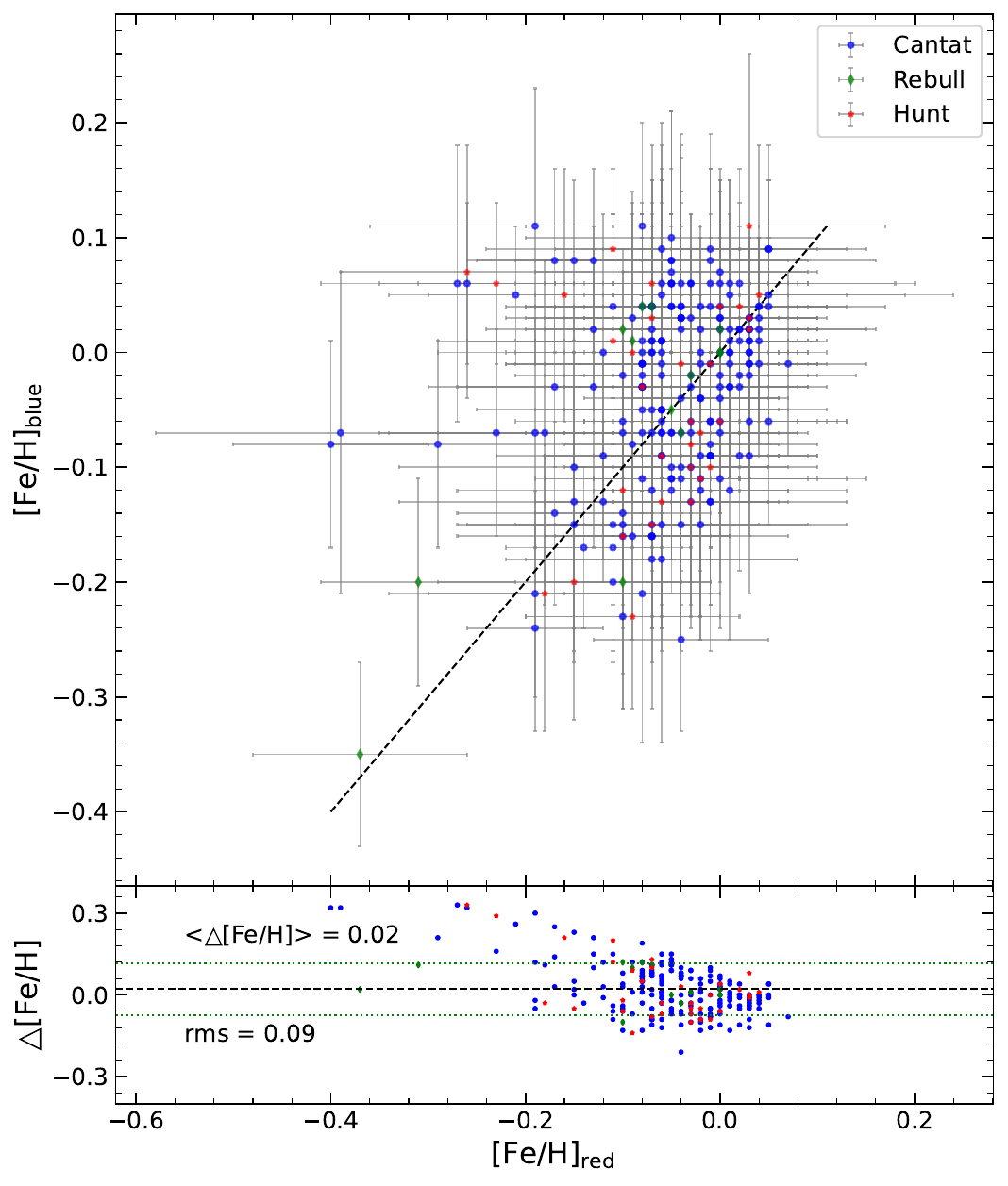}
\caption{Comparison of the metallicity values derived from the blue- and red-arm LAMOST-MRS spectra with ROTFIT. The meaning of lines and symbols is the same as in Fig.~\ref{Fig:RV_blue_red}. }
\label{Fig:feh_blue_red}
\end{figure}

\begin{figure}[htb]
\includegraphics[width=8cm]{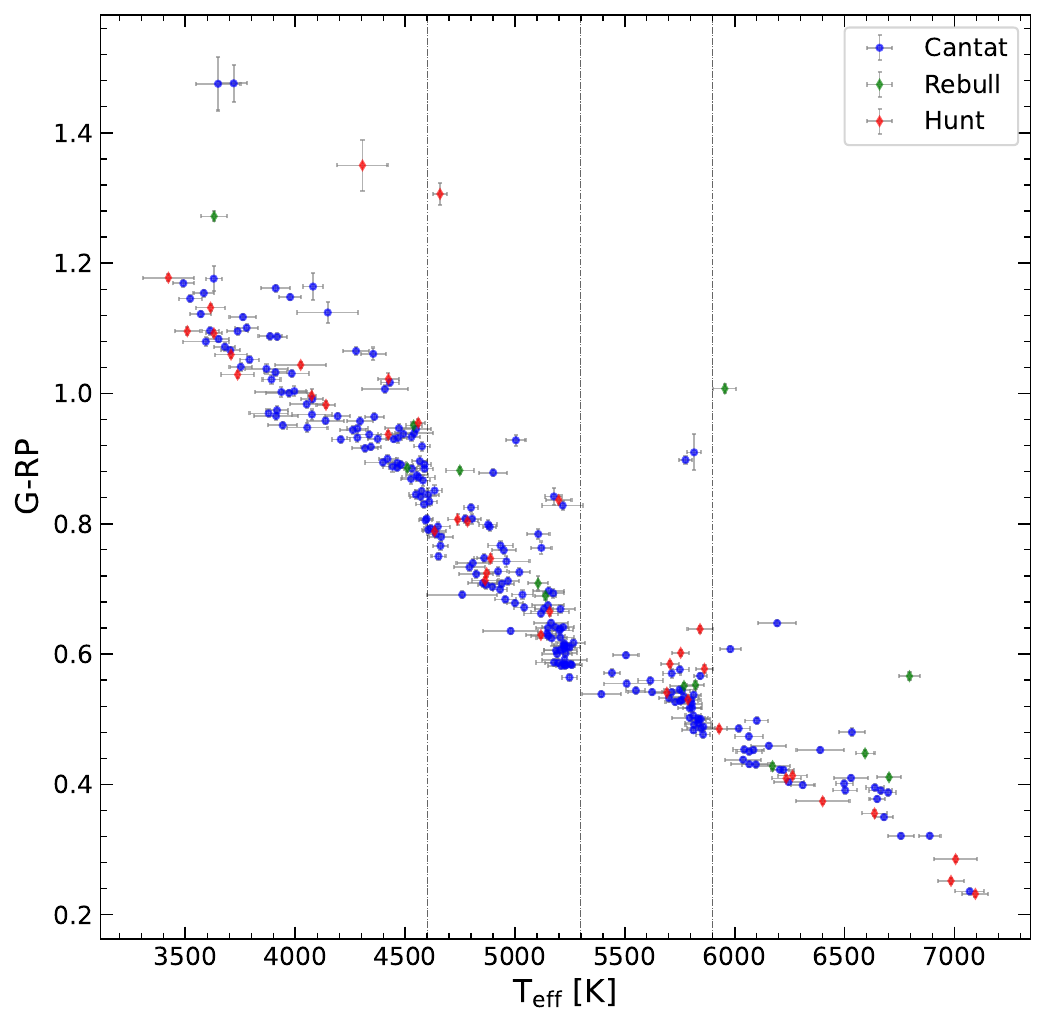}
\caption{Comparison of the \gaia\ $G$--$R_{\rm P}$ color and the \teff\ measured in this work for the stars belonging to the three subsamples as indicated in the legend.}
\label{Fig:Gaia_color}
\end{figure}

\begin{figure*}[htb]
\includegraphics[width=8.1cm]{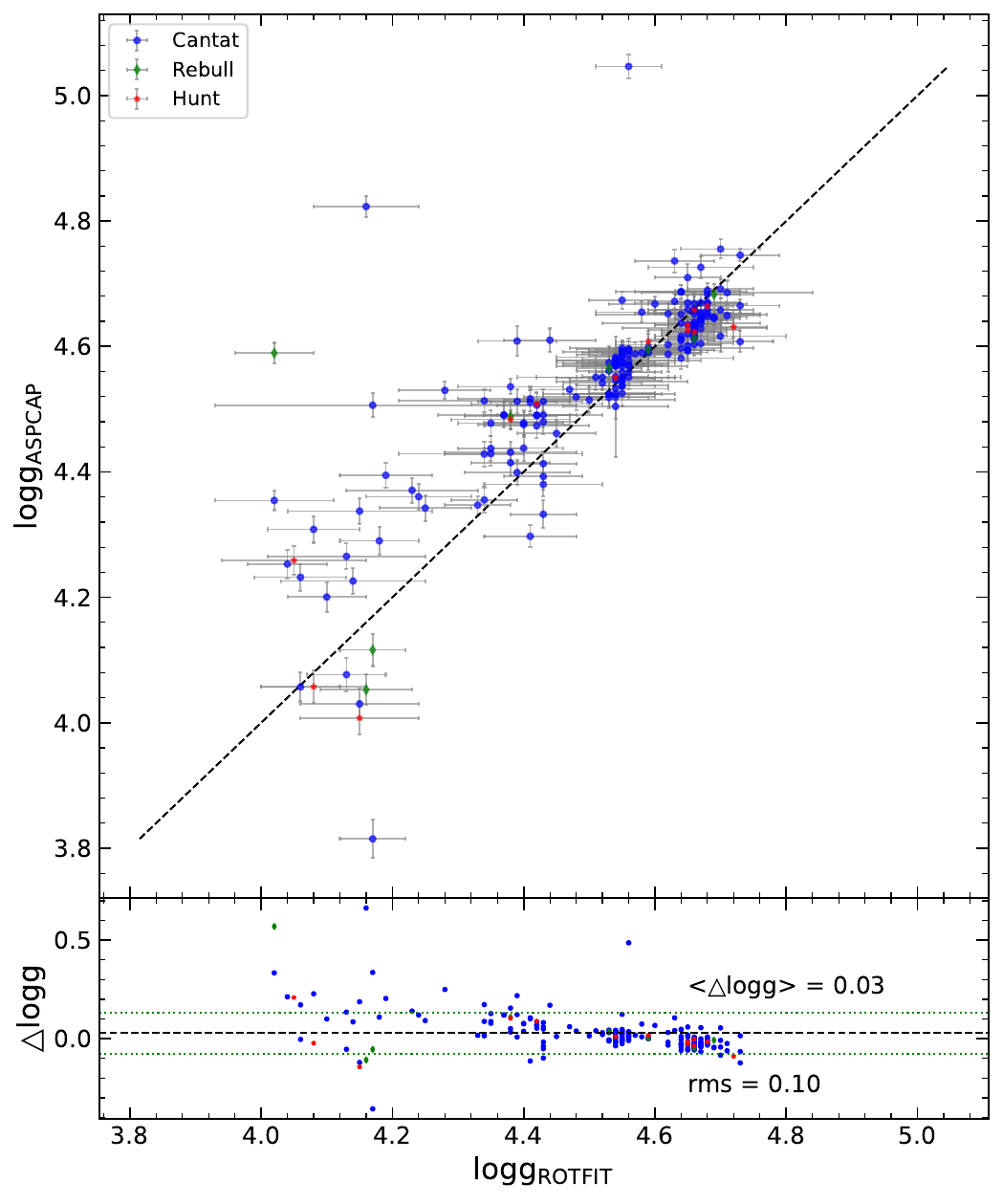}
\hspace{1cm}
\includegraphics[width=8cm]{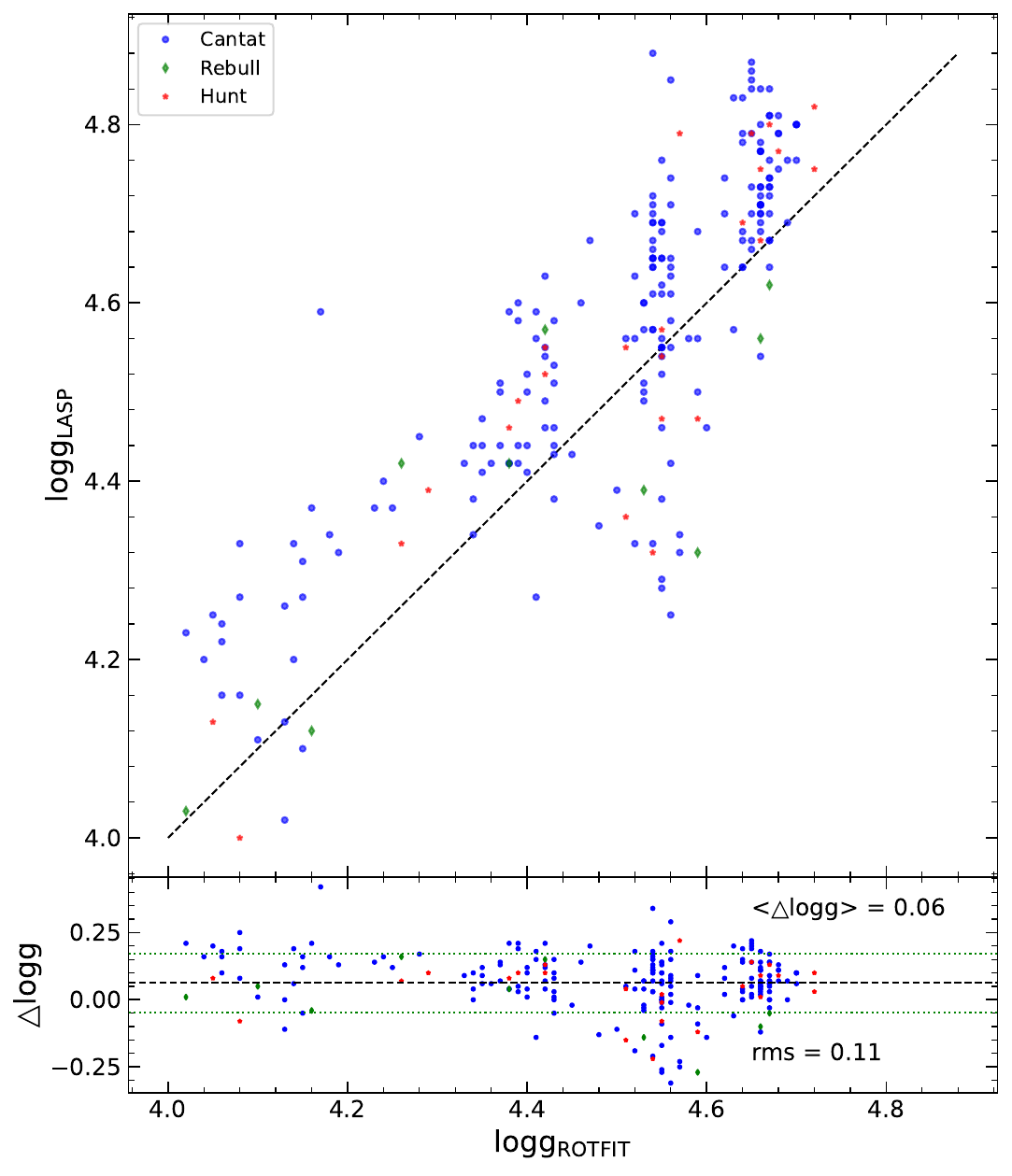}
\caption{Comparison between the \logg\ measured in this work and those found in the literature. Left panel: \rotfit\ versus APOGEE-2 DR17 values \citep{Abdurro2022}. 
The color of symbols distinguishes the three subsamples as in the previous figures and indicated in the legend. The one-to-one relation is shown by the dashed line. The \logg\ differences between \rotfit\ and APOGEE, $\Delta$\logg, are displayed in the lower box.
Right panel: Comparison with the \lasp\ gravity values ({\tt https://www.lamost.org/dr11/v1.0}). The meaning of lines and symbols is as in the left panel. }
\label{Fig:logg_comp}
\end{figure*}

\begin{figure*}[htb]
\includegraphics[width=8cm]{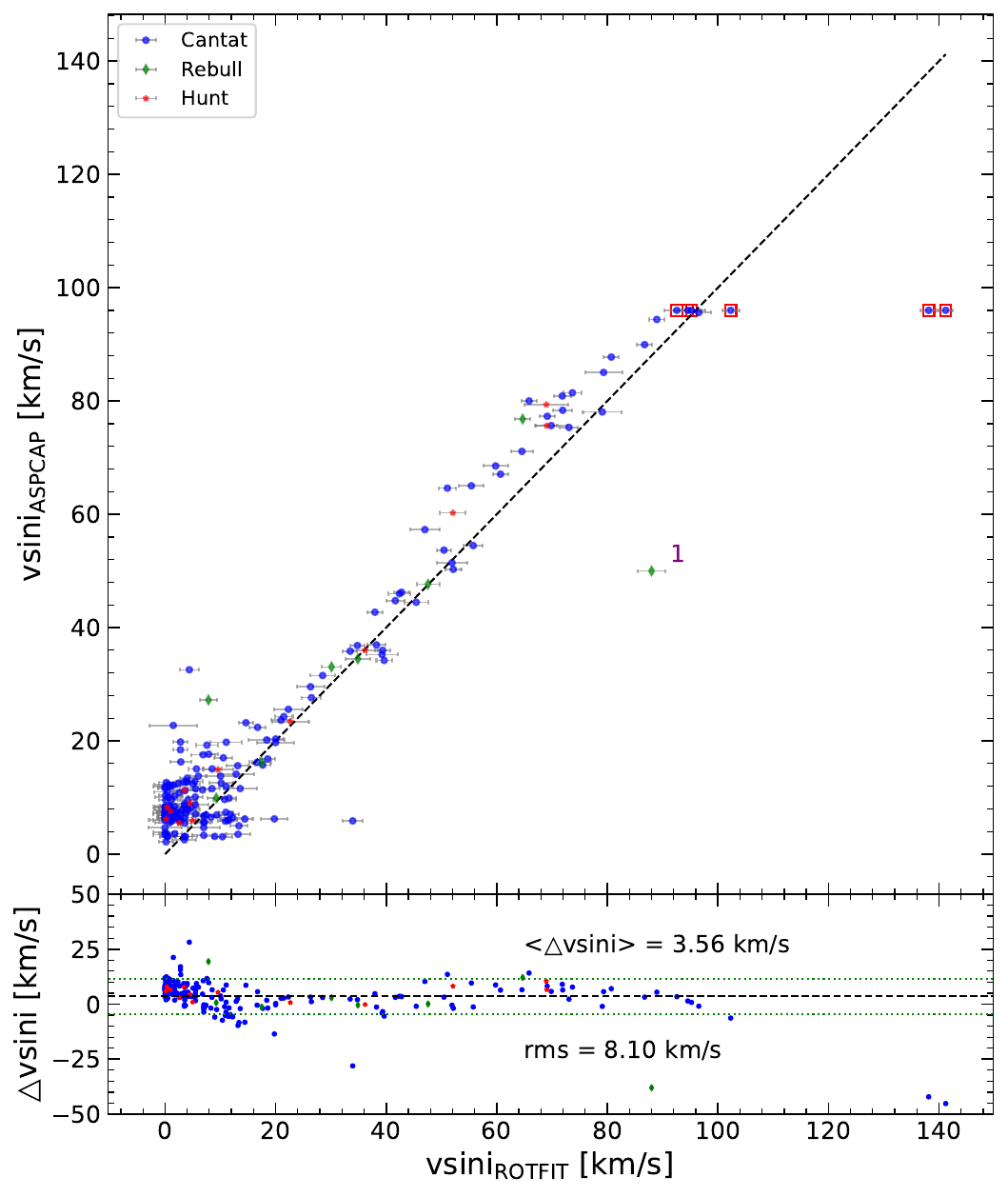}
\hspace{1cm}
\includegraphics[width=8cm]{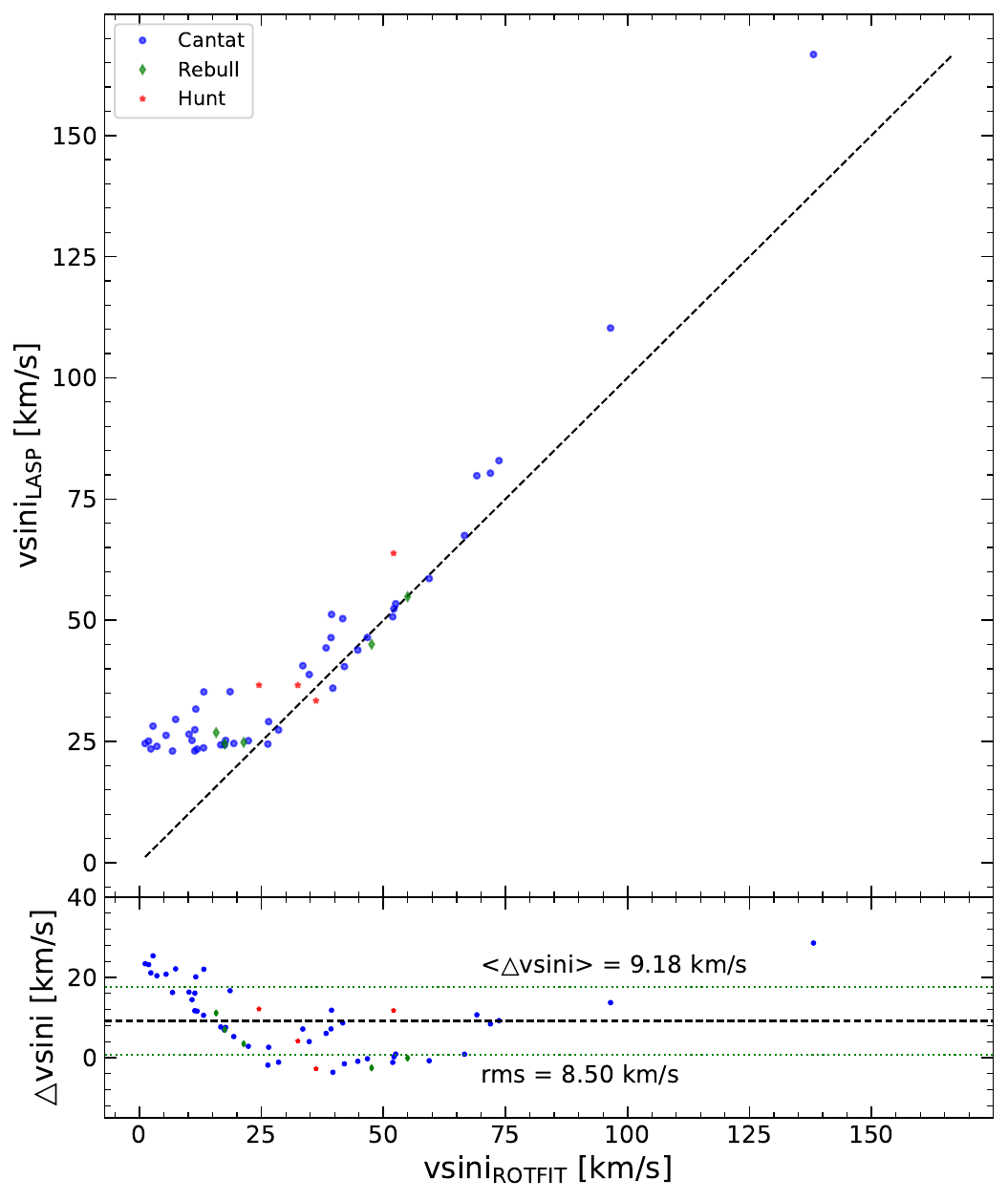}
\includegraphics[width=8cm]{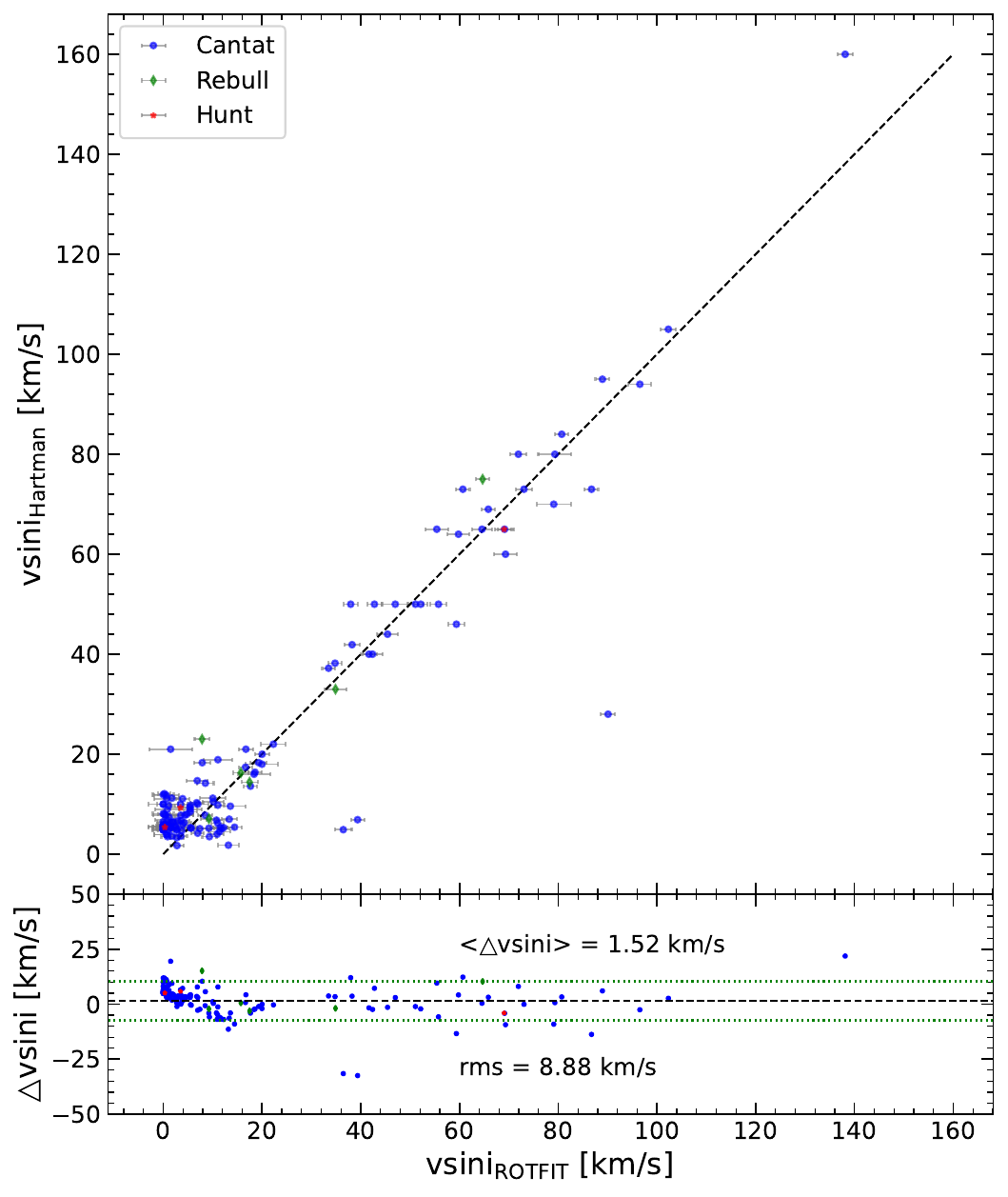}
\hspace{2cm}
\includegraphics[width=8cm]{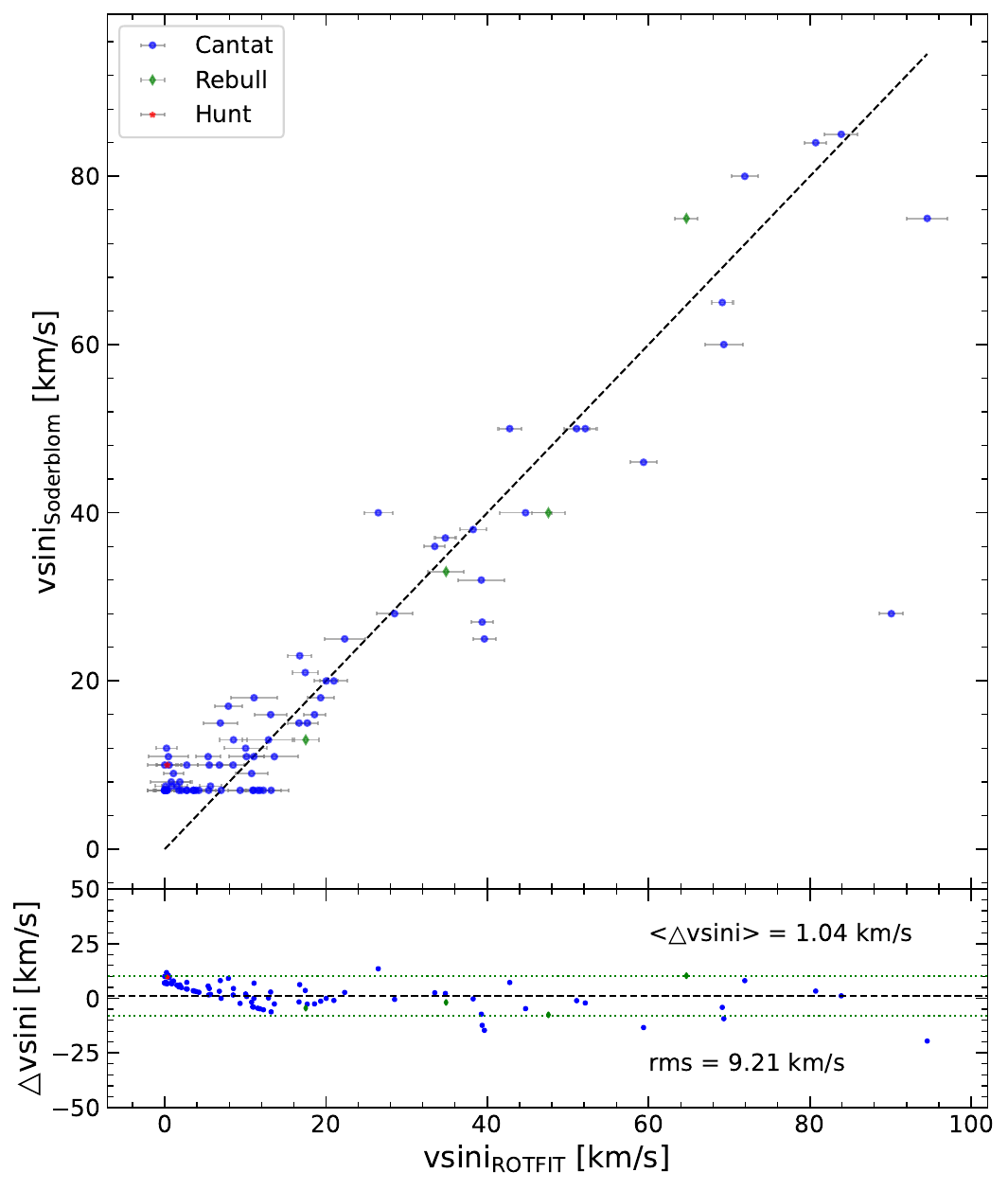}
\caption{Comparison between the \vsini\ values measured in this work and those found in the literature. Top left panel: \rotfit\ versus APOGEE-2 DR17 values \citep{Abdurro2022}. 
The color of symbols distinguishes the three subsamples as in the previous figures and indicated in the legend. The one-to-one relation is shown by the dashed line. The \vsini\ differences between \rotfit\ and APOGEE values, $\Delta$\vsini, are displayed in the lower box.
Top right panel: Comparison with the \lasp\ DR11 v1.0 \vsini\ values ({\tt https://www.lamost.org/dr11/v1.0}). The meaning of lines and symbols is as in the left panel. 
Bottom left panel:  \rotfit\ versus \citep{Hartman2010}  \vsini\ values.
Bottom right panel: \rotfit\ versus \citet{Soderblom1993c} \vsini\ values. }
\label{Fig:vsini_comp}
\end{figure*}

\begin{figure*}
\begin{center}
\includegraphics[width=6.cm,viewport= 10 10 580 400]{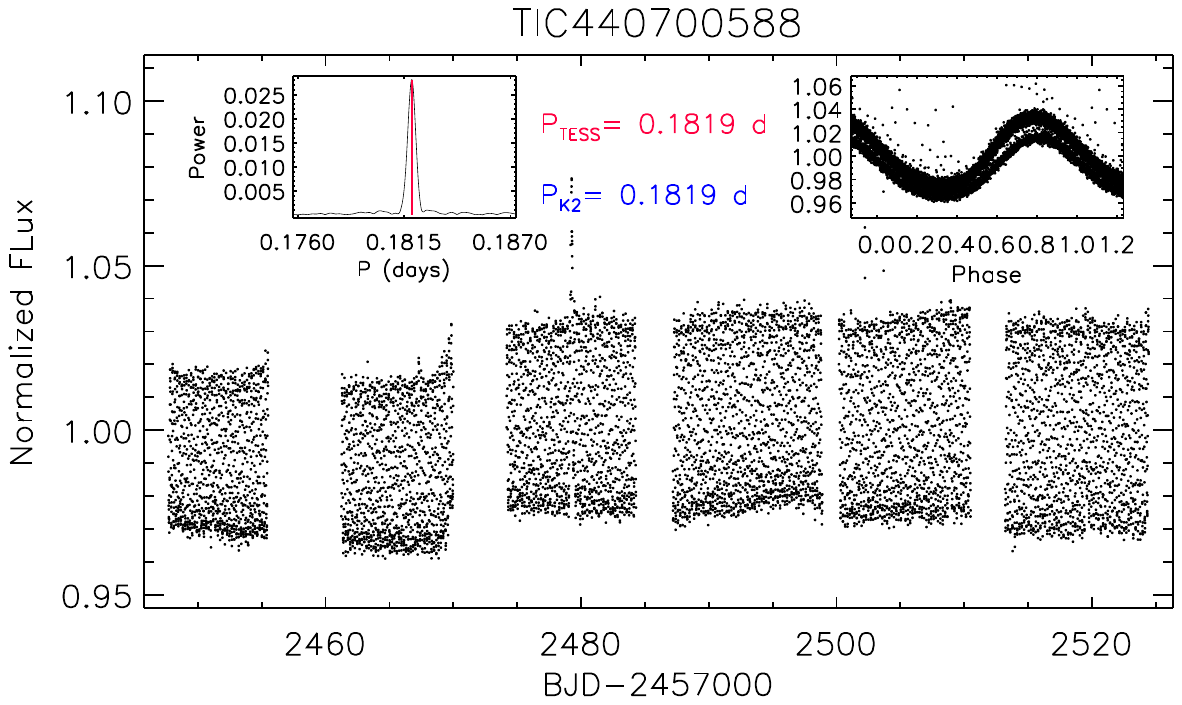}	
\includegraphics[width=6.cm,viewport= 10 10 580 400]{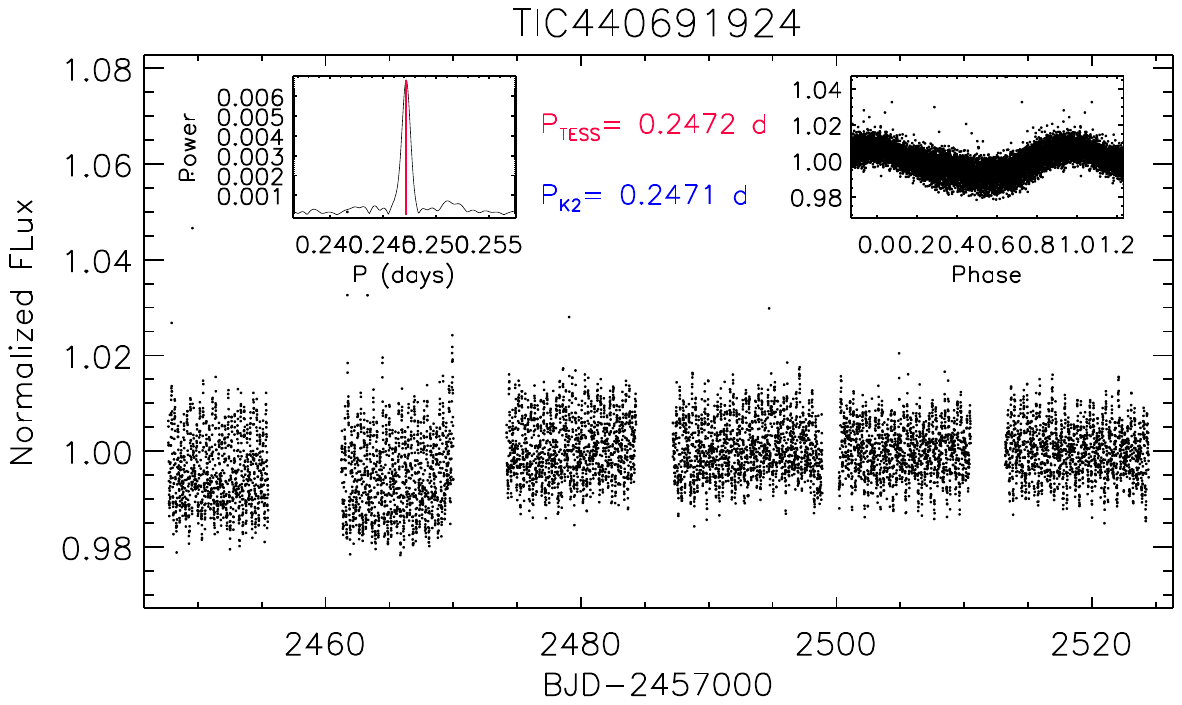}	
\includegraphics[width=6.cm,viewport= 10 10 580 400]{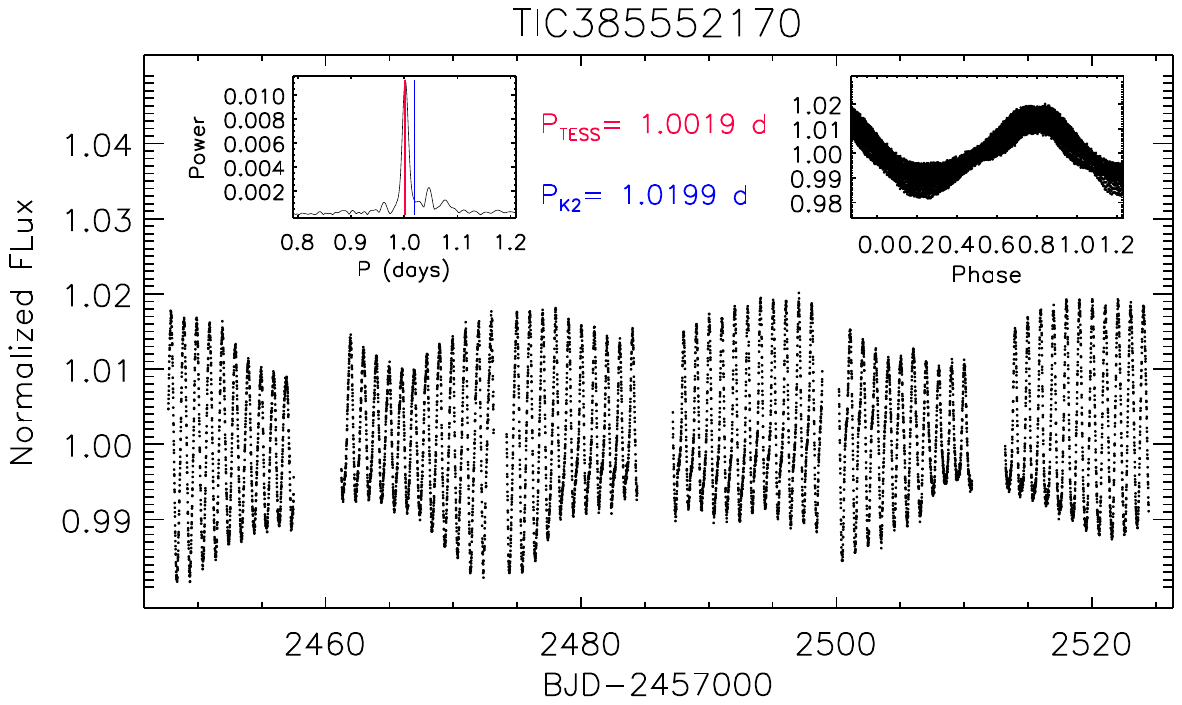}	
\includegraphics[width=6.cm,viewport= 10 10 580 400]{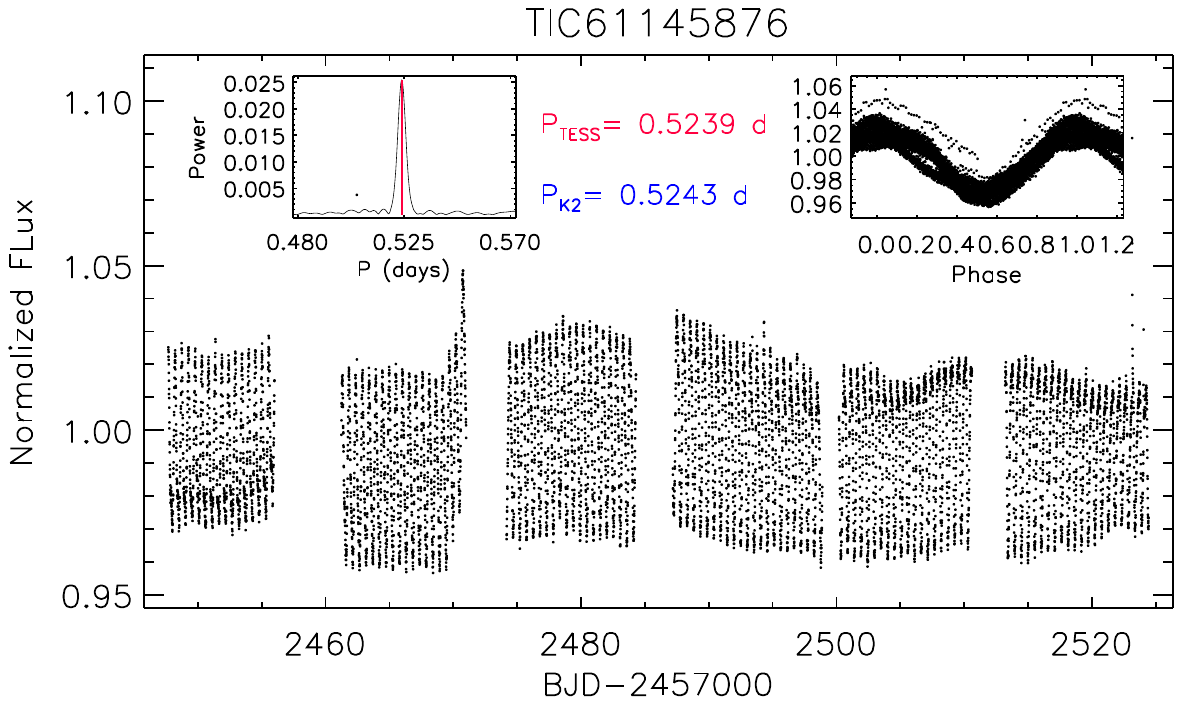}	
\includegraphics[width=6.cm,viewport= 10 10 580 400]{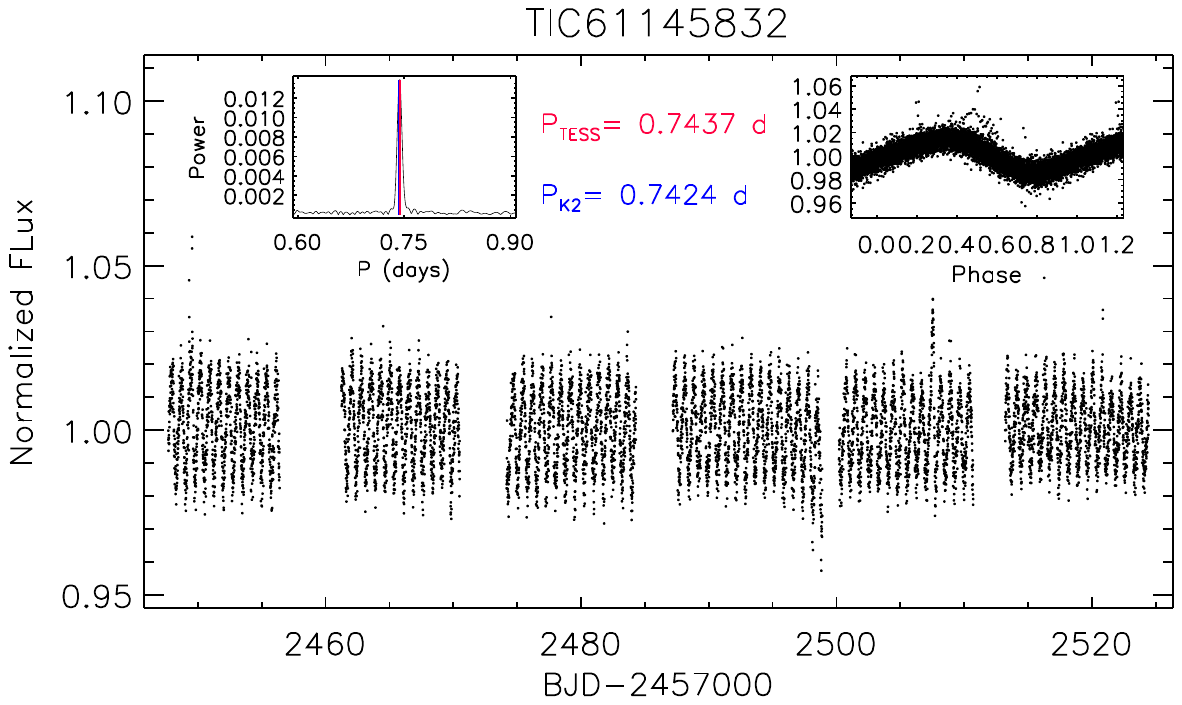}	
\includegraphics[width=6.cm,viewport= 10 10 580 400]{plot_tess_TIC385552170.pdf}	
\includegraphics[width=6.cm,viewport= 10 10 580 400]{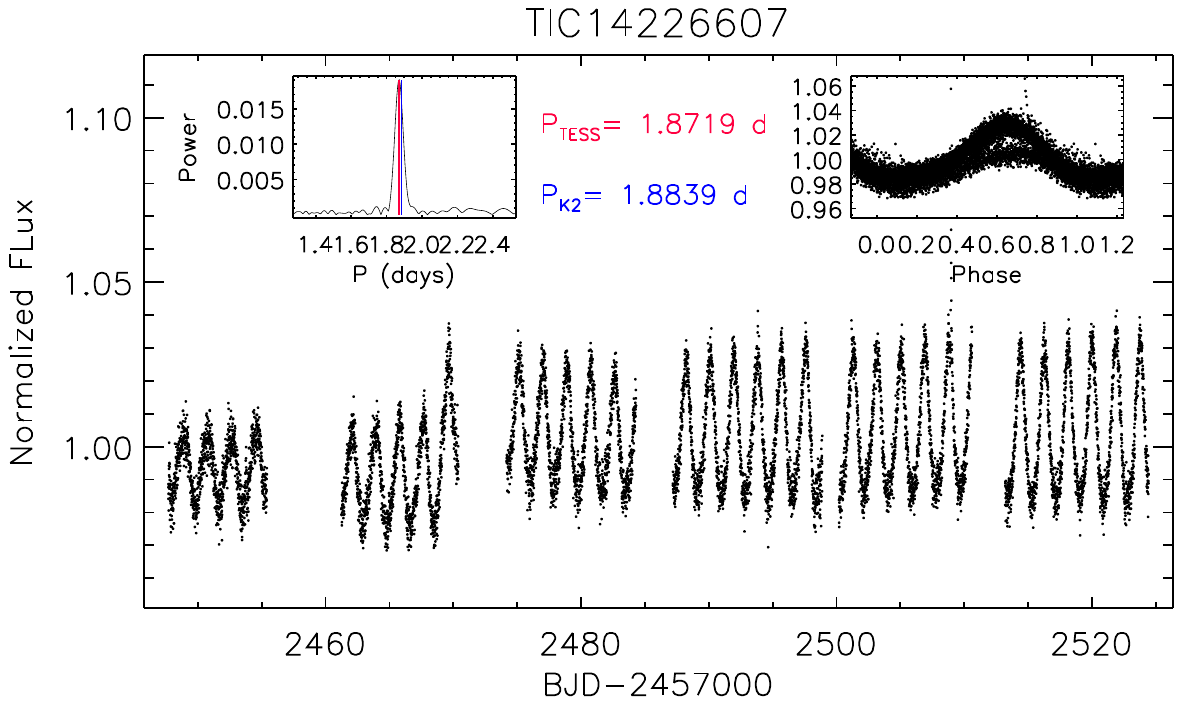}	
\includegraphics[width=6.cm,viewport= 10 10 580 400]{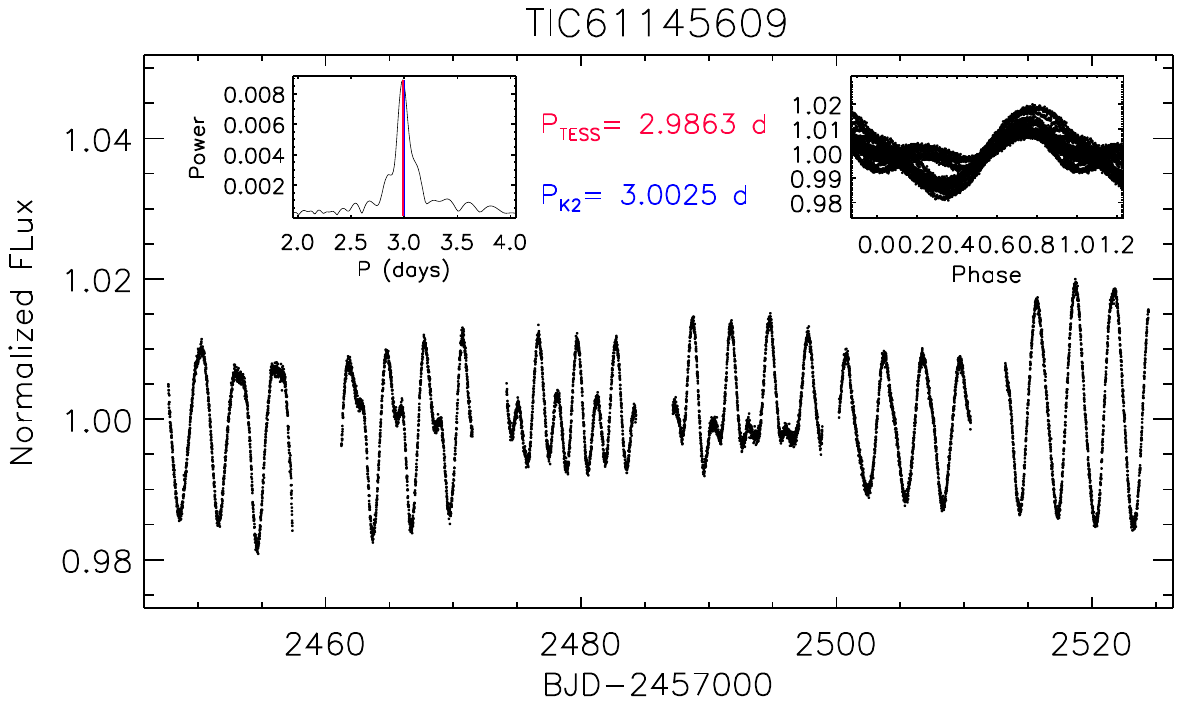}	
\includegraphics[width=6.cm,viewport= 10 10 580 400]{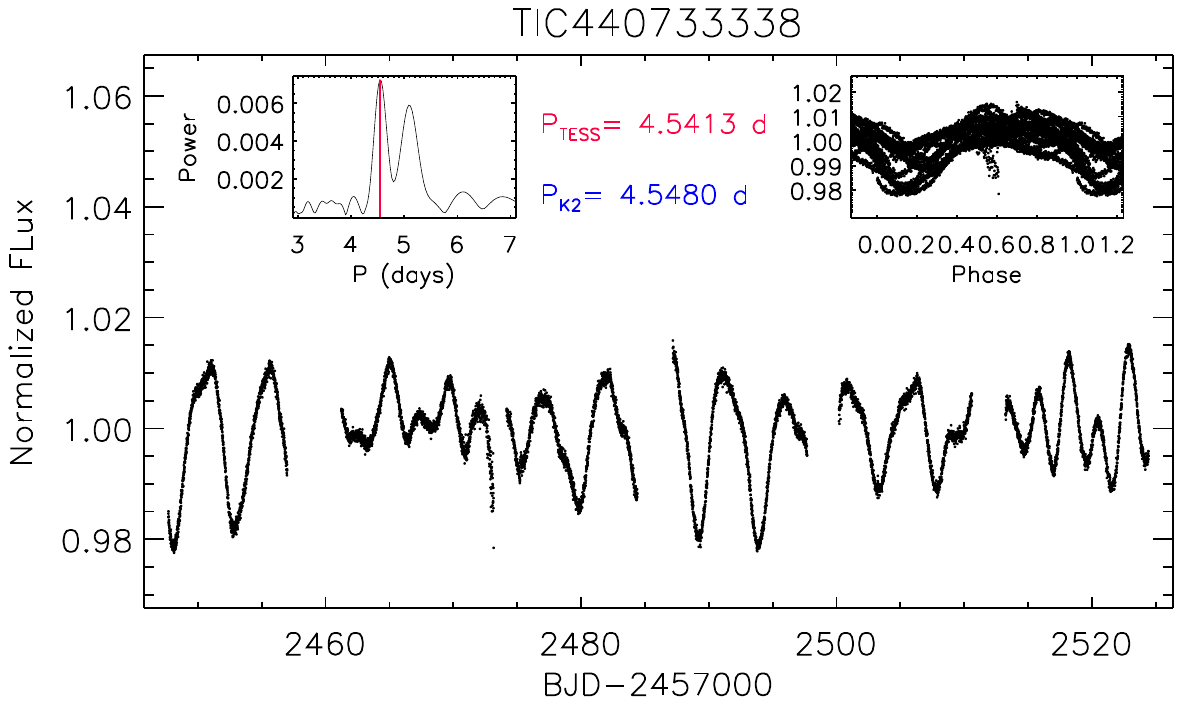}	
\includegraphics[width=6.cm,viewport= 10 10 580 400]{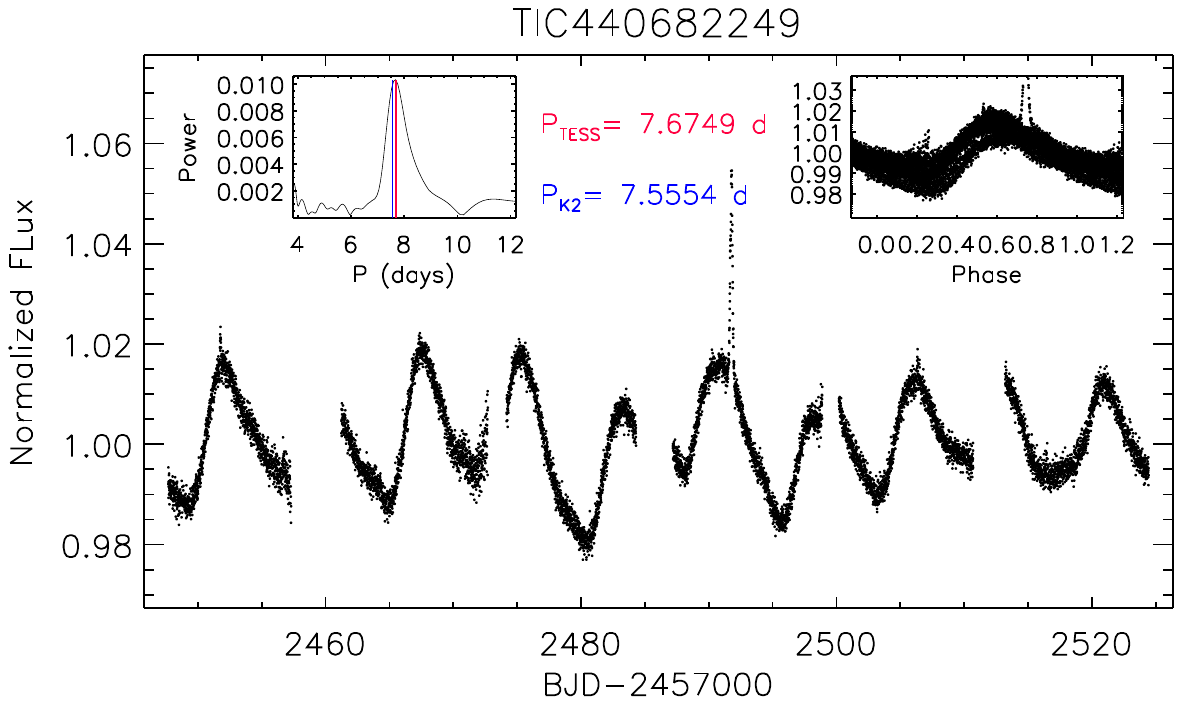}	
\includegraphics[width=6.cm,viewport= 10 10 580 400]{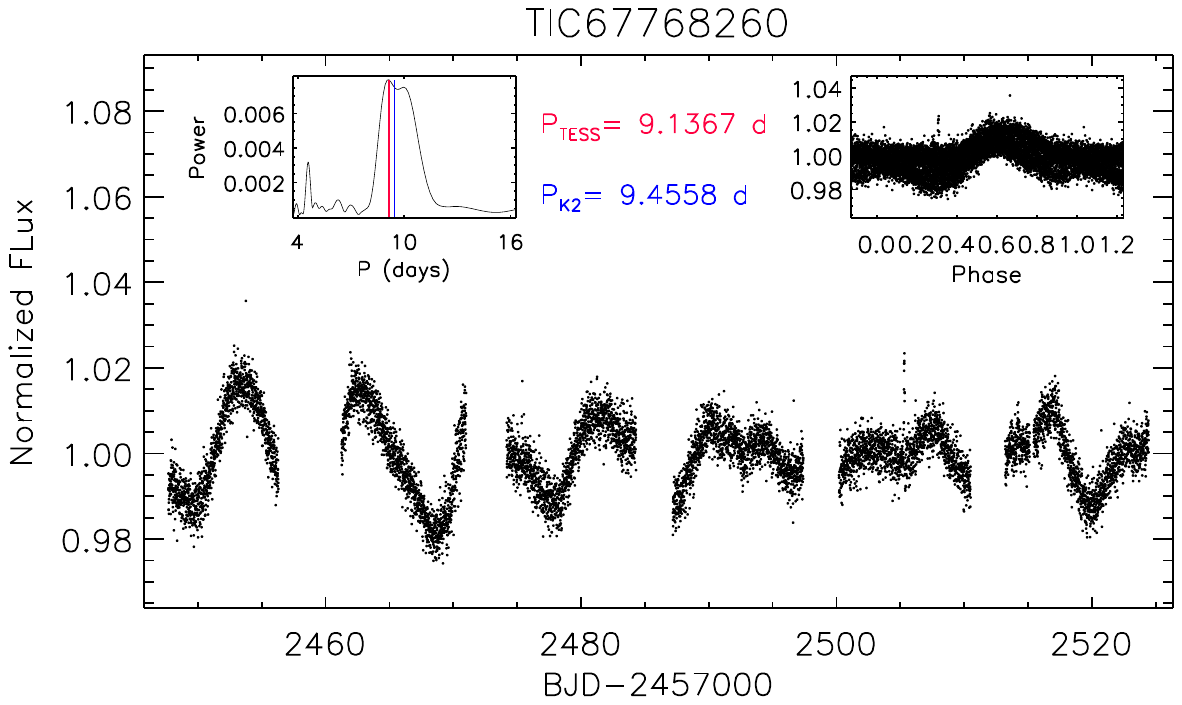}	
\includegraphics[width=6.cm,viewport= 10 10 580 400]{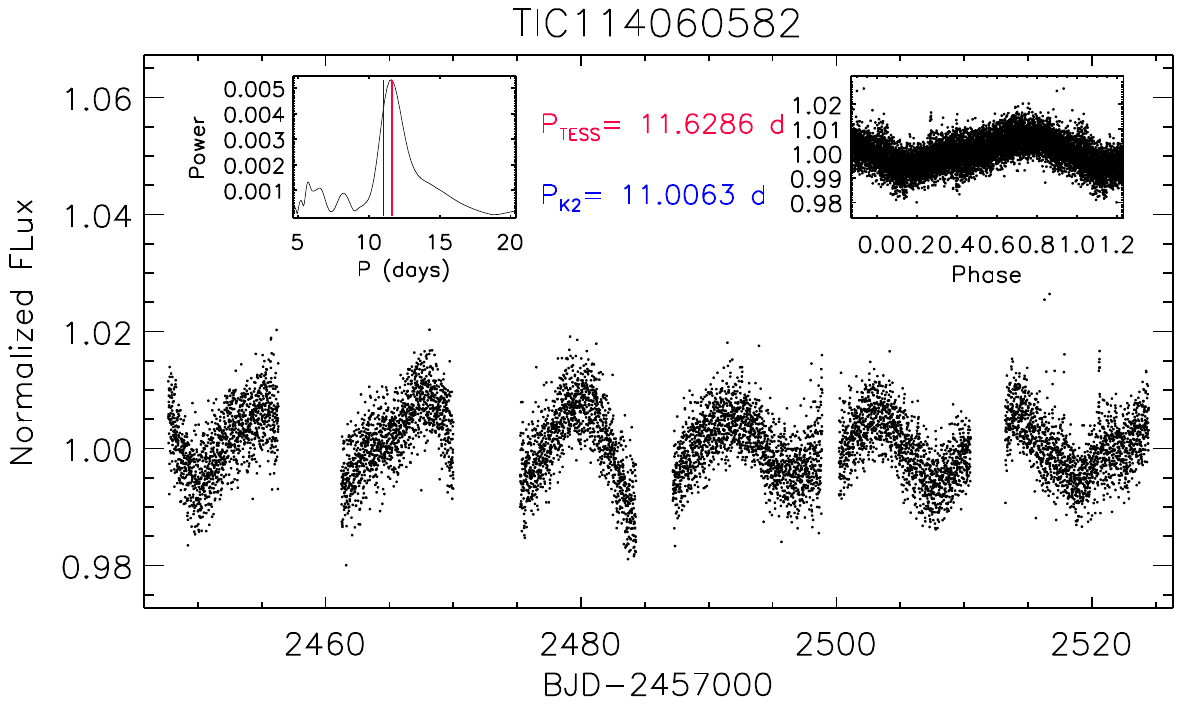}	
\vspace{0.5cm}
\caption{Results of the period analysis of {\it TESS} light curves for a sample of Pleiades members observed also with \kepler-\ktwo. In each panel, the inset in the upper left corner shows the 
cleaned periodogram computed by us in the \tess\ data, with the period marked by a vertical red line and indicated with red characters. The period derived by \citet{Rebull2016} with the \ktwo\ photometry is listed and marked in blue.  The inset in the upper right corner of each panel displays the phased light curve.}
\label{fig:TESS_Phot_comp}
\end{center}
\end{figure*}

\begin{figure*}
\begin{center}
\includegraphics[width=8.cm,viewport= 10 10 580 400]{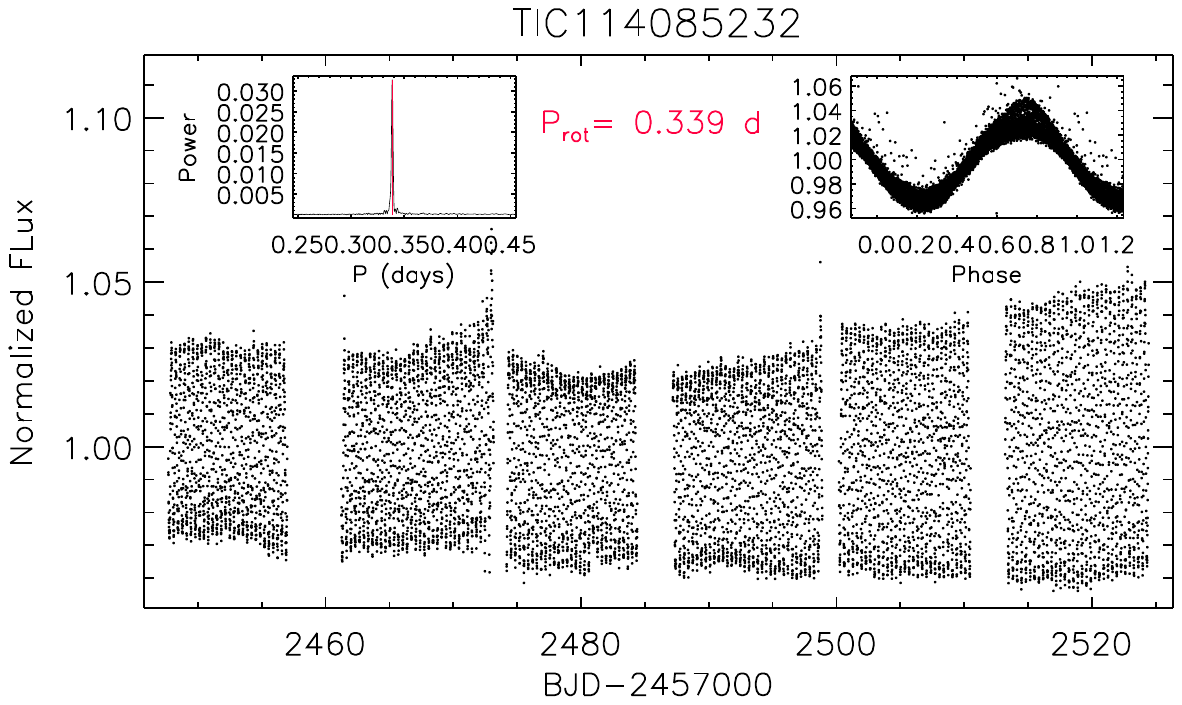}	
\includegraphics[width=8.cm,viewport= 10 10 580 400]{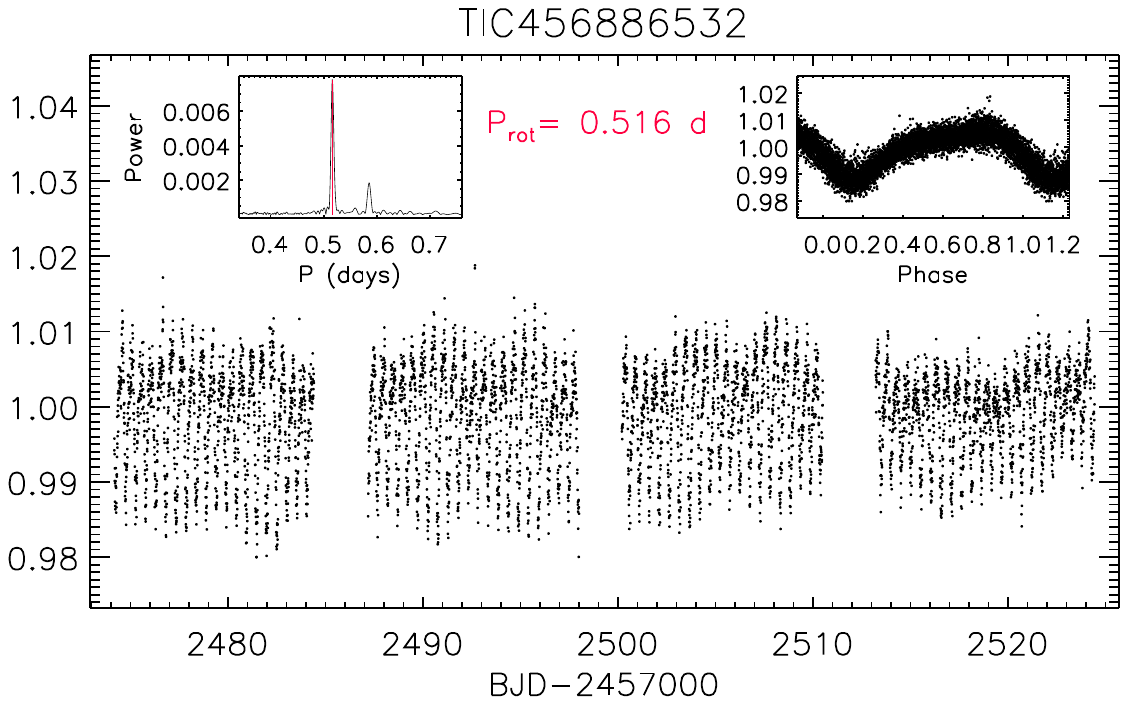}	
\includegraphics[width=8.cm,viewport= 10 10 580 400]{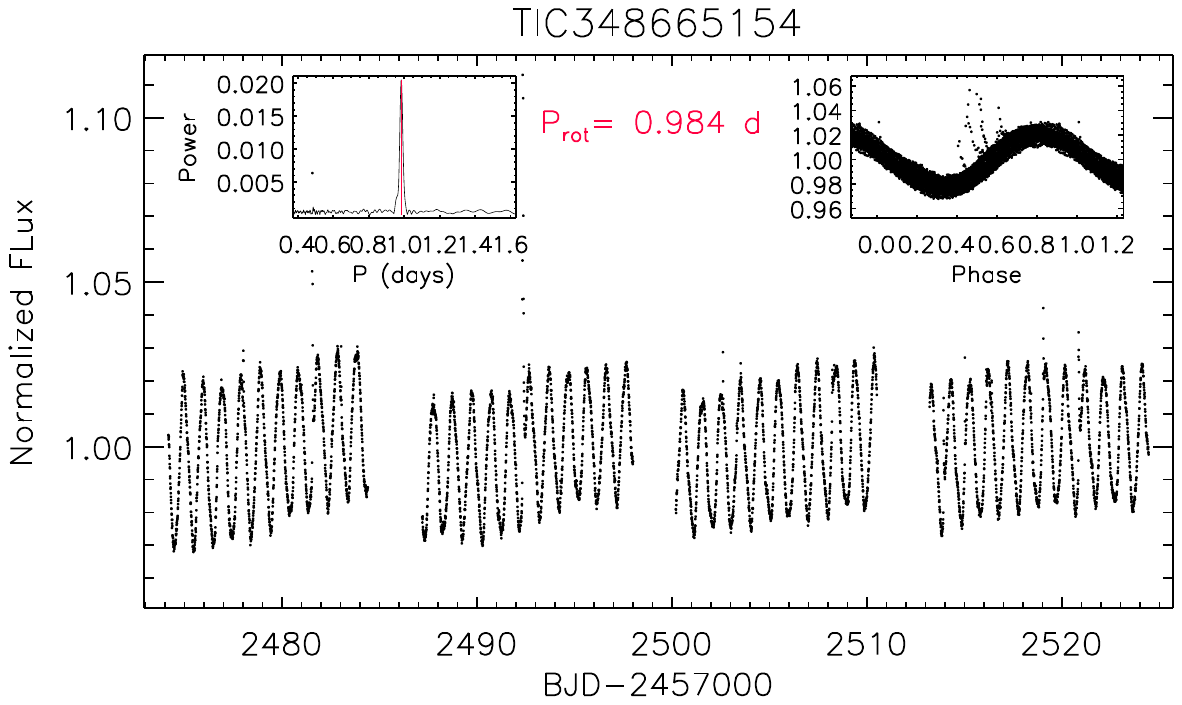}	
\includegraphics[width=8.cm,viewport= 10 10 580 400]{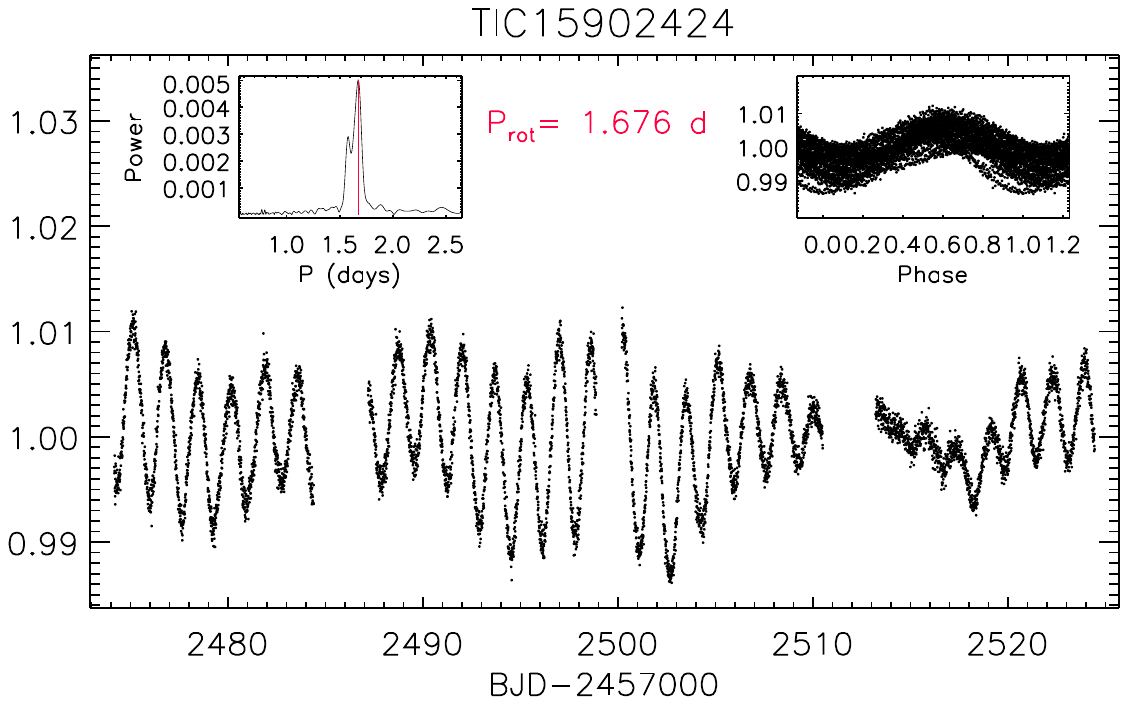}	
\includegraphics[width=8.cm,viewport= 10 10 580 400]{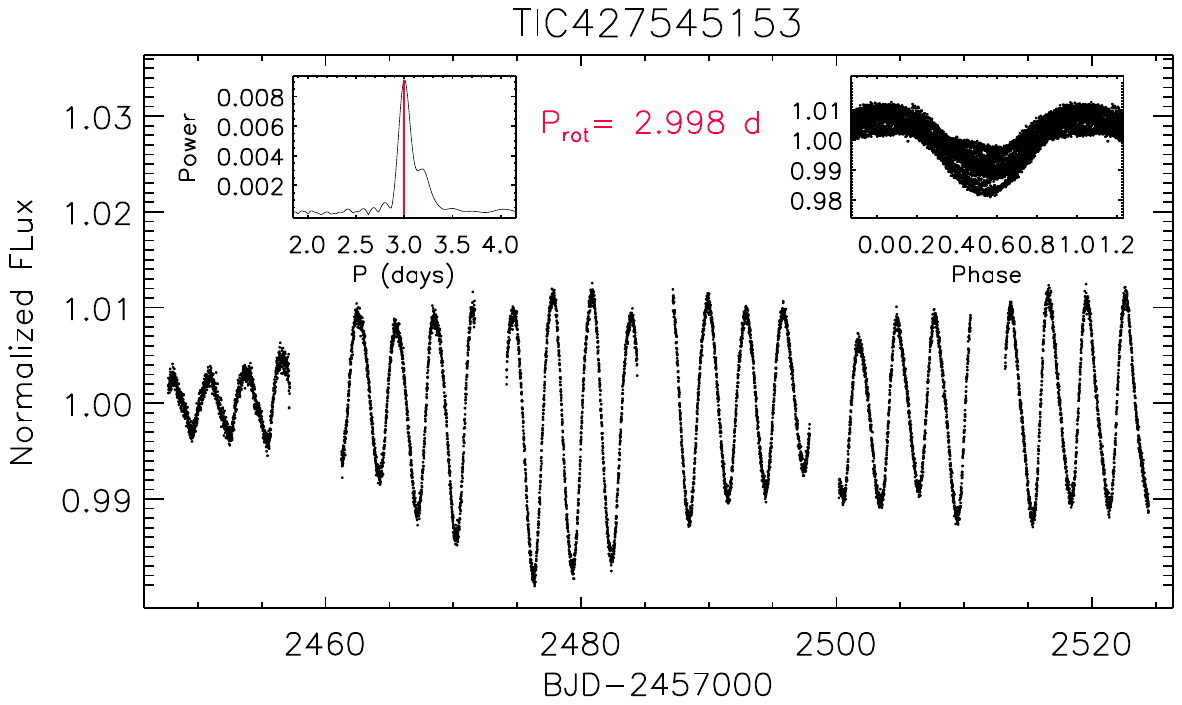}	
\includegraphics[width=8.cm,viewport= 10 10 580 400]{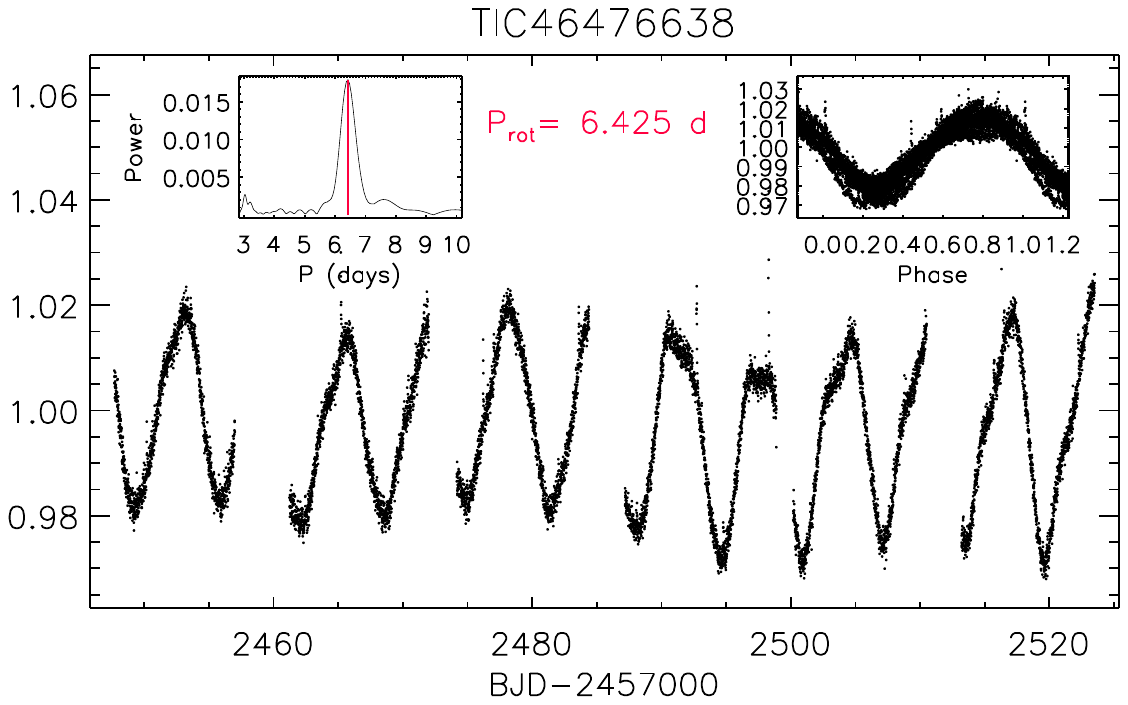}	
\includegraphics[width=8.cm,viewport= 10 10 580 400]{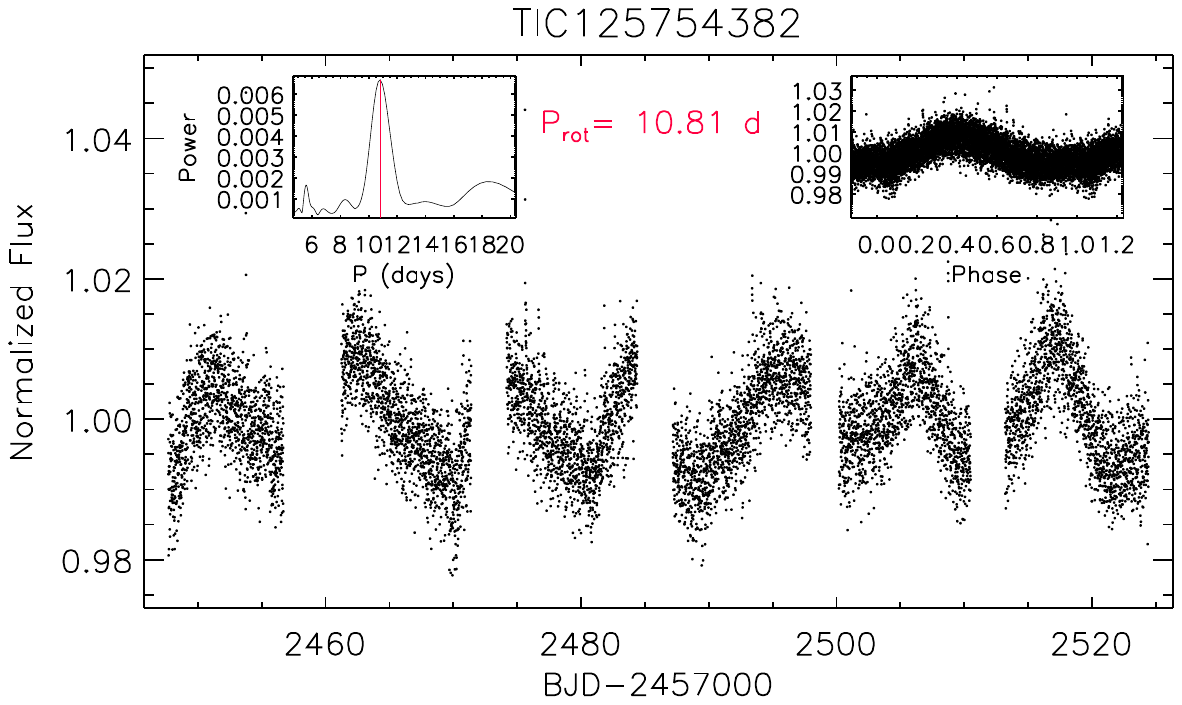}	
\includegraphics[width=8.cm,viewport= 10 10 580 400]{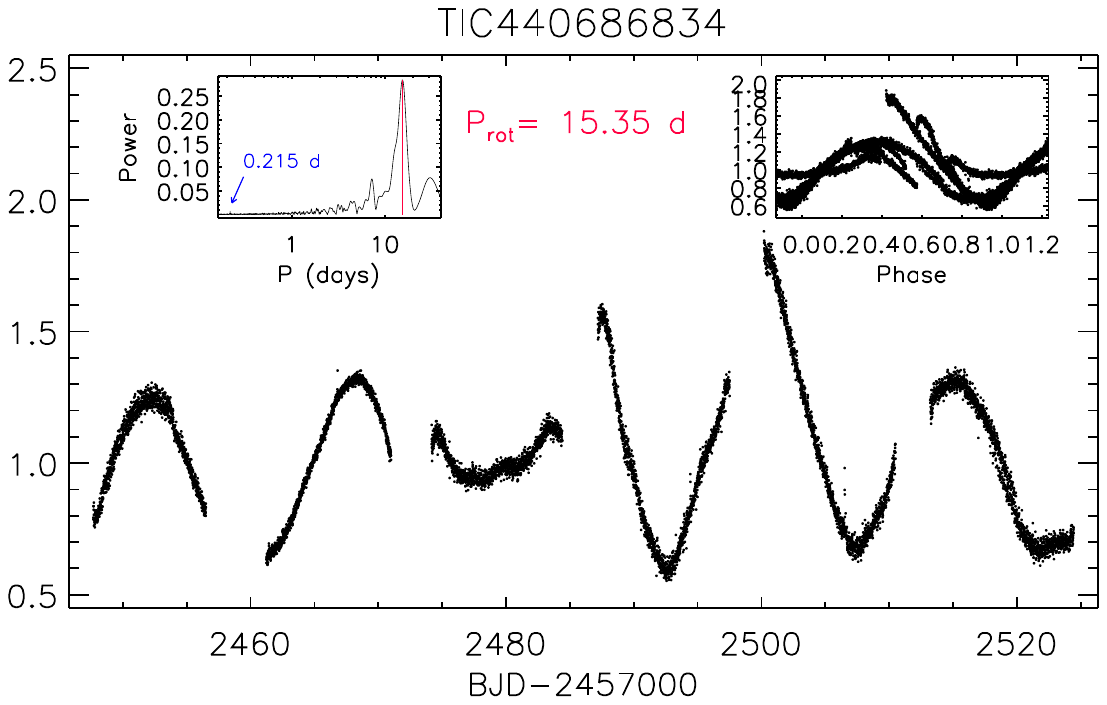}	
\vspace{0.5cm}
\caption{{\it TESS} light curves for a sample of Pleiades members. In each panel, the inset in the upper left corner shows the 
cleaned periodogram, with the period marked by a vertical red line and indicated with red characters. The inset in the upper right corner displays the data phased with this period.}
\label{fig:TESS_Phot}
\end{center}
\end{figure*}

\begin{figure}[th]
\begin{center}
\includegraphics[width=9.5cm,viewport= 0 0 580 380]{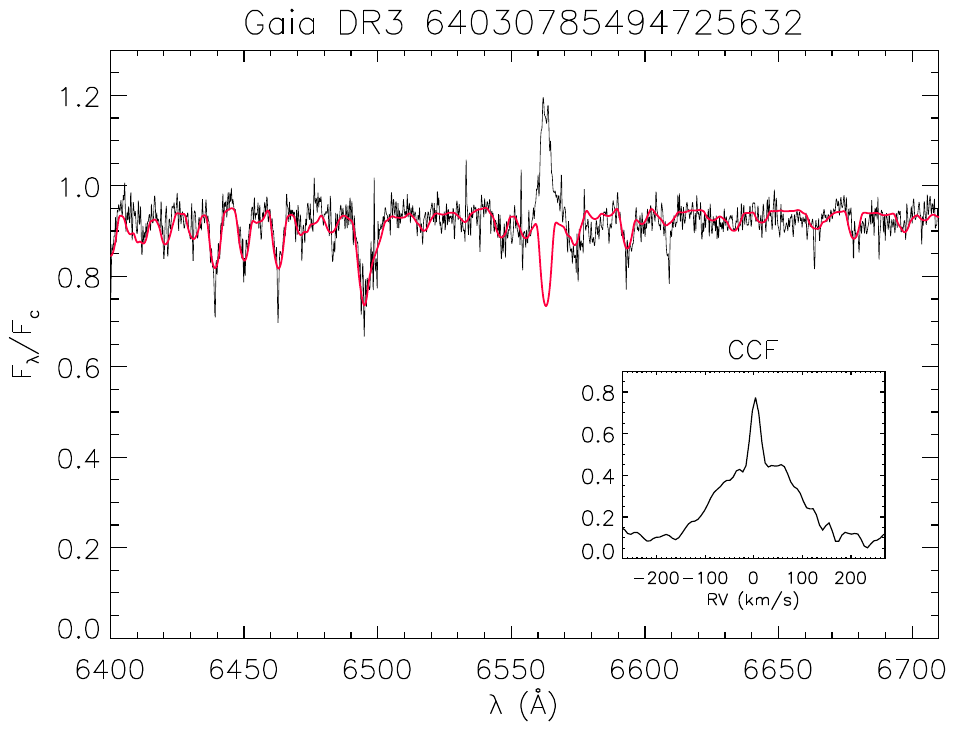}
\caption{Red-arm spectrum of \gaia\ DR3 64030785494725632 (black line) with the best template (red line) found by \rotfit\ that roughly fits the broad spectral lines. The cross-correlation function is displayed in the inset plot and clearly shows a broad and a narrow peak.}
\label{Fig:64030785494725632}
\end{center}
\end{figure}

\begin{figure}
\begin{center}
\includegraphics[width=9.3cm,viewport= 0 0 580 380]{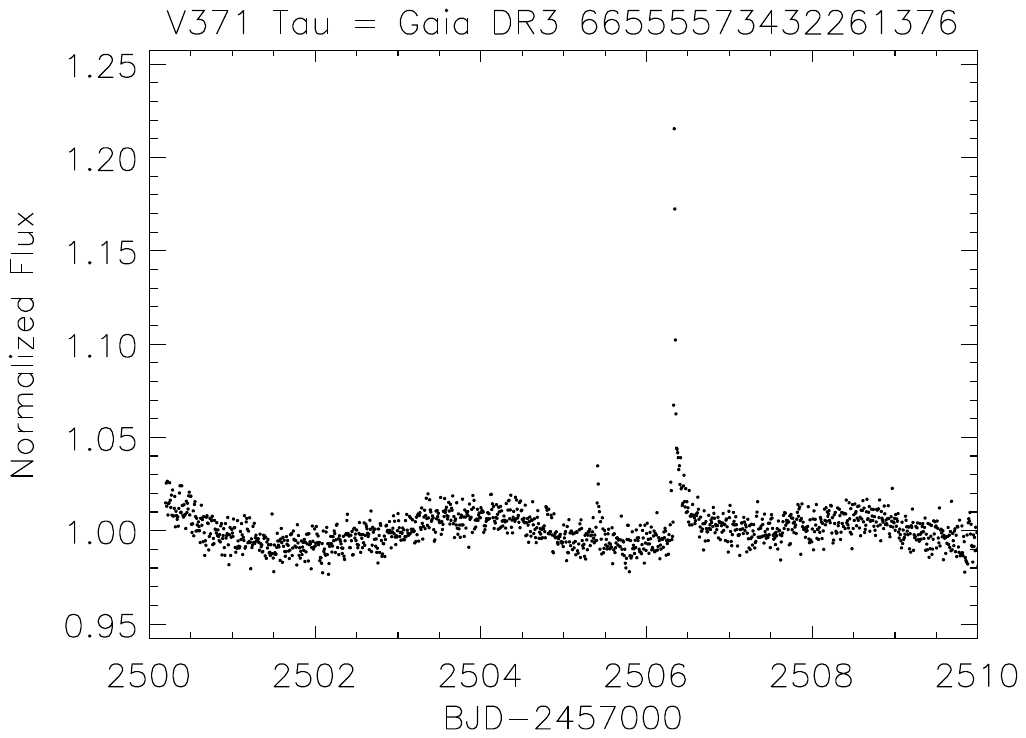}
\caption{A portion of a \tess\ light curve of V371\,Tau that displays two flares standing out above to the rotational modulation.}
\label{fig:Flare_TESS2}
\end{center}
\end{figure}

\begin{figure}
\begin{center}
\includegraphics[width=9.5cm,viewport= 0 10 430 600]{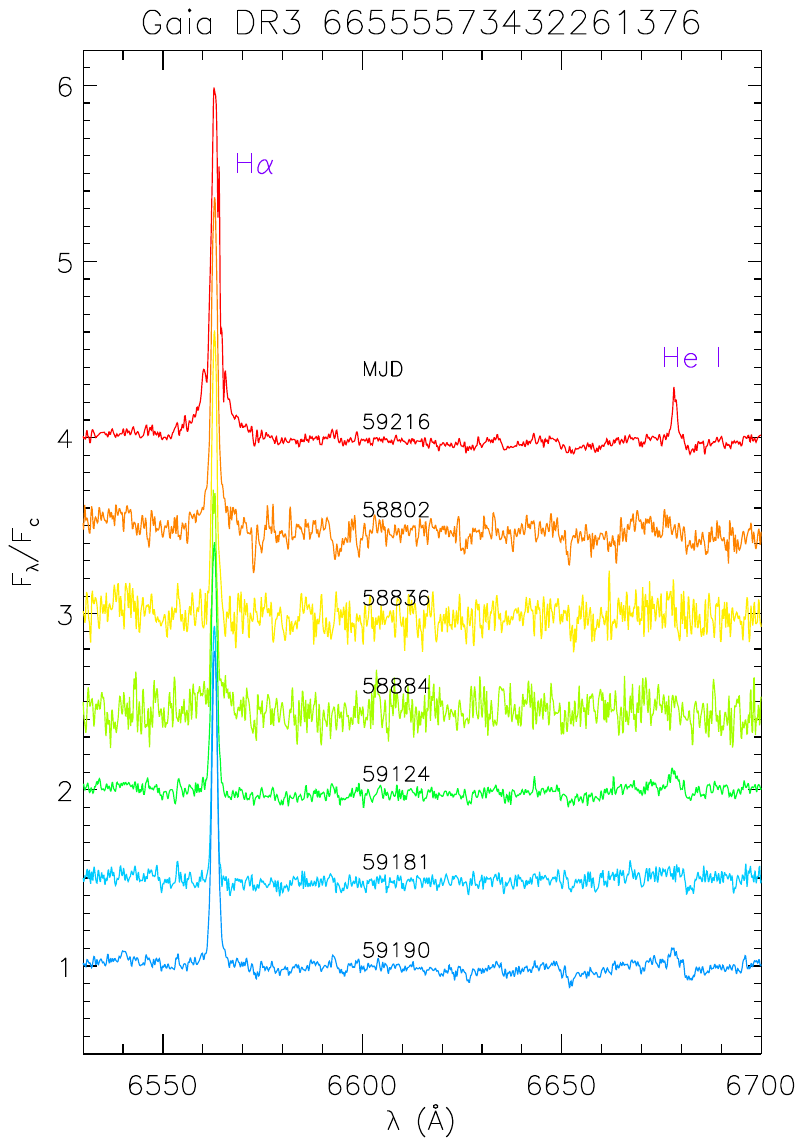}
\caption{Same as Fig.\,\ref{fig:Flare_LOTau} but for V371\,Tau.
}
\label{fig:Flare_V371Tau}
\end{center}
\end{figure}

\begin{figure}
\begin{center}
\includegraphics[width=9.5cm,viewport= 0 10 430 600]{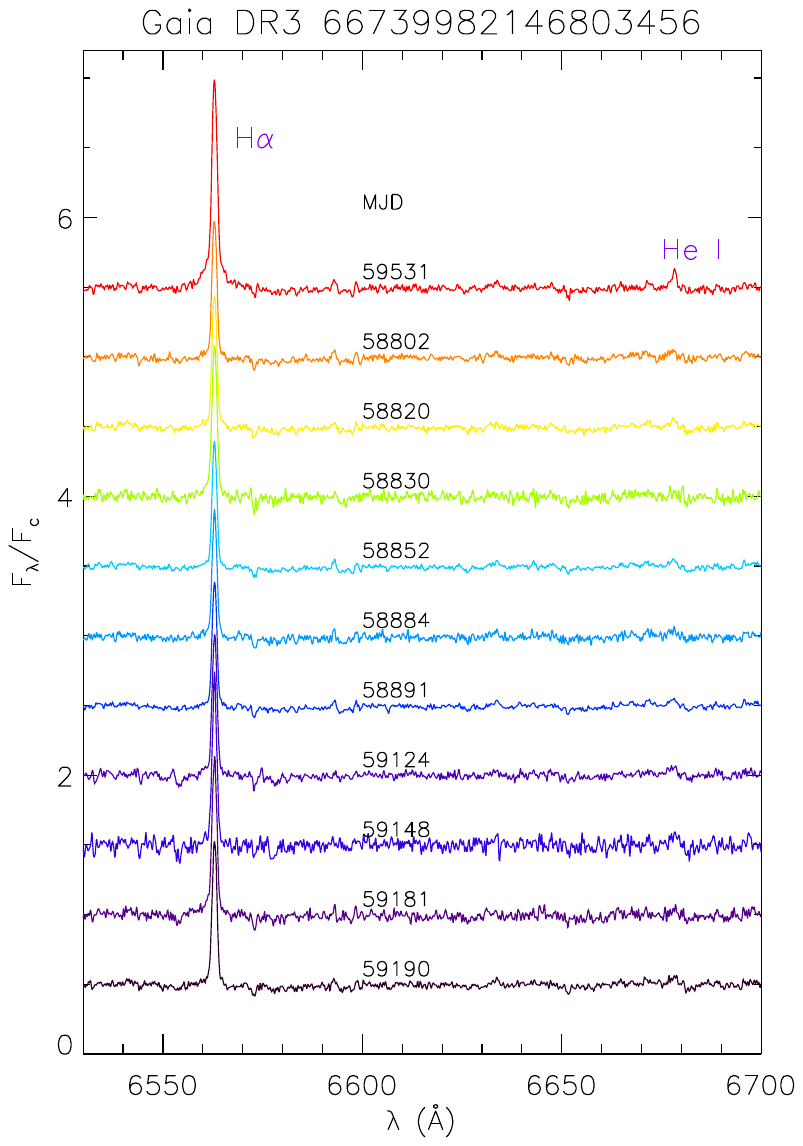}
\caption{Same as Fig.\,\ref{fig:Flare_LOTau} but for V343\,Tau.
}
\label{fig:Flare_V343Tau}
\end{center}
\end{figure}

\begin{table}[tb]
\caption[ ]{RV corrections. }
\label{Tab:RV_corr}
\begin{tabular}{lrcrcr}
\hline
\hline
\noalign{\smallskip}
 RJD--Plate & $\mu_{\rm blue}$ & $\sigma_{\rm blue}$  & $\mu_{\rm red}$ & $\sigma_{\rm red}$   & $N$ \\ 
       & \multicolumn{2}{c}{(\kms)} & \multicolumn{2}{c}{(\kms)}\\  \\ 
\hline
\noalign{\smallskip} 
58802--TD035052N235741K01  &  -1.78   &   2.22  &  0.54   &  0.59  & 7\\ 
58820--TD035052N235741K01  &  -1.07   &   2.23  &  0.24   &  0.30  & 5\\ 
58830--TD035052N235741K01  &  -2.57   &   2.63  & -1.70   &  1.52  & 6\\ 
58836--TD035052N235741K01  &  -1.85   &   1.24  & -0.75   &  1.38  & 6\\ 
58852--TD035052N235741K01  &  -0.83   &   2.39  &  0.18   &  0.32  & 6\\ 
58884--TD035052N235741K01  &   1.07   &   0.86  & -0.57   &  0.57  & 6\\ 
58891--TD035052N235741K01  &   1.68   &   0.96  & -0.09   &  0.36  & 6\\ 
59124--TD035052N235741K01  &  -1.60   &   2.15  & -0.44   &  1.79  & 6\\ 
59148--TD035052N235741K01  &  -3.03   &   2.19  &  0.17   &  0.73  & 4\\ 
59151--NT033426N242751S01  &   1.78   &   0.08  & -1.36   &  0.08  & 4\\ 
59181--TD035052N235741K01  &  -3.02   &   2.32  & -0.80   &  2.37  & 6\\ 
59190--TD035052N235741K01  &  -1.39   &   2.30  & -0.17   &  0.48  & 6\\ 
59216--TD035052N235741K01  &  -2.56   &   0.91  &  0.87   &  0.62  & 4\\ 
59250--NT034921N242251S01  &  -0.49   &   0.95  & -0.20   &  1.49  & 4\\ 
59531--TD035052N235741K01  &  -2.00   &   2.23  & -0.28   &  0.55  & 6\\ 
58030--HIP1849101	   &   3.17   &   1.35  &  2.61   &  1.37  & 26\\ 
58057--HIP1685901	  &   3.92   &   0.77  &  2.24   &  0.84  & 20\\ 
58061--HIP1849001	  &   3.81   &   1.33  &  2.51   &  0.98  & 19\\ 
58093--HIP1673701	  &   4.44   &   1.35  &  3.21   &  1.29  & 30\\ 
58119--kp2\_04\_301	  &   4.23   &   0.88  &  2.54   &  0.94  & 42\\ 
58150--kp2\_04\_401	  &   2.99   &   0.93  &  3.28   &  1.08  & 4\\ 
58441--TD041055N155647B01  &   0.20   &   2.10  & -1.35   &  1.55  & 51\\ 
58853--NT040512N273634M01  &  -0.44   &   1.26  & -1.71   &  1.07  & 95\\ 
59125--NT034818N263521M01  &  -0.25   &   1.63  & -3.89   &  0.00  & 77\\ 
59179--NT031534N265412S01  &  -0.92   &   1.90  & -1.29   &  0.93  & 13\\ 
59186--NT032846N302231S01  &  -0.91   &   0.70  & -1.31   &  0.91  & 10\\ 
59208--TD042836N191049H01  &  -1.01   &   1.93  & -1.15   &  1.49  & 6\\ 
59244--TD042836N191049H01  &  -0.24   &   1.20  & -0.32   &  0.72  & 6\\ 
59248--NT041225N233428M01  &  -0.37   &   1.36  & -0.81   &  0.95  & 77\\ 
59530--TD032020N290254T01  &  -0.60   &   1.29  & -0.36   &  1.21  & 63\\ 
59532--TD042836N191049H01  &  -0.06   &   0.64  & -2.09   &  0.69  & 5\\ 
59541--NT040520N220032S01  &  -0.45   &   2.54  & -1.35   &  3.48  & 35\\ 
59861--TD032020N290254T01  &  -0.18   &   2.94  & -0.66   &  4.00  & 48\\ 
59955--TD031804N245237T01  &  -1.20   &   1.37  & -2.42   &  1.33  & 38\\ 
58027--HIP1966201	  &   4.18   &   1.11  &  2.91   &  1.02  & 21\\ 
58026--HIP1799901	  &   3.88   &   1.41  &  2.77   &  1.47  & 40\\ 
58065--HIP1799901	  &   4.62   &   1.04  &  2.74   &  0.75  & 36\\ 
58084--HIP1799901	  &   4.18   &   1.15  &  3.02   &  1.09  & 41\\ 
58086--HIP1799901	  &   4.30   &   1.15  &  2.82   &  1.01  & 41\\ 
58122--HIP1799901	  &   3.91   &   1.34  &  3.05   &  1.40  & 42\\ 
58153--HIP1799901	  &   3.94   &   1.44  &  3.06   &  1.00  & 41\\ 
58120--kp2\_04\_401	  &   4.26   &   0.84  &  2.90   &  1.00  & 31\\ 

\hline
\noalign{\smallskip} 
\end{tabular}
{\bf Notes.} RJD is the reduced Julian Date. The center and width of the RV difference distributions are indicated with $\mu$ and $\sigma$, respectively. The number of stars is reported in the last column.
\end{table}

\onecolumn

\begin{landscape}
\renewcommand{\tabcolsep}{0.1cm}

\vspace{0cm}
\begin{longtable}{ccccccrrrrrrrrrrrr}
\caption{Radial velocities derived with ROTFIT from the red- and blue-arm MRS \lamost\ spectra and the final corrected RV values.}\\
\scriptsize
\hspace{-2cm}
\begin{tabular}{ccccccrrrrrrrrrrrr}
\hline
\hline
\noalign{\smallskip}
Designation & \gaia\ DR3 ID &  RA  &   DEC   & HJD & Spectrum   & $RV_{\rm red}$ & err & $RV_{\rm blue}$ & err & $RV_{\rm corr}$ & err &  Sub \\ 
        &      & (J2000) & (J2000) &($-$2\,400\,000) &     & \multicolumn{2}{c}{(\kms)} & \multicolumn{2}{c}{(\kms)} &  \multicolumn{2}{c}{(\kms)}  & \\ 
\hline
\noalign{\smallskip} 

   J035427.83+212323.0  &  51742471745296768  &   58.615971  &  21.389748  & 58061.24010  &	     med-58061-HIP1849001\_sp04-154   &   7.04  &    0.61  &    3.54  &	  0.74   &   8.66   &	0.47   &  C	  \\
   J035922.89+223416.9  &  53335045618385536  &   59.845394  &  22.571371  & 58802.24172  &  med-58802-TD035052N235741K01\_sp07-149   &   6.08  &    0.50  &    4.82  &	  1.92   &   6.39   &	0.48   &  C	  \\
   J035922.89+223416.9  &  53335045618385536  &   59.845394  &  22.571371  & 58820.15217  &  med-58820-TD035052N235741K01\_sp07-149   &   7.12  &    0.72  &    8.81  &	  2.04   &   7.40   &	0.68   &  C	  \\
   J035922.89+223416.9  &  53335045618385536  &   59.845394  &  22.571371  & 58830.14153  &  med-58830-TD035052N235741K01\_sp07-149   &  -0.31  &    0.69  &   -1.35  &	  1.73   &  -2.27   &	0.64   &  C	  \\
   J035922.89+223416.9  &  53335045618385536  &   59.845394  &  22.571371  & 58836.03299  &  med-58836-TD035052N235741K01\_sp07-149   &   4.73  &    1.73  &   16.22  &	  2.83   &   6.81   &	1.48   &  C	  \\
   J035922.89+223416.9  &  53335045618385536  &   59.845394  &  22.571371  & 58852.08909  &  med-58852-TD035052N235741K01\_sp07-149   &   7.15  &    0.72  &   11.02  &	  2.45   &   7.56   &	0.69   &  C	  \\
   J035922.89+223416.9  &  53335045618385536  &   59.845394  &  22.571371  & 58883.99410  &  med-58884-TD035052N235741K01\_sp07-149   &  -1.88  &    1.12  &   -9.90  &	  2.30   &  -3.67   &	1.01   &  C	  \\
   J035922.89+223416.9  &  53335045618385536  &   59.845394  &  22.571371  & 58890.98578  &  med-58891-TD035052N235741K01\_sp07-149   &   7.41  &    0.78  &    8.43  &	  2.34   &   7.60   &	0.74   &  C	  \\
   J035922.89+223416.9  &  53335045618385536  &   59.845394  &  22.571371  & 59124.32563  &  med-59124-TD035052N235741K01\_sp07-149   &   8.35  &    0.88  &    8.48  &	  1.84   &   7.72   &	0.79   &  C	  \\
\noalign{\smallskip}
\hline 
\noalign{\medskip}
\end{tabular}
~\\
{\bf Notes.} The full table is available at the CDS. $RV_{\rm red}$ and $RV_{\rm blue}$ are measured values for the respective arm without correction. 
\label{Tab:RV_data}
\end{longtable}

\end{landscape}

\normalsize

\twocolumn

\renewcommand{\tabcolsep}{0.2cm}

\begin{table}[ht]
\caption{Rotation periods derived with \tess\ and \ktwo.}
\begin{tabular}{crrr}
\hline
\hline
\noalign{\smallskip}
\gaia-DR3        &  TIC   & \prot$^{T}$  & \prot$^{K2}$ \\
\hline
\noalign{\smallskip}
64148738180119680 & 440700588  &  0.1819 &  0.1819 \\
63958801843006208 & 440691924  &  0.2472 &  0.2471 \\
65296357738731008 & 385552372  &  0.3280 &  0.3279 \\
64980278208557696 & 61145876   &  0.5239 &  0.5243 \\
64977705525131904 & 61145832   &  0.7437 &  0.7424 \\
65222759179728640 & 385552170  &  1.0019 &  1.0199 \\
53335045618385536 & 14226607   &  1.8719 &  1.8839 \\
64899017426872960 & 61145609   &  2.9863 &  3.0025 \\
51742471745296768 & 440733338  &  4.5413 &  4.5480 \\
63916431989200256 & 440682249  &  7.6749 &  7.5554 \\
65089473457126784 & 67768260   &  9.1367 &  9.4558 \\
65151531440914560 & 114060582  & 11.6286 & 11.0063 \\
\noalign{\smallskip}
\hline \\
\end{tabular}
{\bf Notes.}  \prot$^{T}$ = period measured in this work with \tess\ photometry;  \prot$^{K2}$ = period measured by \citet{Rebull2016} with \ktwo\ photometry.
\label{Tab:Comp_Prot}
\end{table}

\renewcommand{\tabcolsep}{0.15cm}

\begin{table}[ht]
\caption{Heliocentric RVs of  DH\,794.}
\begin{tabular}{lrcrcl}
\hline
\hline
\noalign{\smallskip}
 HJD     &  RV$_{\rm 1}$   &  $\sigma_{\rm RV_1}$ &  RV$_{\rm 2}$   &  $\sigma_{\rm RV_2}$ & Instrument \\
(2\,400\,000+) &  \multicolumn{2}{c}{(\kms)} & \multicolumn{2}{c}{(\kms)} & \\
 \hline
\noalign{\smallskip}
  58121.695   &   5.48   &  1.26 &     5.48 & 1.26  &  APOGEE \\
  58146.631   & $-$16.61 &  1.85 &    38.04 & 5.30  &  APOGEE \\
  58151.673   & $-$32.47 &  2.32 &    60.22 & 6.42  &  APOGEE \\
  58386.930   &    31.41 &  1.90 & $-$21.71 & 4.91  &  APOGEE \\
  58802.24174 &    13.86 &  0.42 &    \dots & \dots &  LAMOST \\
  58820.15215 &    44.00 &  0.51 & $-$39.74 & 2.67  &  LAMOST \\
  58830.14149 &  $-$9.65 &  1.04 &    \dots & \dots &  LAMOST \\
  58836.03294 &  $-$2.43 &  1.71 &    \dots & \dots &  LAMOST \\
  58890.98565 & $-$17.40 &  0.49 &    44.33 & 4.54  &  LAMOST \\
  59216.09464 & $-$32.12 &  0.69 &    55.05 & 3.28  &  LAMOST \\
\noalign{\smallskip}
\hline \\
\end{tabular}
{\bf Notes.} The RV of the single CCF peak observed near the conjunctions has been assigned to both components.
\label{Tab:RV_DH794}
\end{table}

\begin{table}[ht]
\caption{Heliocentric RVs of  HII\,761.}
\begin{tabular}{lrcrcl}
\hline
\hline
\noalign{\smallskip}
 HJD     &  RV$_{\rm 1}$   &  $\sigma_{\rm RV_1}$ &  RV$_{\rm 2}$   &  $\sigma_{\rm RV_2}$ & Instrument \\
(2\,400\,000+) &  \multicolumn{2}{c}{(\kms)} & \multicolumn{2}{c}{(\kms)} & \\
 \hline
\noalign{\smallskip}
  57652.945   &  $-$44.58 & 2.01 &     81.37 & 8.38 &  APOGEE \\
  57408.659   &  $-$33.71 & 1.96 &     61.47 & 6.60 &  APOGEE \\
  57649.914   &  $-$41.64 & 1.98 &     76.54 & 7.82 &  APOGEE \\
  57764.659   &      5.76 & 1.28 &      5.76 & 1.28 &  APOGEE \\
  57684.851   &     42.54 & 1.87 &  $-$45.43 & 6.96 &  APOGEE \\
  58037.940   &     43.85 & 1.81 &  $-$48.24 & 6.83 &  APOGEE \\
  55851.863   &     47.05 & 2.06 &  $-$52.09 & 6.70 &  APOGEE \\
  55847.869   &  $-$13.23 & 1.79 &     32.67 & 4.90 &  APOGEE \\
  55854.888   &     24.70 & 2.11 &  $-$21.97 & 4.88 &  APOGEE \\
  58802.24175 &     57.25 & 0.54 &  $-$63.47 & 6.07 &  LAMOST \\
  58820.15212 &  $-$34.00 & 0.64 &     56.41 & 3.39 &  LAMOST \\
  58830.14144 &  $-$39.16 & 0.53 &     60.12 & 4.71 &  LAMOST \\
  58836.03287 &     22.02 & 1.20 &  $-$30.52 & 7.53 &  LAMOST \\
  59124.32584 &  $-$26.57 & 0.63 &     51.88 & 2.78 &  LAMOST \\
  59148.24953 &  $-$35.19 & 0.56 &     61.57 & 2.50 &  LAMOST \\
  59181.12024 &  $-$44.08 & 0.52 &     72.66 & 5.97 &  LAMOST \\
  59190.15132 &      2.89 & 0.62 &      2.89 & 0.62 &  LAMOST \\	  
  59216.09454 &     43.83 & 0.71 &  $-$51.06 & 3.14 &  LAMOST \\
\noalign{\smallskip}
\hline \\
\end{tabular}
{\bf Notes.} The RV of the single CCF peak observed near the conjunctions has been assigned to both components.
\label{Tab:RV_HII761}
\end{table}

\begin{table}[ht]
\caption{Heliocentric RVs of HCG\,495.}
\begin{tabular}{lrcrcl}
\hline
\hline
\noalign{\smallskip}
 HJD     &  RV$_{\rm 1}$   &  $\sigma_{\rm RV_1}$ &  RV$_{\rm 2}$   &  $\sigma_{\rm RV_2}$ & Instrument \\
(2\,400\,000+) &  \multicolumn{2}{c}{(\kms)} & \multicolumn{2}{c}{(\kms)} & \\
 \hline
\noalign{\smallskip}
  57408.659	&     5.40  &  0.81 &	  5.40 & 0.81 & APOGEE \\ 
  57649.914	& $-$18.37  &  2.30 &	 30.83 & 2.64 & APOGEE \\
  57652.945	& $-$14.10  &  1.94 &	 25.76 & 2.18 & APOGEE \\ 
  58084.17582	&    57.74  &  0.80 & $-$49.17 & 0.88 & LAMOST \\
\noalign{\smallskip}
\hline \\
\end{tabular}
{\bf Notes.} The RV of the single CCF peak observed near the conjunctions has been assigned to both components.
\label{Tab:RV_HCG495}
\end{table}

\section{Discrepant objects}
\label{Appendix:discrepant}

The seven discrepant objects in Fig.~\ref{Fig:RV_comp} are \#1 (=\gaia-DR3\,64148738180119680), 
\#2 (=\gaia-DR3\,64928605459180416 = HII\,2406), \#3 (=\gaia-DR3\,65641913628380288 = V399~Tau), 
\#4 (=\gaia-DR3\,66771146429454592 = HII\,2172 = HD\,282965), \#5 (=\gaia-DR3\,66957994687842176 = V1091~Tau),  
\#6 (=\gaia-DR3\,69864313155605120 = HII\,571), and \#7  (=\gaia-DR3\,65292234570088064=HD\,23246).
 The objects \#2, \#4, and \#6 are already known binaries \citep[e.g.,][]{Torres2021}. They are further investigated in Sect.\,\ref{Sec:SB} and their RV curves are shown in Fig.\,\ref{Fig:RV_HII571}. For two of these sources (\#3 and \#6) there is only one \lamost\ spectrum. The remaining three have a few \lamost\ measures and do not show a clear RV variation in them. Further RV data are needed to check if they are SB1 systems. 

The stars with discrepant rotation periods with respect to \citet{Hartman2010}, which are discussed in Sect.~\ref{Sec:Prot}, are reported in Table~\ref{Tab:Discrepant_Prot}.

\renewcommand{\tabcolsep}{0.15cm}

\begin{table}[h]
\caption{Objects with discrepant rotation period with respect to \citet{Hartman2010}.}
\begin{tabular}{lcrrrr}
\hline
\hline
\noalign{\smallskip}
ID  & \gaia-DR3        &  TIC   & \prot$^P$  &  ref$^*$  & \prot$^H$ \\
\hline
\noalign{\smallskip}
\#1  & 65241313435901568 & 348639247 & 17.28 & R & 0.9407 \\
\#2  & 66808869124393600 & 125736708 & 7.18  & R & 0.8670 \\ 
\#3  & 66499154741977216 & 440691238 & 5.71  & R & 2.933  \\ 
\#4  & 64923279699744256 & 440690776 & 6.15  & R & 3.057  \\ 
\#5  & 69847610026687104 & 125736522 & 3.424 & R & 1.721  \\ 
\#6  & 66846291177113088 & 405489929 & 3.694 & R & 1.367  \\ 
\#7  & 65249250535404928 & 348639169 & 2.054 & R & 3.180  \\ 
\#8  & 66903293984399872 & 84331646  & 7.57  & R & 0.333  \\ 
\#9  & 65011412428446592 & 440681387 & 0.845 & R & 1.207  \\ 
\#10 & 65222759179728640 & 385552170 & 1.020 & R & 0.294  \\ 
\#11 & 66802654309459712 & 385552466 & 1.139 & R & 0.532  \\ 
\#12 & 69816346960886784 & 385552629 & 1.578 & R & 0.917  \\ 
\#13 & 65214409762926720 &  67829860 & 7.04  & T & 0.904  \\ 
\#14 & 66733552578791296 & 125736899 & 4.24  & T & 7.50   \\ 
\#15 & 64981931772948480 &  61139504 & 1.307 & T & 4.08   \\ 
\#16 & 64030785494725632 & 440686834 & 15.3  & T & 0.215  \\
\noalign{\smallskip}
\hline \\
\end{tabular}
{\bf Notes.}  \prot$^P$ = period adopted or measured in this work;  \prot$^H$ = period measured by \citet{Hartman2010}.\\   $^(*)$ R = \ktwo\ \citep{Rebull2016}; T = \tess\ (this work).
\label{Tab:Discrepant_Prot}
\end{table}

\end{appendix}

\end{document}